\documentclass[%
 reprint, twocolumn,
 superscriptaddress,
 amsmath,amssymb,
 aps,
 prl,
floatfix,
]{revtex4-2}

\usepackage{graphicx}
\usepackage{dcolumn}
\usepackage{bm}
\usepackage{xcolor}
\usepackage{hyperref}
\hypersetup{colorlinks=true,citecolor=blue,filecolor=blue,urlcolor=blue,}
\usepackage[noabbrev]{cleveref}

\usepackage{xparse} 

\newcommand{\mnras}{{\em Mon. Not. Roy. Astron. Soc. }}
\newcommand{\apjl}{{\em Astrophys. J. Let. }}
\newcommand{\apjs}{ApJS}
\newcommand{\jcap}{JCAP}

\newcommand{\aap}{{\em Astron. Astrophys. }}
\newcommand{\aj}{AJ }

\def\prd{{\em Phys. Rev. }{\bf D }}
\def\prl{{\em Phys. Rev. }{\bf L }}

\newcommand{\lya}{Ly$\alpha$}
\newcommand{\lyaf}{Ly$\alpha$ forest}
\newcommand{\lyb}{Ly$\beta$}

\newcommand{\lyalyalyalya}{Ly$\alpha$(Ly$\alpha$)$\times$Ly$\alpha$(Ly$\alpha$)}
\newcommand{\lyalyalyalyb}{Ly$\alpha$(Ly$\alpha$)$\times$Ly$\alpha$(Ly$\beta$)}
\newcommand{\lyalyaqso}{Ly$\alpha$(Ly$\alpha$)$\times$QSO}
\newcommand{\lyalybqso}{Ly$\alpha$(Ly$\beta$)$\times$QSO}

\newcommand{\hMpc}{\ $h^{-1}\text{Mpc}$}
\newcommand{\Hunits}{km/s/Mpc}

\newcommand{\lcdm}{{$\Lambda$CDM}}

\newcommand{\phip}{{$\phi_\mathrm{p}$}}
\newcommand{\phis}{{$\phi_\mathrm{s}$}}
\newcommand{\phif}{{$\phi_f$}}
\newcommand{\alphap}{{$\alpha_\mathrm{p}$}}
\newcommand{\alphas}{{$\alpha_\mathrm{s}$}}

\newcommand{\vega}{\texttt{Vega}}
\newcommand{\polychord}{\texttt{PolyChord}}

\begin{document}


\title{Constraints on the cosmic expansion rate at redshift 2.3 from the Lyman-$\alpha$ forest}

\author{Andrei Cuceu}
 \email{cuceu.1@osu.edu}
 \affiliation{Center for Cosmology and Astro-Particle Physics, The Ohio State University, Columbus, Ohio 43210, USA}
 \affiliation{Department of Astronomy, The Ohio State University, Columbus, Ohio 43210, USA}
 \affiliation{Department of Physics, The Ohio State University, Columbus, Ohio 43210, USA}
 \affiliation{Department of Physics and Astronomy, University College London, Gower Street, London WC1E 6BT, United Kingdom}
\author{Andreu Font-Ribera}%
 \affiliation{Institut de Física d’Altes Energies, The Barcelona Institute of Science and Technology, Campus UAB, 08193 Bellaterra (Barcelona), Spain}
 \affiliation{Department of Physics and Astronomy, University College London, Gower Street, London WC1E 6BT, United Kingdom}
\author{Seshadri Nadathur}%
 \affiliation{Institute of Cosmology and Gravitation, University of Portsmouth, Burnaby Road, Portsmouth, PO1 3FX, United Kingdom}
\author{Benjamin Joachimi}%
 \affiliation{Department of Physics and Astronomy, University College London, Gower Street, London WC1E 6BT, United Kingdom}
\author{Paul Martini}%
 \affiliation{Department of Astronomy, The Ohio State University, Columbus, Ohio 43210, USA}
 \affiliation{Center for Cosmology and Astro-Particle Physics, The Ohio State University, Columbus, Ohio 43210, USA}

\date{\today}

\begin{abstract}
We determine the product of the expansion rate and angular-diameter distance at redshift $z=2.3$ from the anisotropy of Lyman-$\alpha$ (\lya) forest correlations measured by the Sloan Digital Sky Survey (SDSS). Our result is the most precise from large-scale structure at $z>1$. In flat \lcdm\ we determine the matter density to be $\Omega_\mathrm{m}=0.36^{+0.03}_{-0.04}$ from \lya\ alone. This is a factor of two tighter than baryon acoustic oscillation results from the same data due to our use of a wide range of scales ($25<r<180$\hMpc). Using a nucleosynthesis prior, we measure the Hubble constant to be $H_0=63.2\pm2.5$ \Hunits. In combination with other SDSS tracers, we find $H_0=67.2\pm0.9$ \Hunits\ and measure the dark energy equation-of-state parameter to be $w=-0.90\pm0.12$. Our work opens a new avenue for constraining cosmology at high redshift.

\end{abstract}

\maketitle


\emph{Introduction.}---Over the last few decades the $\Lambda$ cold dark matter model (\lcdm) has become the standard model of cosmology. However, increased precision in cosmological measurements gave rise to tensions between different probes, highlighting potential shortcomings in the standard model. The most discussed of these is the $\sim5\sigma$ discrepancy between the value of the Hubble constant, $H_0$, inferred from cosmic microwave background (CMB) measurements by the Planck satellite \cite{Planck:2020} (which assume \lcdm), and local measurements using the cosmic distance ladder \cite{Riess:2021}. Furthermore, the most important component of \lcdm\ at the present time, dark energy, is still not theoretically understood. On the observational side, the main way to address these challenges is to measure the expansion rate with greater precision through different probes, and at different stages of the evolution of the Universe.

Probes of large-scale structure (LSS) currently provide some of the tightest constraints on the expansion rate after recombination (e.g. \cite{Alam:2017,GilMarin:2020,Bautista:2021,Hou:2021,Neveux:2020,Alam:2021,Abbott:2022,DES:2022,Nadathur:2020,Brieden:2022}). While many of these measurements focus on the baryon acoustic oscillation (BAO) feature as a standard ruler \cite{Eisenstein:2005,Cole:2005}, the clustering of LSS tracers contains important information beyond BAO. At low redshift, $z < 1.5$, analyses of the two-point statistics of galaxies improve on BAO constraints by using information from a wider range of scales, at the expense of including more model assumptions (e.g. \cite{Satpathy:2017,Beutler:2017,GilMarin:2020,Bautista:2021,Hou:2021,Neveux:2020}). Usually these analyses take advantage of the Alcock-Paczy\'nski (AP; \cite{Alcock:1979}) effect, the power of which has been demonstrated using voids by \cite{Nadathur:2019,Nadathur:2020}. This effect adds an anisotropy on all scales in the LSS distribution if the fiducial cosmology used to compute comoving distances from angles and redshifts is different from the truth. Therefore, a measurement of this apparent anisotropy can help determine the true cosmology. This effect is usually also measured in BAO analyses, but only using a small fraction of the available information (around the BAO peak). At high redshifts, $z > 1.5$, where the Lyman-$\alpha$ (\lya) forest is used as a continuous tracer of the LSS, the best constraints currently come from the BAO scale alone \cite{duMasdesBourboux:2020,Alam:2021}.

We measure the AP effect for the first time using a broad range of scales in the three-dimensional \lyaf\ correlation functions. We use data from the Sloan Digital Sky Survey (SDSS) data release 16 (DR16; \cite{Ahumada:2020}), which includes measurements from the Baryon Oscillation Spectroscopic Survey (BOSS; \cite{Eisenstein:2011}), and its successor, extended BOSS (eBOSS; \cite{Dawson:2016}). Furthermore, we examine the cosmological constraints derived from this measurement alone and in combination with other SDSS tracers at lower redshift. 

\emph{Methods and data.}---We use the \lyaf\ three-dimensional correlation functions computed by the eBOSS collaboration using SDSS Data Release 16 (DR16; \cite{duMasdesBourboux:2020}), and focus on extracting the AP information from the anisotropy in these correlations. Our tracers include \lya\ flux in the \lya\ region (between the \lya\ and \lyb\ peaks), denoted \lya(\lya); \lya\ flux in the Lyman-$\beta$ (\lyb) region (blue-ward of the \lyb\ peak), denoted \lya(\lyb); and the quasar distribution (QSO; \cite{Lyke:2020}). In total we use four correlation functions, which for modelling purposes are categorised into two \lya\ auto-correlations, \lyalyalyalya\ and \lyalyalyalyb, and two \lya-QSO cross-correlations, \lyalyaqso\ and \lyalybqso. We perform a joint fit of the full shapes of these four correlations in order to extract the AP effect. We use the term \textit{full-shape} to refer to the extraction of cosmological information from a broad range of scales in the correlation function (that includes BAO), rather than just the BAO peak alone.

Our model of the \lya\ correlation functions broadly follows the framework used by BOSS and eBOSS in past BAO analyses. We use a template approach based on a fiducial cosmology, starting from an isotropic linear matter power spectrum, computed using CAMB \cite{Lewis:1999}. This template is decomposed into a peak and a smooth component following \cite{Kirkby:2013}. The \lya\ modelling is applied independently to the two components, and they are combined only at the end to build the full correlation model. The key ingredients of this model are the linear \lya\ and QSO biases and redshift space distortions (RSD), along with the AP effect (see \cite{Kirkby:2013,duMasdesBourboux:2020,Cuceu:2022}).

We compute the correlation models and the Gaussian likelihood with the \vega\ package \footnote{\url{https://github.com/andreicuceu/vega}}, using the same models and parameters for the contaminants as in eBOSS \cite{duMasdesBourboux:2020,Cuceu:2022}. See the supplemental material for more details on the data, model and likelihood, which includes Refs. \cite{Gorski:2005,Bautista:2017,duMasdesBourboux:2017,Bourboux:2021,Ramirez:2022,Farr:2020,deSainteAgathe:2019,Arinyo:2015,Givans:2022,Percival:2009,Cuceu:2021,Gontcho:2014,Rogers:2018,FontRibera:2013,Youles:2022,Abareshi:2022,Torrado:2019,Torrado:2021}. We sample posterior distributions using the Nested Sampler \polychord\ \cite{Handley:2015a,Handley:2015b}.

In BAO analyses, the coordinates of the peak component are allowed to vary anisotropically, in order to fit both the BAO scale and the AP effect from the peak position. Here we also vary the coordinates of the smooth component in order to extract AP from the full correlation. As the AP effect introduces an anisotropy in the correlation function, we follow \cite{Cuceu:2021} and introduce the parameters:
\begin{equation}
    \phi(z) \equiv \frac{q_\bot(z)}{q_{||}(z)} \;\;\text{and}\;\; \alpha(z) \equiv \sqrt{q_\bot(z)q_{||}(z)},
\end{equation}
where $q_{||}$ and $q_\bot$ rescale the comoving coordinates, $r_{||}$ and $r_\bot$, along and across the line-of-sight, respectively. $\phi$ measures the anisotropy, and therefore the AP effect, while $\alpha$ measures the isotropic scale \footnote{Note that the choice of isotropic scale parameter is an arbitrary one and not based on arguments about the optimal parameter to measure.}. As we have two components, we have the option of using two sets of these parameters, one for the peak component, (\phip, \alphap), and one for the smooth component, (\phis, \alphas). This is appropriate for $\alpha$, where the cosmological information comes from the BAO scale (\alphap), while \alphas\ is treated as a nuisance. On the other hand, both \phip\ and \phis\ measure the same effect (AP), and therefore our baseline analysis uses one coherent parameter for both components, denoted \phif. However, the two $\phi$ parameters are still useful for distinguishing the BAO measurement from the broadband information, as they are affected by different contaminants \cite{Cuceu:2021}.

The main contaminants affecting \lyaf\ correlations are high column density (HCD) absorbers, metal absorbers, and the distortion due to quasar continuum fitting. We model these contaminants following the approach used by BOSS and eBOSS BAO analyses \cite{Bautista:2017,duMasdesBourboux:2017,duMasdesBourboux:2020}. Following \cite{duMasdesBourboux:2020}, we model deviations from linear theory in the \lya\ auto-correlation using the multiplicative correction proposed by \cite{Arinyo:2015}. For the \lya-QSO cross-correlation, we also follow \cite{duMasdesBourboux:2020}, and use a Lorentzian damping term \cite{Percival:2009} to model both the redshift errors and the Finger-of-God effect in the cross-correlation.

In order to validate our method, we performed a detailed analysis on synthetic data using one hundred eBOSS mock realizations. The results on synthetic data and a detailed description of our methodology are presented in \cite{Cuceu:2022} (also see supplemental material). The main \lyaf\ contaminants are simulated in this synthetic data, and we found that our method results in unbiased AP measurements when fitting scales between $25$\hMpc\ and $180$\hMpc\ \cite{Cuceu:2022}. Based on these results, we use a minimum scale $r_\mathrm{min}=25$\hMpc. This restricts our analysis to large scales, reducing the impact of non-linearities.

We discussed additional sources of contamination, not currently included in state-of-the-art \lyaf\ mocks, in \cite{Cuceu:2022}. The most important of these are deviations from linear theory, and realistic redshift errors \cite{Cuceu:2022}. \lyaf\ non-linearities have been recently studied by \cite{Givans:2022} using hydrodynamical simulations, and in the case of the \lya\ auto-correlation were found to be well modelled by the empirical relation introduced by \cite{Arinyo:2015}. On the other hand, for the \lya-QSO cross-correlation the QSO redshift errors generally dominate on small scales. We discuss this in more detail in the supplemental material.

We also tested the sensitivity of our result to different analysis choices, focusing on robustness tests for \phis, as tests for \alphap\ and \phip\ (using a different parametrization) were done by \cite{duMasdesBourboux:2020}. We blinded our analysis by adding a random value to our \phis\ measurement, such that we did not know the true result until we chose the exact configuration of our baseline model. We tested a variety of modelling options, presented in the supplemental material, and found that our result is robust to these changes. The only noteworthy deviation happens when changing the model for QSO redshift errors. While our baseline analysis uses a Lorentzian damping term, we found a $\sim0.5\sigma$ shift in our \phis\ measurement when using a Gaussian damping term instead. As the distribution of quasar redshift errors has long tails \cite{Lyke:2020}, we followed \cite{duMasdesBourboux:2020} and chose the Lorentzian damping in our baseline analysis.

\begin{figure}
\includegraphics[width=1.0\linewidth]{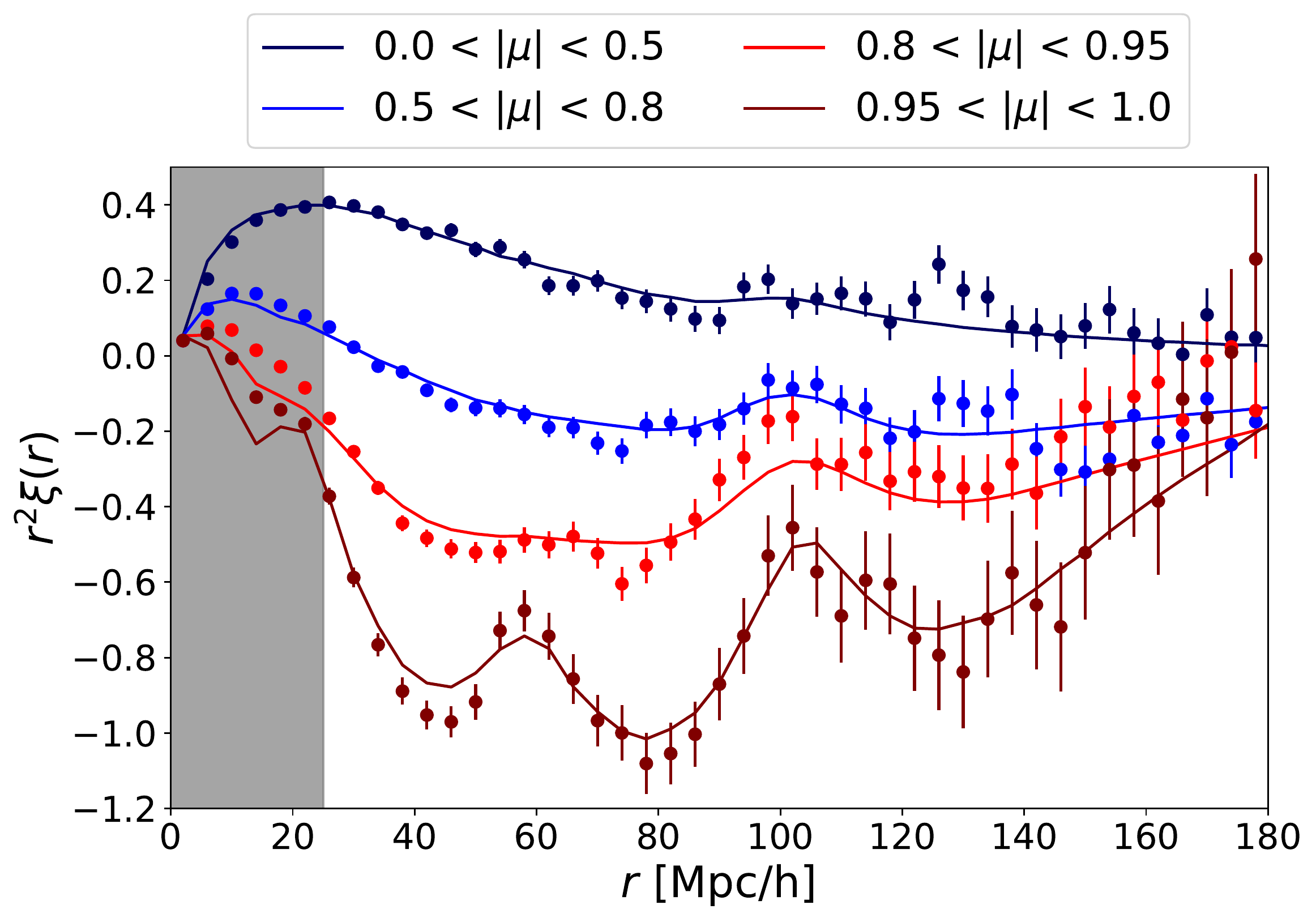}
\caption{The eBOSS \lyaf\ auto-correlation function $\xi$ (points with errorbars) and the best-fit model, as a function of comoving separation, $r$. For illustration, the correlation function is compressed into four bins in the cosine of the line-of-sight angle, $\mu$. The grey area shows the small scales which we exclude from the fit. The Alcock-Paczy\'nski effect is extracted from the anisotropy in the correlation (its dependence on $\mu$).}
\label{fig:auto_corr}
\end{figure}

The joint fit of the four \lya\ correlation functions using the baseline model produces a good fit, with $\chi^2_\mathrm{min}=9391.85$ for $9363$ degrees of freedom ($\mathrm{PTE}=0.41$). The fit quality is consistent between mocks and data, with $38\%$ of mock realisations giving a lower $\chi^2$ value. \Cref{fig:auto_corr} shows the data and best-fit model for our main correlation, \lyalyalyalya. As our data and model are 2D functions of $r_{||}$ and $r_\bot$, we compress them into four $\mu=r_{||}/r$ bins, and show them as a function of isotropic separation $r=\sqrt{r_{||}^2+r_\bot^2}$, for visualization purposes.

\begin{figure}
\includegraphics[width=1\linewidth]{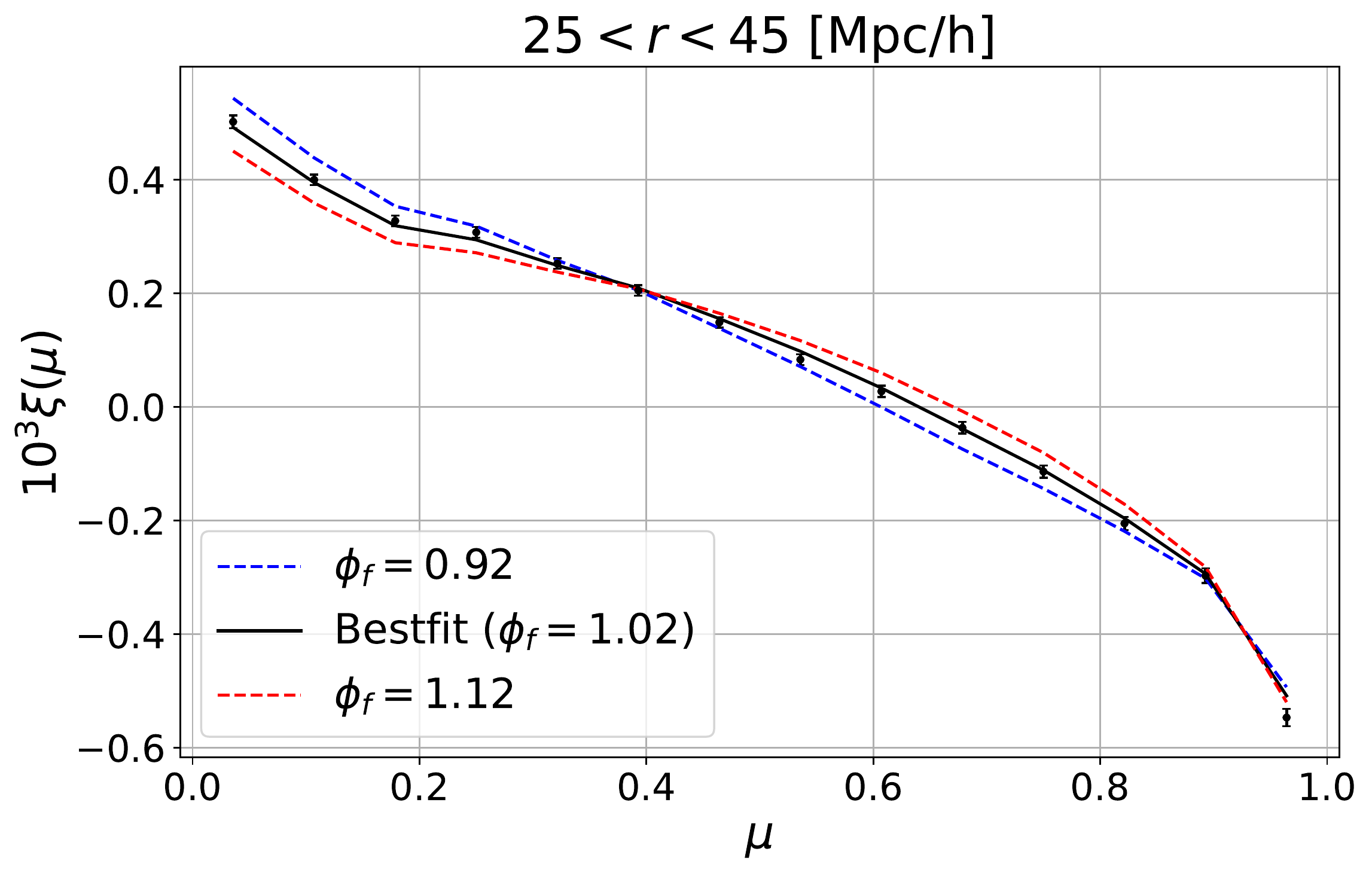}
\caption{\lyaf\ auto-correlation function (points with errorbars) in a shell at separation $r$, shown as a function of the cosine of the line-of-sight angle, $\mu$. We compare the best fit model (black) with two other models where the \phif\ parameter, which measures the AP effect relative to the fiducial cosmology, takes different values (red and blue). This illustrates how the AP effect is constrained.}
\label{fig:shell_phi}
\end{figure}

To better understand the AP effect and where this information comes from, we also compress the correlation function into shells in $r$, and plot them as a function of $\mu$. One such example is shown in \Cref{fig:shell_phi}, where we choose the smallest separation shell ($25<r<45$\hMpc) of the \lyalyalyalya\ correlation. This figure illustrates how the model changes for different values of \phif, showing that the AP information comes in large part from intermediate and small values of $\mu$, rather than the line-of-sight region ($\mu=1$) where most \lyaf\ contaminants have an impact. Shells for larger separations and also showing the impact of RSD are included in the supplemental material.

\emph{Results}---After unblinding, we can compare $\phi$ results when fitting two separate parameters for the peak and smooth components (\phip\ and \phis), and when we only fit one shared parameter (\phif). The main results of this work are the AP measurements from the broadband (\phis), and the full-shape measurement (\phif). The three results are:
\begin{align}
    \text{BAO AP: }& \phi_\mathrm{p} = 0.933 \pm 0.041, \\
    \text{Broadband AP: }& \phi_\mathrm{s} = 1.037 \pm 0.020, \\
    \text{Full shape AP: }& \phi_{f} = 1.021 \pm 0.019.
\end{align}

As our fiducial cosmology is based on Planck CMB results, $\phi=1$ represents the best-fit Planck cosmology. \lya\ BAO measurements from BOSS and eBOSS have resulted in $\phi$ values smaller than $1$ \cite{duMasdesBourboux:2020}. Our BAO AP measurement is $1.7\sigma$ lower than the Planck best-fit value, consistent with the result obtained by \cite{duMasdesBourboux:2020} using the same dataset and matching scale cuts. On the other hand, our broadband AP result gives a value $1.9\sigma$ higher than Planck \footnote{Tension metrics were computed from the full posterior distributions by summing over probability densities.}. 

While individually both \phip\ and \phis\ are consistent with Planck, they are in $2.3\sigma$ tension with each other. Our result for \phis\ is robust to changes in the modelling (see Supplemental Material). The biggest shift, coming from changing the model for QSO redshift errors, only reduces the tension with BAO from $2.3$ to $2.0\sigma$. As our baseline model for the errors is the more realistic, we kept the pre-unblinding model choices when inferring cosmology. 


Our final measurement is given by the \phif\ constraint, which combines BAO and broadband information. Fitting the full-shape of \lya\ correlation functions results in a factor of two improvement in constraining power over the BAO-only constraint measured by \cite{duMasdesBourboux:2020}.

In addition to \phif, we also infer cosmology from the isotropic BAO scale, which we measure to be $\alpha_\mathrm{p}=1.017\pm0.014$. This differs slightly from the value reported by \cite{duMasdesBourboux:2020}, $\alpha_\mathrm{p}=1.000\pm0.014$, due to the use of different scale cuts and the correlation between \phif\ and \alphap\ (Pearson correlation coefficient of $0.21$). The two-dimensional posterior of these parameters is well approximated by a multivariate Gaussian, and therefore, we use this Gaussian result in the analysis below.

\emph{Inferring cosmology}---Our \phif\ and \alphap\ measurements represent ratios between distances measured using the assumed fiducial cosmological model and the true cosmology. Following \cite{Cuceu:2021}, these are given by:
\begin{align}
    \mathrm{AP}: \phi &= \frac{D_\mathrm{M}(z) H(z)}{\big[D_\mathrm{M}(z) H(z)\big]_\mathrm{fid}}, \label{eq:phi} \\
    \mathrm{BAO}: \alpha &= \sqrt{\frac{D_\mathrm{M}(z) D_H(z)/r_d^2}{\big[D_\mathrm{M}(z) D_H(z)/r_d^2\big]_\mathrm{fid}}}, \label{eq:alpha}
\end{align}
where $D_\mathrm{M}$ is the transverse comoving distance, $H(z)$ is the Hubble parameter, $r_d$ is the size of the acoustic scale, $D_H=c/H$, with the speed of light, $c$. The effective redshift of our measurement is $z=2.33$. We use the Gaussian likelihood in $\phi$ and $\alpha$ above together with \Cref{eq:phi,eq:alpha} to derive constraints on cosmological parameters through Monte Carlo Markov Chain (MCMC) sampling of model posteriors with \texttt{Cobaya} \cite{Torrado:2019,Torrado:2021}. 

For completeness, we also report our measurement in terms of the usual ratios: $D_\mathrm{H}(z=2.33)/r_d=8.65\pm0.13$ and $D_\mathrm{M}(z=2.33)/r_d=40.26\pm0.71$. Using the Planck measured value of $r_d=147.05\pm0.30$ Mpc \cite{Planck:2020}, we determine the expansion rate at redshift $2.33$ to be $H(z=2.33)=235.9\pm3.5\text{ km s}^{-1}\,\text{Mpc}^{-1}$.

We also compare and combine our measurement with results from other datasets, including the full-shape results from lower-redshift SDSS tracers and CMB anisotropy measurements by Planck. We sample the posteriors for combinations of these likelihoods using parameter choices and priors following \cite{Alam:2021}. For SDSS, we use the ``BAO-plus" likelihoods which include BAO and full-shape information from BOSS and eBOSS, including the latest data from DR16 \cite{Alam:2021}. We also use the public Planck chains \footnote{\url{https://pla.esac.esa.int//#cosmology}} where appropriate for comparison purposes.

\emph{Flat \lcdm.}---In flat \lcdm, measurements of the isotropic BAO scale (\alphap) constrain a combination of $\Omega_\mathrm{m}$ and $H_0 r_d$. This degeneracy can be broken by the AP measurement of $\phi$, which directly translates to a constraint on $\Omega_\mathrm{m}$. Our \lya\ full-shape result, \phif, corresponds to $\Omega_\mathrm{m}=0.36^{+0.03}_{-0.04}$. This is a factor of two higher precision than obtained from previous BAO-only results \cite{duMasdesBourboux:2020} and produces our substantially improved cosmological constraints.

In order to measure the Hubble constant, $H_0$, from the degenerate combination $H_0 r_d$, we require the value of $r_d$. This depends on the total matter density, the baryon density and the neutrino density \cite{Aubourg:2015}. We fix the neutrino density by assuming degenerate mass eigenstates with total mass $\Sigma m_\nu=0.06$ eV$/c^2$, and assume a prior on the baryon density, $\Omega_\mathrm{b}h^2 = 0.02233\pm0.00036$, based on measurements of the primordial deuterium to hydrogen ratio and big bang nucleosynthesis (BBN; \cite{Cooke:2018,Mossa:2020}). With these priors, we measure $H_0$ from our results of \alphap\ and \phif, independently of information from Planck CMB anisotropies.

\begin{figure}
\includegraphics[width=1.0\linewidth]{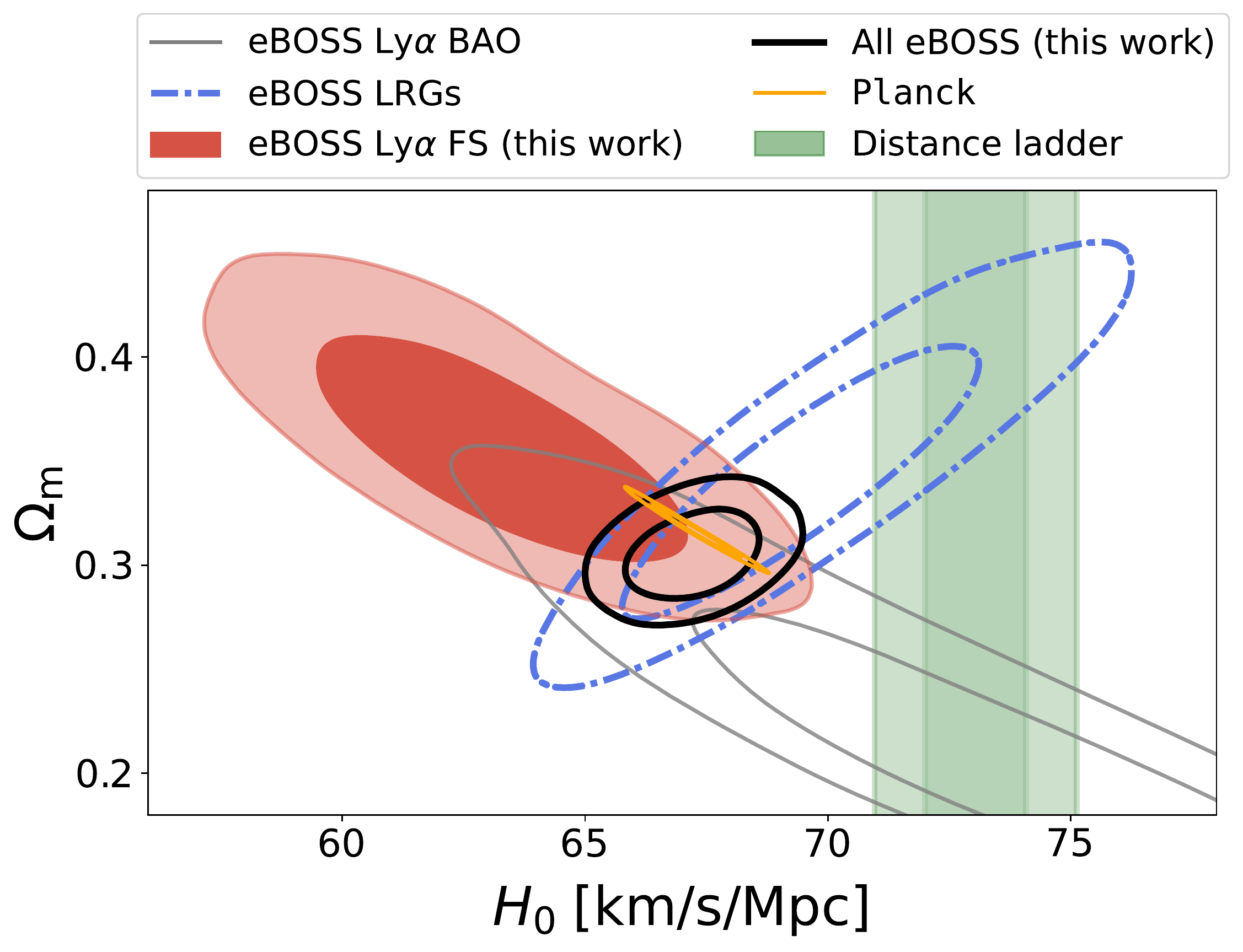}
\caption{Marginalized $68\%$ and $95\%$ contours on $\Omega_\mathrm{m}$ and $H_0$, assuming \lcdm. The eBOSS results assume a BBN prior on the baryon density. These results are independent of CMB anisotropy measurements, but are consistent with Planck and in tension with the distance ladder results of \cite{Riess:2021}.}
\label{fig:lcdm}
\end{figure}

In \Cref{fig:lcdm} we show posterior distributions of $\Omega_\mathrm{m}$ and $H_0$. The two main cosmological constraints in this work are:
\begin{align}
    \text{Ly}\alpha\text{ FS: } H_0 &= 63.2 \pm 2.5 \text{ km s}^{-1}\,\text{Mpc}^{-1},\\
    \text{All eBOSS: } H_0 &= 67.23 \pm 0.91 \text{ km s}^{-1}\,\text{Mpc}^{-1}.
\end{align}
\lyaf\ constraints alone result in a tight measurement of $H_0$, comparable to that from the most powerful single probe in SDSS, luminous red galaxies (LRG). Interestingly, we obtain a low measurement of the Hubble constant, which is in $3.6\sigma$ tension with the direct result from the distance ladder \cite{Riess:2021}, but still compatible with Planck at the $1.6\sigma$ level. When combining our measurement with results from other eBOSS tracers, we obtain a tight posterior that is compatible with Planck and in $4.2\sigma$ tension with the distance ladder. Finally, the improvement in constraining power when performing a full-shape analysis of the \lya\ correlations can be seen from the difference between the grey and the red posteriors in \Cref{fig:lcdm}.

\begin{figure}
\includegraphics[width=1.0\linewidth]{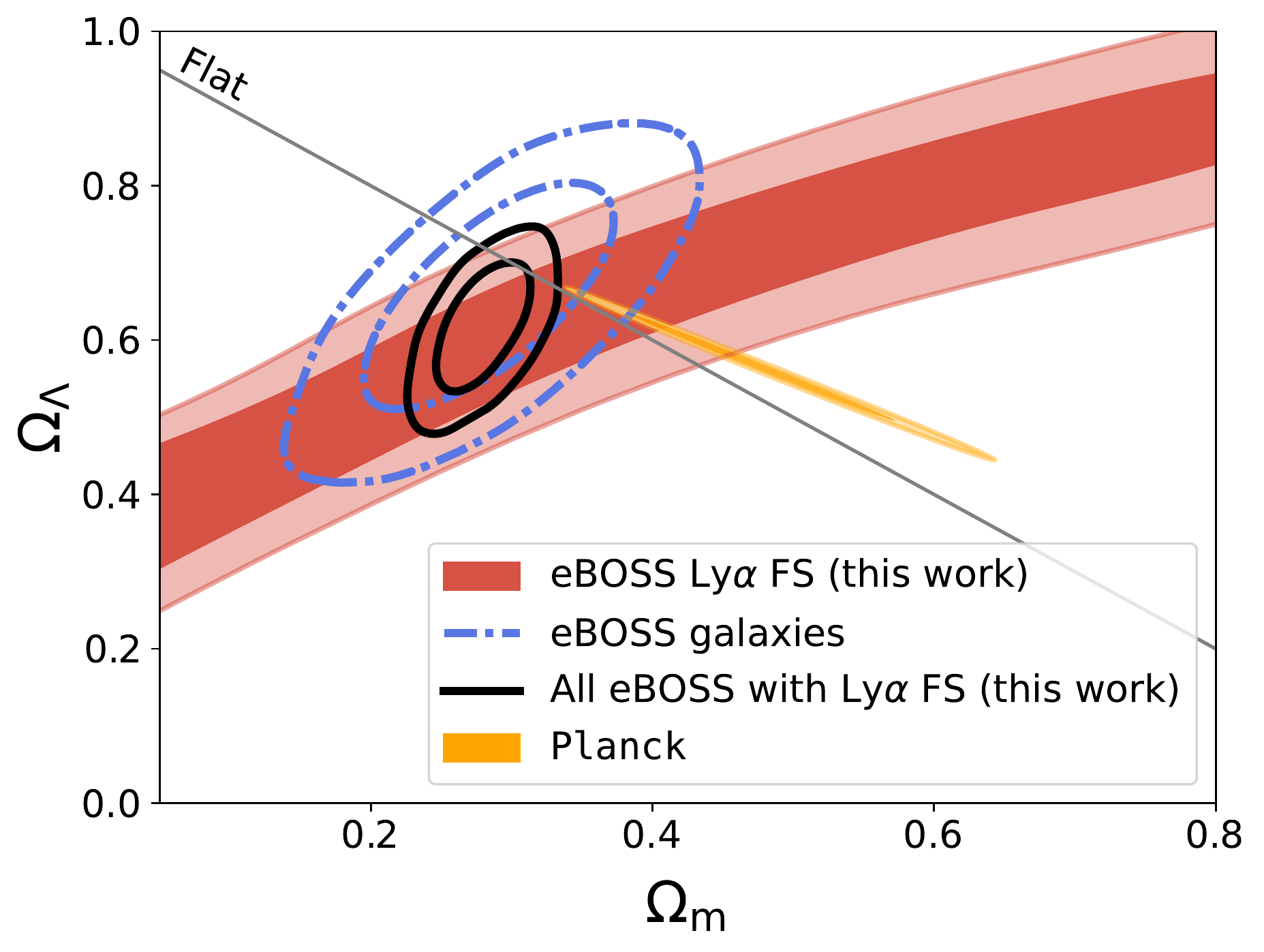}
\caption{Marginalized constraints on the dark energy density parameter, $\Omega_\Lambda$, versus $\Omega_\mathrm{m}$ from \lya, eBOSS galaxies, and Planck, under the assumption that dark energy is described by the cosmological constant. The addition of the \lya\ measurement improves on the eBOSS galaxies constraint by a factor of two and results in a $>12\sigma$ detection of the late-time accelerated expansion of the Universe.}
\label{fig:omega_lambda}
\end{figure}

\emph{Dark energy and curvature.}---We next relax the assumption that the Universe is flat and allow the curvature parameter, $\Omega_\mathrm{k}$, to vary. We still assume dark energy is described by the cosmological constant, which means its equation of state parameter is $w=-1$. We show our measurements of $\Omega_\Lambda$ versus $\Omega_\mathrm{m}$ in \Cref{fig:omega_lambda}, where we do not include any external prior, as we directly sample the combination $H_0 r_d$. The \lyaf\ constrains a degenerate posterior between $\Omega_\mathrm{m}$ and $\Omega_\Lambda$ that nevertheless excludes $\Omega_\Lambda=0$ at $>4\sigma$ level.

When combined with the other eBOSS tracers, we measure $\Omega_\Lambda = 0.62^{+0.06}_{-0.05}$, and $\Omega_\mathrm{k} = 0.10 \pm 0.07$. Our results are compatible with the Universe today being flat and dominated by dark energy. We find direct geometrical evidence of late-time acceleration due to dark energy at the $>12\sigma$ level.

\begin{figure}
\includegraphics[width=1.0\linewidth]{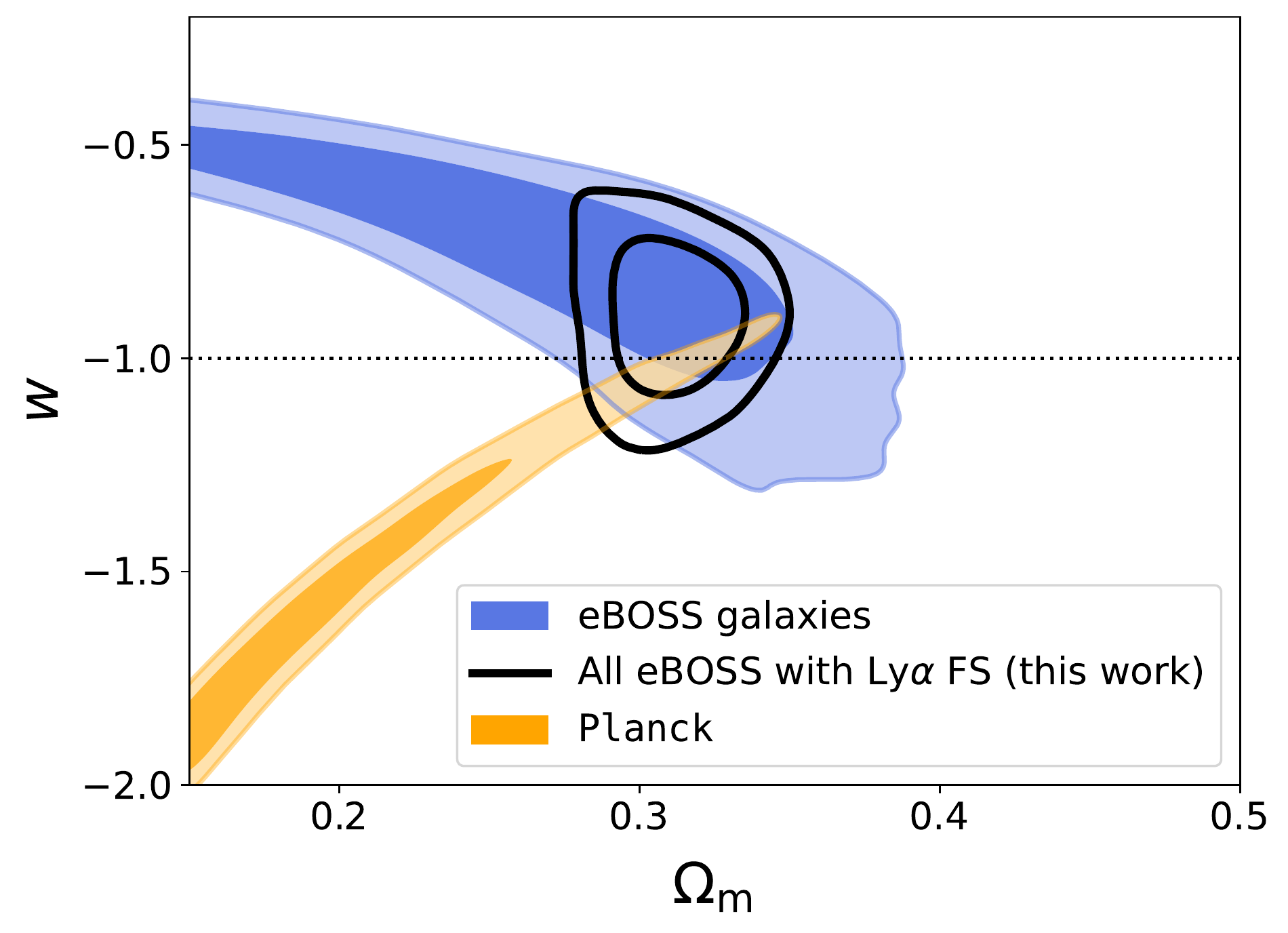}
\caption{Constraints on the dark energy equation of state parameter $w$ from eBOSS galaxies alone and combined with our \lya\ full-shape measurement. Adding \lya\ results in a $\sim12\%$ measurement of $w$ from SDSS alone. This is independent of Planck constraints.}
\label{fig:wcdm}
\end{figure}

Finally, we consider models where $w$ is allowed to vary, and $\Omega_\mathrm{k}$ is fixed to zero. We present our results in \Cref{fig:wcdm}, where we focus on SDSS measurements. The eBOSS galaxies full-shape measurements result in a partially degenerate posterior between $w$ and $\Omega_\mathrm{m}$. Adding our \lya\ full-shape measurement breaks this degeneracy and we measure $w = -0.90\pm0.12$, consistent with the late-time acceleration being caused by a cosmological constant dark energy. We emphasize that this result only depends on SDSS data, which sets this experiment apart in its ability to constrain dark energy without needing other datasets. 

\emph{Conclusions.}---We present the first cosmological measurement from a broad range of spatial scales in the Lyman-$\alpha$ (\lya) forest three-dimensional correlations, through the Alcock-Paczy\'nski effect. Using eBOSS DR16 data, we obtain the tightest cosmological constraints to date from large-scale structure at $z>1$, shown through our constraints on $H_0$ and the late-time accelerated expansion.

Key areas of future improvement include better modelling of quasar redshift errors and a better understanding of the impact of non-linearities. Furthermore, \lyaf\ correlation functions could also be used to measure the growth of cosmic structure as proposed by \cite{Cuceu:2021,Gerardi:2022}.  Our measurement opens a new avenue for constraining cosmology at high redshifts ($2 < z < 4$) with future surveys such as the ongoing Dark Energy Spectroscopic Instrument \cite{Abareshi:2022}.\newline

\begin{acknowledgments}
We thank James Rich, David H. Weinberg and Naim G{\" o}ksel Kara{\c c}ayl{\i} for useful discussion and comments during the preparation of this manuscript. AC and PM acknowledge support from the United States Department of Energy, Office of High Energy Physics under Award Number DE-SC-0011726. AFR acknowledges support through the program Ramon y Cajal (RYC-2018-025210) of the Spanish Ministry of Science and Innovation and from the European Union’s Horizon Europe research and innovation programme (COSMO-LYA, grant agreement 101044612). IFAE is partially funded by the CERCA program of the Generalitat de Catalunya.
SN acknowledges support from an STFC Ernest Rutherford Fellowship, grant reference ST/T005009/2. BJ acknowledges support by STFC Consolidated Grant ST/V000780/1. We acknowledge the use of the \texttt{GetDist} \cite{Lewis:2019} and \texttt{numpy} \cite{Harris:2020} packages. For the purpose of open access, the authors have applied a CC BY public copyright licence to any Author Accepted Manuscript version arising.

\end{acknowledgments}

\bibliography{main}

\begin{thebibliography}{58}%
\makeatletter
\providecommand \@ifxundefined [1]{%
 \@ifx{#1\undefined}
}%
\providecommand \@ifnum [1]{%
 \ifnum #1\expandafter \@firstoftwo
 \else \expandafter \@secondoftwo
 \fi
}%
\providecommand \@ifx [1]{%
 \ifx #1\expandafter \@firstoftwo
 \else \expandafter \@secondoftwo
 \fi
}%
\providecommand \natexlab [1]{#1}%
\providecommand \enquote  [1]{``#1''}%
\providecommand \bibnamefont  [1]{#1}%
\providecommand \bibfnamefont [1]{#1}%
\providecommand \citenamefont [1]{#1}%
\providecommand \href@noop [0]{\@secondoftwo}%
\providecommand \href [0]{\begingroup \@sanitize@url \@href}%
\providecommand \@href[1]{\@@startlink{#1}\@@href}%
\providecommand \@@href[1]{\endgroup#1\@@endlink}%
\providecommand \@sanitize@url [0]{\catcode `\\12\catcode `\$12\catcode
  `\&12\catcode `\#12\catcode `\^12\catcode `\_12\catcode `\%12\relax}%
\providecommand \@@startlink[1]{}%
\providecommand \@@endlink[0]{}%
\providecommand \url  [0]{\begingroup\@sanitize@url \@url }%
\providecommand \@url [1]{\endgroup\@href {#1}{\urlprefix }}%
\providecommand \urlprefix  [0]{URL }%
\providecommand \Eprint [0]{\href }%
\providecommand \doibase [0]{https://doi.org/}%
\providecommand \selectlanguage [0]{\@gobble}%
\providecommand \bibinfo  [0]{\@secondoftwo}%
\providecommand \bibfield  [0]{\@secondoftwo}%
\providecommand \translation [1]{[#1]}%
\providecommand \BibitemOpen [0]{}%
\providecommand \bibitemStop [0]{}%
\providecommand \bibitemNoStop [0]{.\EOS\space}%
\providecommand \EOS [0]{\spacefactor3000\relax}%
\providecommand \BibitemShut  [1]{\csname bibitem#1\endcsname}%
\let\auto@bib@innerbib\@empty
\bibitem [{\citenamefont {{Planck Collaboration}}\ \emph
  {et~al.}(2020)\citenamefont {{Planck Collaboration}}, \citenamefont
  {{Aghanim}}, \citenamefont {{Akrami}}, \citenamefont {{Ashdown}},
  \citenamefont {{Aumont}}, \citenamefont {{Baccigalupi}}, \citenamefont
  {{Ballardini}}, \citenamefont {{Banday}}, \citenamefont {{Barreiro}},
  \citenamefont {{Bartolo}}, \citenamefont {{Basak}}, \citenamefont {{Battye}},
  \citenamefont {{Benabed}}, \citenamefont {{Bernard}}, \citenamefont
  {{Bersanelli}}, \citenamefont {{Bielewicz}}, \citenamefont {{Bock}},
  \citenamefont {{Bond}}, \citenamefont {{Borrill}}, \citenamefont {{Bouchet}},
  \citenamefont {{Boulanger}}, \citenamefont {{Bucher}}, \citenamefont
  {{Burigana}}, \citenamefont {{Butler}}, \citenamefont {{Calabrese}},
  \citenamefont {{Cardoso}}, \citenamefont {{Carron}}, \citenamefont
  {{Challinor}}, \citenamefont {{Chiang}}, \citenamefont {{Chluba}},
  \citenamefont {{Colombo}}, \citenamefont {{Combet}}, \citenamefont
  {{Contreras}}, \citenamefont {{Crill}}, \citenamefont {{Cuttaia}},
  \citenamefont {{de Bernardis}}, \citenamefont {{de Zotti}}, \citenamefont
  {{Delabrouille}}, \citenamefont {{Delouis}}, \citenamefont {{Di Valentino}},
  \citenamefont {{Diego}}, \citenamefont {{Dor{\'e}}}, \citenamefont
  {{Douspis}}, \citenamefont {{Ducout}}, \citenamefont {{Dupac}}, \citenamefont
  {{Dusini}}, \citenamefont {{Efstathiou}}, \citenamefont {{Elsner}},
  \citenamefont {{En{\ss}lin}}, \citenamefont {{Eriksen}}, \citenamefont
  {{Fantaye}}, \citenamefont {{Farhang}}, \citenamefont {{Fergusson}},
  \citenamefont {{Fernandez-Cobos}}, \citenamefont {{Finelli}}, \citenamefont
  {{Forastieri}}, \citenamefont {{Frailis}}, \citenamefont {{Fraisse}},
  \citenamefont {{Franceschi}}, \citenamefont {{Frolov}}, \citenamefont
  {{Galeotta}}, \citenamefont {{Galli}}, \citenamefont {{Ganga}}, \citenamefont
  {{G{\'e}nova-Santos}}, \citenamefont {{Gerbino}}, \citenamefont {{Ghosh}},
  \citenamefont {{Gonz{\'a}lez-Nuevo}}, \citenamefont {{G{\'o}rski}},
  \citenamefont {{Gratton}}, \citenamefont {{Gruppuso}}, \citenamefont
  {{Gudmundsson}}, \citenamefont {{Hamann}}, \citenamefont {{Handley}},
  \citenamefont {{Hansen}}, \citenamefont {{Herranz}}, \citenamefont
  {{Hildebrandt}}, \citenamefont {{Hivon}}, \citenamefont {{Huang}},
  \citenamefont {{Jaffe}}, \citenamefont {{Jones}}, \citenamefont {{Karakci}},
  \citenamefont {{Keih{\"a}nen}}, \citenamefont {{Keskitalo}}, \citenamefont
  {{Kiiveri}}, \citenamefont {{Kim}}, \citenamefont {{Kisner}}, \citenamefont
  {{Knox}}, \citenamefont {{Krachmalnicoff}}, \citenamefont {{Kunz}},
  \citenamefont {{Kurki-Suonio}}, \citenamefont {{Lagache}}, \citenamefont
  {{Lamarre}}, \citenamefont {{Lasenby}}, \citenamefont {{Lattanzi}},
  \citenamefont {{Lawrence}}, \citenamefont {{Le Jeune}}, \citenamefont
  {{Lemos}}, \citenamefont {{Lesgourgues}}, \citenamefont {{Levrier}},
  \citenamefont {{Lewis}}, \citenamefont {{Liguori}}, \citenamefont {{Lilje}},
  \citenamefont {{Lilley}}, \citenamefont {{Lindholm}}, \citenamefont
  {{L{\'o}pez-Caniego}}, \citenamefont {{Lubin}}, \citenamefont {{Ma}},
  \citenamefont {{Mac{\'\i}as-P{\'e}rez}}, \citenamefont {{Maggio}},
  \citenamefont {{Maino}}, \citenamefont {{Mandolesi}}, \citenamefont
  {{Mangilli}}, \citenamefont {{Marcos-Caballero}}, \citenamefont {{Maris}},
  \citenamefont {{Martin}}, \citenamefont {{Martinelli}}, \citenamefont
  {{Mart{\'\i}nez-Gonz{\'a}lez}}, \citenamefont {{Matarrese}}, \citenamefont
  {{Mauri}}, \citenamefont {{McEwen}}, \citenamefont {{Meinhold}},
  \citenamefont {{Melchiorri}}, \citenamefont {{Mennella}}, \citenamefont
  {{Migliaccio}}, \citenamefont {{Millea}}, \citenamefont {{Mitra}},
  \citenamefont {{Miville-Desch{\^e}nes}}, \citenamefont {{Molinari}},
  \citenamefont {{Montier}}, \citenamefont {{Morgante}}, \citenamefont
  {{Moss}}, \citenamefont {{Natoli}}, \citenamefont {{N{\o}rgaard-Nielsen}},
  \citenamefont {{Pagano}}, \citenamefont {{Paoletti}}, \citenamefont
  {{Partridge}}, \citenamefont {{Patanchon}}, \citenamefont {{Peiris}},
  \citenamefont {{Perrotta}}, \citenamefont {{Pettorino}}, \citenamefont
  {{Piacentini}}, \citenamefont {{Polastri}}, \citenamefont {{Polenta}},
  \citenamefont {{Puget}}, \citenamefont {{Rachen}}, \citenamefont
  {{Reinecke}}, \citenamefont {{Remazeilles}}, \citenamefont {{Renzi}},
  \citenamefont {{Rocha}}, \citenamefont {{Rosset}}, \citenamefont {{Roudier}},
  \citenamefont {{Rubi{\~n}o-Mart{\'\i}n}}, \citenamefont {{Ruiz-Granados}},
  \citenamefont {{Salvati}}, \citenamefont {{Sandri}}, \citenamefont
  {{Savelainen}}, \citenamefont {{Scott}}, \citenamefont {{Shellard}},
  \citenamefont {{Sirignano}}, \citenamefont {{Sirri}}, \citenamefont
  {{Spencer}}, \citenamefont {{Sunyaev}}, \citenamefont {{Suur-Uski}},
  \citenamefont {{Tauber}}, \citenamefont {{Tavagnacco}}, \citenamefont
  {{Tenti}}, \citenamefont {{Toffolatti}}, \citenamefont {{Tomasi}},
  \citenamefont {{Trombetti}}, \citenamefont {{Valenziano}}, \citenamefont
  {{Valiviita}}, \citenamefont {{Van Tent}}, \citenamefont {{Vibert}},
  \citenamefont {{Vielva}}, \citenamefont {{Villa}}, \citenamefont
  {{Vittorio}}, \citenamefont {{Wandelt}}, \citenamefont {{Wehus}},
  \citenamefont {{White}}, \citenamefont {{White}}, \citenamefont {{Zacchei}},\
  and\ \citenamefont {{Zonca}}}]{Planck:2020}%
  \BibitemOpen
  \bibfield  {author} {\bibinfo {author} {\bibnamefont {{Planck
  Collaboration}}}, \bibinfo {author} {\bibfnamefont {N.}~\bibnamefont
  {{Aghanim}}}, \bibinfo {author} {\bibfnamefont {Y.}~\bibnamefont {{Akrami}}},
  \bibinfo {author} {\bibfnamefont {M.}~\bibnamefont {{Ashdown}}}, \bibinfo
  {author} {\bibfnamefont {J.}~\bibnamefont {{Aumont}}}, \bibinfo {author}
  {\bibfnamefont {C.}~\bibnamefont {{Baccigalupi}}}, \bibinfo {author}
  {\bibfnamefont {M.}~\bibnamefont {{Ballardini}}}, \bibinfo {author}
  {\bibfnamefont {A.~J.}\ \bibnamefont {{Banday}}}, \bibinfo {author}
  {\bibfnamefont {R.~B.}\ \bibnamefont {{Barreiro}}}, \bibinfo {author}
  {\bibfnamefont {N.}~\bibnamefont {{Bartolo}}}, \bibinfo {author}
  {\bibfnamefont {S.}~\bibnamefont {{Basak}}}, \bibinfo {author} {\bibfnamefont
  {R.}~\bibnamefont {{Battye}}}, \bibinfo {author} {\bibfnamefont
  {K.}~\bibnamefont {{Benabed}}}, \bibinfo {author} {\bibfnamefont {J.~P.}\
  \bibnamefont {{Bernard}}}, \bibinfo {author} {\bibfnamefont {M.}~\bibnamefont
  {{Bersanelli}}}, \bibnamefont {and~others},\ }\href
  {https://doi.org/10.1051/0004-6361/201833910} {\bibfield  {journal} {\bibinfo
   {journal} {\aap}\ }\textbf {\bibinfo {volume} {641}},\ \bibinfo {eid} {A6}
  (\bibinfo {year} {2020})},\ \Eprint {https://arxiv.org/abs/1807.06209}
  {arXiv:1807.06209 [astro-ph.CO]} \BibitemShut {NoStop}%
\bibitem [{\citenamefont {{Riess}}\ \emph {et~al.}(2022)\citenamefont
  {{Riess}}, \citenamefont {{Yuan}}, \citenamefont {{Macri}}, \citenamefont
  {{Scolnic}}, \citenamefont {{Brout}}, \citenamefont {{Casertano}},
  \citenamefont {{Jones}}, \citenamefont {{Murakami}}, \citenamefont {{Anand}},
  \citenamefont {{Breuval}}, \citenamefont {{Brink}}, \citenamefont
  {{Filippenko}}, \citenamefont {{Hoffmann}}, \citenamefont {{Jha}},
  \citenamefont {{D'arcy Kenworthy}}, \citenamefont {{Mackenty}}, \citenamefont
  {{Stahl}},\ and\ \citenamefont {{Zheng}}}]{Riess:2021}%
  \BibitemOpen
  \bibfield  {author} {\bibinfo {author} {\bibfnamefont {A.~G.}\ \bibnamefont
  {{Riess}}}, \bibinfo {author} {\bibfnamefont {W.}~\bibnamefont {{Yuan}}},
  \bibinfo {author} {\bibfnamefont {L.~M.}\ \bibnamefont {{Macri}}}, \bibinfo
  {author} {\bibfnamefont {D.}~\bibnamefont {{Scolnic}}}, \bibinfo {author}
  {\bibfnamefont {D.}~\bibnamefont {{Brout}}}, \bibinfo {author} {\bibfnamefont
  {S.}~\bibnamefont {{Casertano}}}, \bibinfo {author} {\bibfnamefont {D.~O.}\
  \bibnamefont {{Jones}}}, \bibinfo {author} {\bibfnamefont {Y.}~\bibnamefont
  {{Murakami}}}, \bibinfo {author} {\bibfnamefont {G.~S.}\ \bibnamefont
  {{Anand}}}, \bibinfo {author} {\bibfnamefont {L.}~\bibnamefont {{Breuval}}},
  \bibinfo {author} {\bibfnamefont {T.~G.}\ \bibnamefont {{Brink}}}, \bibinfo
  {author} {\bibfnamefont {A.~V.}\ \bibnamefont {{Filippenko}}}, \bibinfo
  {author} {\bibfnamefont {S.}~\bibnamefont {{Hoffmann}}}, \bibinfo {author}
  {\bibfnamefont {S.~W.}\ \bibnamefont {{Jha}}}, \bibinfo {author}
  {\bibfnamefont {W.}~\bibnamefont {{D'arcy Kenworthy}}}, \bibnamefont
  {and~others},\ }\href {https://doi.org/10.3847/2041-8213/ac5c5b} {\bibfield
  {journal} {\bibinfo  {journal} {\apjl}\ }\textbf {\bibinfo {volume} {934}},\
  \bibinfo {eid} {L7} (\bibinfo {year} {2022})},\ \Eprint
  {https://arxiv.org/abs/2112.04510} {arXiv:2112.04510 [astro-ph.CO]}
  \BibitemShut {NoStop}%
\bibitem [{\citenamefont {{Alam}}\ \emph {et~al.}(2017)\citenamefont {{Alam}},
  \citenamefont {{Ata}}, \citenamefont {{Bailey}}, \citenamefont {{Beutler}},
  \citenamefont {{Bizyaev}}, \citenamefont {{Blazek}}, \citenamefont
  {{Bolton}}, \citenamefont {{Brownstein}}, \citenamefont {{Burden}},
  \citenamefont {{Chuang}}, \citenamefont {{Comparat}}, \citenamefont
  {{Cuesta}}, \citenamefont {{Dawson}}, \citenamefont {{Eisenstein}},
  \citenamefont {{Escoffier}}, \citenamefont {{Gil-Mar{\'\i}n}}, \citenamefont
  {{Grieb}}, \citenamefont {{Hand}}, \citenamefont {{Ho}}, \citenamefont
  {{Kinemuchi}}, \citenamefont {{Kirkby}}, \citenamefont {{Kitaura}},
  \citenamefont {{Malanushenko}}, \citenamefont {{Malanushenko}}, \citenamefont
  {{Maraston}}, \citenamefont {{McBride}}, \citenamefont {{Nichol}},
  \citenamefont {{Olmstead}}, \citenamefont {{Oravetz}}, \citenamefont
  {{Padmanabhan}}, \citenamefont {{Palanque-Delabrouille}}, \citenamefont
  {{Pan}}, \citenamefont {{Pellejero-Ibanez}}, \citenamefont {{Percival}},
  \citenamefont {{Petitjean}}, \citenamefont {{Prada}}, \citenamefont
  {{Price-Whelan}}, \citenamefont {{Reid}}, \citenamefont
  {{Rodr{\'\i}guez-Torres}}, \citenamefont {{Roe}}, \citenamefont {{Ross}},
  \citenamefont {{Ross}}, \citenamefont {{Rossi}}, \citenamefont
  {{Rubi{\~n}o-Mart{\'\i}n}}, \citenamefont {{Saito}}, \citenamefont
  {{Salazar-Albornoz}}, \citenamefont {{Samushia}}, \citenamefont
  {{S{\'a}nchez}}, \citenamefont {{Satpathy}}, \citenamefont {{Schlegel}},
  \citenamefont {{Schneider}}, \citenamefont {{Sc{\'o}ccola}}, \citenamefont
  {{Seo}}, \citenamefont {{Sheldon}}, \citenamefont {{Simmons}}, \citenamefont
  {{Slosar}}, \citenamefont {{Strauss}}, \citenamefont {{Swanson}},
  \citenamefont {{Thomas}}, \citenamefont {{Tinker}}, \citenamefont
  {{Tojeiro}}, \citenamefont {{Maga{\~n}a}}, \citenamefont {{Vazquez}},
  \citenamefont {{Verde}}, \citenamefont {{Wake}}, \citenamefont {{Wang}},
  \citenamefont {{Weinberg}}, \citenamefont {{White}}, \citenamefont
  {{Wood-Vasey}}, \citenamefont {{Y{\`e}che}}, \citenamefont {{Zehavi}},
  \citenamefont {{Zhai}},\ and\ \citenamefont {{Zhao}}}]{Alam:2017}%
  \BibitemOpen
  \bibfield  {author} {\bibinfo {author} {\bibfnamefont {S.}~\bibnamefont
  {{Alam}}}, \bibinfo {author} {\bibfnamefont {M.}~\bibnamefont {{Ata}}},
  \bibinfo {author} {\bibfnamefont {S.}~\bibnamefont {{Bailey}}}, \bibinfo
  {author} {\bibfnamefont {F.}~\bibnamefont {{Beutler}}}, \bibinfo {author}
  {\bibfnamefont {D.}~\bibnamefont {{Bizyaev}}}, \bibinfo {author}
  {\bibfnamefont {J.~A.}\ \bibnamefont {{Blazek}}}, \bibinfo {author}
  {\bibfnamefont {A.~S.}\ \bibnamefont {{Bolton}}}, \bibinfo {author}
  {\bibfnamefont {J.~R.}\ \bibnamefont {{Brownstein}}}, \bibinfo {author}
  {\bibfnamefont {A.}~\bibnamefont {{Burden}}}, \bibinfo {author}
  {\bibfnamefont {C.-H.}\ \bibnamefont {{Chuang}}}, \bibinfo {author}
  {\bibfnamefont {J.}~\bibnamefont {{Comparat}}}, \bibinfo {author}
  {\bibfnamefont {A.~J.}\ \bibnamefont {{Cuesta}}}, \bibinfo {author}
  {\bibfnamefont {K.~S.}\ \bibnamefont {{Dawson}}}, \bibinfo {author}
  {\bibfnamefont {D.~J.}\ \bibnamefont {{Eisenstein}}}, \bibinfo {author}
  {\bibfnamefont {S.}~\bibnamefont {{Escoffier}}}, \bibnamefont {and~others},\
  }\href {https://doi.org/10.1093/mnras/stx721} {\bibfield  {journal} {\bibinfo
   {journal} {\mnras}\ }\textbf {\bibinfo {volume} {470}},\ \bibinfo {pages}
  {2617} (\bibinfo {year} {2017})},\ \Eprint {https://arxiv.org/abs/1607.03155}
  {arXiv:1607.03155 [astro-ph.CO]} \BibitemShut {NoStop}%
\bibitem [{\citenamefont {{Gil-Mar{\'\i}n}}\ \emph {et~al.}(2020)\citenamefont
  {{Gil-Mar{\'\i}n}}, \citenamefont {{Bautista}}, \citenamefont {{Paviot}},
  \citenamefont {{Vargas-Maga{\~n}a}}, \citenamefont {{de la Torre}},
  \citenamefont {{Fromenteau}}, \citenamefont {{Alam}}, \citenamefont
  {{{\'A}vila}}, \citenamefont {{Burtin}}, \citenamefont {{Chuang}},
  \citenamefont {{Dawson}}, \citenamefont {{Hou}}, \citenamefont {{de Mattia}},
  \citenamefont {{Mohammad}}, \citenamefont {{M{\"u}ller}}, \citenamefont
  {{Nadathur}}, \citenamefont {{Neveux}}, \citenamefont {{Percival}},
  \citenamefont {{Raichoor}}, \citenamefont {{Rezaie}}, \citenamefont {{Ross}},
  \citenamefont {{Rossi}}, \citenamefont {{Ruhlmann-Kleider}}, \citenamefont
  {{Smith}}, \citenamefont {{Tamone}}, \citenamefont {{Tinker}}, \citenamefont
  {{Tojeiro}}, \citenamefont {{Wang}}, \citenamefont {{Zhao}}, \citenamefont
  {{Zhao}}, \citenamefont {{Brinkmann}}, \citenamefont {{Brownstein}},
  \citenamefont {{Choi}}, \citenamefont {{Escoffier}}, \citenamefont {{de la
  Macorra}}, \citenamefont {{Moon}}, \citenamefont {{Newman}}, \citenamefont
  {{Schneider}}, \citenamefont {{Seo}},\ and\ \citenamefont
  {{Vivek}}}]{GilMarin:2020}%
  \BibitemOpen
  \bibfield  {author} {\bibinfo {author} {\bibfnamefont {H.}~\bibnamefont
  {{Gil-Mar{\'\i}n}}}, \bibinfo {author} {\bibfnamefont {J.~E.}\ \bibnamefont
  {{Bautista}}}, \bibinfo {author} {\bibfnamefont {R.}~\bibnamefont
  {{Paviot}}}, \bibinfo {author} {\bibfnamefont {M.}~\bibnamefont
  {{Vargas-Maga{\~n}a}}}, \bibinfo {author} {\bibfnamefont {S.}~\bibnamefont
  {{de la Torre}}}, \bibinfo {author} {\bibfnamefont {S.}~\bibnamefont
  {{Fromenteau}}}, \bibinfo {author} {\bibfnamefont {S.}~\bibnamefont
  {{Alam}}}, \bibinfo {author} {\bibfnamefont {S.}~\bibnamefont {{{\'A}vila}}},
  \bibinfo {author} {\bibfnamefont {E.}~\bibnamefont {{Burtin}}}, \bibinfo
  {author} {\bibfnamefont {C.-H.}\ \bibnamefont {{Chuang}}}, \bibinfo {author}
  {\bibfnamefont {K.~S.}\ \bibnamefont {{Dawson}}}, \bibinfo {author}
  {\bibfnamefont {J.}~\bibnamefont {{Hou}}}, \bibinfo {author} {\bibfnamefont
  {A.}~\bibnamefont {{de Mattia}}}, \bibinfo {author} {\bibfnamefont {F.~G.}\
  \bibnamefont {{Mohammad}}}, \bibinfo {author} {\bibfnamefont {E.-M.}\
  \bibnamefont {{M{\"u}ller}}}, \bibnamefont {and~others},\ }\href
  {https://doi.org/10.1093/mnras/staa2455} {\bibfield  {journal} {\bibinfo
  {journal} {\mnras}\ }\textbf {\bibinfo {volume} {498}},\ \bibinfo {pages}
  {2492} (\bibinfo {year} {2020})},\ \Eprint {https://arxiv.org/abs/2007.08994}
  {arXiv:2007.08994 [astro-ph.CO]} \BibitemShut {NoStop}%
\bibitem [{\citenamefont {{Bautista}}\ \emph {et~al.}(2021)\citenamefont
  {{Bautista}}, \citenamefont {{Paviot}}, \citenamefont {{Vargas Maga{\~n}a}},
  \citenamefont {{de la Torre}}, \citenamefont {{Fromenteau}}, \citenamefont
  {{Gil-Mar{\'\i}n}}, \citenamefont {{Ross}}, \citenamefont {{Burtin}},
  \citenamefont {{Dawson}}, \citenamefont {{Hou}}, \citenamefont {{Kneib}},
  \citenamefont {{de Mattia}}, \citenamefont {{Percival}}, \citenamefont
  {{Rossi}}, \citenamefont {{Tojeiro}}, \citenamefont {{Zhao}}, \citenamefont
  {{Zhao}}, \citenamefont {{Alam}}, \citenamefont {{Brownstein}}, \citenamefont
  {{Chapman}}, \citenamefont {{Choi}}, \citenamefont {{Chuang}}, \citenamefont
  {{Escoffier}}, \citenamefont {{de la Macorra}}, \citenamefont {{du Mas des
  Bourboux}}, \citenamefont {{Mohammad}}, \citenamefont {{Moon}}, \citenamefont
  {{M{\"u}ller}}, \citenamefont {{Nadathur}}, \citenamefont {{Newman}},
  \citenamefont {{Schneider}}, \citenamefont {{Seo}},\ and\ \citenamefont
  {{Wang}}}]{Bautista:2021}%
  \BibitemOpen
  \bibfield  {author} {\bibinfo {author} {\bibfnamefont {J.~E.}\ \bibnamefont
  {{Bautista}}}, \bibinfo {author} {\bibfnamefont {R.}~\bibnamefont
  {{Paviot}}}, \bibinfo {author} {\bibfnamefont {M.}~\bibnamefont {{Vargas
  Maga{\~n}a}}}, \bibinfo {author} {\bibfnamefont {S.}~\bibnamefont {{de la
  Torre}}}, \bibinfo {author} {\bibfnamefont {S.}~\bibnamefont {{Fromenteau}}},
  \bibinfo {author} {\bibfnamefont {H.}~\bibnamefont {{Gil-Mar{\'\i}n}}},
  \bibinfo {author} {\bibfnamefont {A.~J.}\ \bibnamefont {{Ross}}}, \bibinfo
  {author} {\bibfnamefont {E.}~\bibnamefont {{Burtin}}}, \bibinfo {author}
  {\bibfnamefont {K.~S.}\ \bibnamefont {{Dawson}}}, \bibinfo {author}
  {\bibfnamefont {J.}~\bibnamefont {{Hou}}}, \bibinfo {author} {\bibfnamefont
  {J.-P.}\ \bibnamefont {{Kneib}}}, \bibinfo {author} {\bibfnamefont
  {A.}~\bibnamefont {{de Mattia}}}, \bibinfo {author} {\bibfnamefont {W.~J.}\
  \bibnamefont {{Percival}}}, \bibinfo {author} {\bibfnamefont
  {G.}~\bibnamefont {{Rossi}}}, \bibinfo {author} {\bibfnamefont
  {R.}~\bibnamefont {{Tojeiro}}}, \bibnamefont {and~others},\ }\href
  {https://doi.org/10.1093/mnras/staa2800} {\bibfield  {journal} {\bibinfo
  {journal} {\mnras}\ }\textbf {\bibinfo {volume} {500}},\ \bibinfo {pages}
  {736} (\bibinfo {year} {2021})},\ \Eprint {https://arxiv.org/abs/2007.08993}
  {arXiv:2007.08993 [astro-ph.CO]} \BibitemShut {NoStop}%
\bibitem [{\citenamefont {{Hou}}\ \emph {et~al.}(2021)\citenamefont {{Hou}},
  \citenamefont {{S{\'a}nchez}}, \citenamefont {{Ross}}, \citenamefont
  {{Smith}}, \citenamefont {{Neveux}}, \citenamefont {{Bautista}},
  \citenamefont {{Burtin}}, \citenamefont {{Zhao}}, \citenamefont
  {{Scoccimarro}}, \citenamefont {{Dawson}}, \citenamefont {{de Mattia}},
  \citenamefont {{de la Macorra}}, \citenamefont {{du Mas des Bourboux}},
  \citenamefont {{Eisenstein}}, \citenamefont {{Gil-Mar{\'\i}n}}, \citenamefont
  {{Lyke}}, \citenamefont {{Mohammad}}, \citenamefont {{Mueller}},
  \citenamefont {{Percival}}, \citenamefont {{Rossi}}, \citenamefont {{Vargas
  Maga{\~n}a}}, \citenamefont {{Zarrouk}}, \citenamefont {{Zhao}},
  \citenamefont {{Brinkmann}}, \citenamefont {{Brownstein}}, \citenamefont
  {{Chuang}}, \citenamefont {{Myers}}, \citenamefont {{Newman}}, \citenamefont
  {{Schneider}},\ and\ \citenamefont {{Vivek}}}]{Hou:2021}%
  \BibitemOpen
  \bibfield  {author} {\bibinfo {author} {\bibfnamefont {J.}~\bibnamefont
  {{Hou}}}, \bibinfo {author} {\bibfnamefont {A.~G.}\ \bibnamefont
  {{S{\'a}nchez}}}, \bibinfo {author} {\bibfnamefont {A.~J.}\ \bibnamefont
  {{Ross}}}, \bibinfo {author} {\bibfnamefont {A.}~\bibnamefont {{Smith}}},
  \bibinfo {author} {\bibfnamefont {R.}~\bibnamefont {{Neveux}}}, \bibinfo
  {author} {\bibfnamefont {J.}~\bibnamefont {{Bautista}}}, \bibinfo {author}
  {\bibfnamefont {E.}~\bibnamefont {{Burtin}}}, \bibinfo {author}
  {\bibfnamefont {C.}~\bibnamefont {{Zhao}}}, \bibinfo {author} {\bibfnamefont
  {R.}~\bibnamefont {{Scoccimarro}}}, \bibinfo {author} {\bibfnamefont {K.~S.}\
  \bibnamefont {{Dawson}}}, \bibinfo {author} {\bibfnamefont {A.}~\bibnamefont
  {{de Mattia}}}, \bibinfo {author} {\bibfnamefont {A.}~\bibnamefont {{de la
  Macorra}}}, \bibinfo {author} {\bibfnamefont {H.}~\bibnamefont {{du Mas des
  Bourboux}}}, \bibinfo {author} {\bibfnamefont {D.~J.}\ \bibnamefont
  {{Eisenstein}}}, \bibinfo {author} {\bibfnamefont {H.}~\bibnamefont
  {{Gil-Mar{\'\i}n}}}, \bibnamefont {and~others},\ }\href
  {https://doi.org/10.1093/mnras/staa3234} {\bibfield  {journal} {\bibinfo
  {journal} {\mnras}\ }\textbf {\bibinfo {volume} {500}},\ \bibinfo {pages}
  {1201} (\bibinfo {year} {2021})},\ \Eprint {https://arxiv.org/abs/2007.08998}
  {arXiv:2007.08998 [astro-ph.CO]} \BibitemShut {NoStop}%
\bibitem [{\citenamefont {{Neveux}}\ \emph {et~al.}(2020)\citenamefont
  {{Neveux}}, \citenamefont {{Burtin}}, \citenamefont {{de Mattia}},
  \citenamefont {{Smith}}, \citenamefont {{Ross}}, \citenamefont {{Hou}},
  \citenamefont {{Bautista}}, \citenamefont {{Brinkmann}}, \citenamefont
  {{Chuang}}, \citenamefont {{Dawson}}, \citenamefont {{Gil-Mar{\'\i}n}},
  \citenamefont {{Lyke}}, \citenamefont {{de la Macorra}}, \citenamefont {{du
  Mas des Bourboux}}, \citenamefont {{Mohammad}}, \citenamefont {{M{\"u}ller}},
  \citenamefont {{Myers}}, \citenamefont {{Newman}}, \citenamefont
  {{Percival}}, \citenamefont {{Rossi}}, \citenamefont {{Schneider}},
  \citenamefont {{Vivek}}, \citenamefont {{Zarrouk}}, \citenamefont {{Zhao}},\
  and\ \citenamefont {{Zhao}}}]{Neveux:2020}%
  \BibitemOpen
  \bibfield  {author} {\bibinfo {author} {\bibfnamefont {R.}~\bibnamefont
  {{Neveux}}}, \bibinfo {author} {\bibfnamefont {E.}~\bibnamefont {{Burtin}}},
  \bibinfo {author} {\bibfnamefont {A.}~\bibnamefont {{de Mattia}}}, \bibinfo
  {author} {\bibfnamefont {A.}~\bibnamefont {{Smith}}}, \bibinfo {author}
  {\bibfnamefont {A.~J.}\ \bibnamefont {{Ross}}}, \bibinfo {author}
  {\bibfnamefont {J.}~\bibnamefont {{Hou}}}, \bibinfo {author} {\bibfnamefont
  {J.}~\bibnamefont {{Bautista}}}, \bibinfo {author} {\bibfnamefont
  {J.}~\bibnamefont {{Brinkmann}}}, \bibinfo {author} {\bibfnamefont {C.-H.}\
  \bibnamefont {{Chuang}}}, \bibinfo {author} {\bibfnamefont {K.~S.}\
  \bibnamefont {{Dawson}}}, \bibinfo {author} {\bibfnamefont {H.}~\bibnamefont
  {{Gil-Mar{\'\i}n}}}, \bibinfo {author} {\bibfnamefont {B.~W.}\ \bibnamefont
  {{Lyke}}}, \bibinfo {author} {\bibfnamefont {A.}~\bibnamefont {{de la
  Macorra}}}, \bibinfo {author} {\bibfnamefont {H.}~\bibnamefont {{du Mas des
  Bourboux}}}, \bibinfo {author} {\bibfnamefont {F.~G.}\ \bibnamefont
  {{Mohammad}}}, \bibnamefont {and~others},\ }\href
  {https://doi.org/10.1093/mnras/staa2780} {\bibfield  {journal} {\bibinfo
  {journal} {\mnras}\ }\textbf {\bibinfo {volume} {499}},\ \bibinfo {pages}
  {210} (\bibinfo {year} {2020})},\ \Eprint {https://arxiv.org/abs/2007.08999}
  {arXiv:2007.08999 [astro-ph.CO]} \BibitemShut {NoStop}%
\bibitem [{\citenamefont {{Alam}}\ \emph {et~al.}(2021)\citenamefont {{Alam}},
  \citenamefont {{Aubert}}, \citenamefont {{Avila}}, \citenamefont {{Balland}},
  \citenamefont {{Bautista}}, \citenamefont {{Bershady}}, \citenamefont
  {{Bizyaev}}, \citenamefont {{Blanton}}, \citenamefont {{Bolton}},
  \citenamefont {{Bovy}}, \citenamefont {{Brinkmann}}, \citenamefont
  {{Brownstein}}, \citenamefont {{Burtin}}, \citenamefont {{Chabanier}},
  \citenamefont {{Chapman}}, \citenamefont {{Choi}}, \citenamefont {{Chuang}},
  \citenamefont {{Comparat}}, \citenamefont {{Cousinou}}, \citenamefont
  {{Cuceu}}, \citenamefont {{Dawson}}, \citenamefont {{de la Torre}},
  \citenamefont {{de Mattia}}, \citenamefont {{Agathe}}, \citenamefont {{des
  Bourboux}}, \citenamefont {{Escoffier}}, \citenamefont {{Etourneau}},
  \citenamefont {{Farr}}, \citenamefont {{Font-Ribera}}, \citenamefont
  {{Frinchaboy}}, \citenamefont {{Fromenteau}}, \citenamefont
  {{Gil-Mar{\'\i}n}}, \citenamefont {{Le Goff}}, \citenamefont
  {{Gonzalez-Morales}}, \citenamefont {{Gonzalez-Perez}}, \citenamefont
  {{Grabowski}}, \citenamefont {{Guy}}, \citenamefont {{Hawken}}, \citenamefont
  {{Hou}}, \citenamefont {{Kong}}, \citenamefont {{Parker}}, \citenamefont
  {{Klaene}}, \citenamefont {{Kneib}}, \citenamefont {{Lin}}, \citenamefont
  {{Long}}, \citenamefont {{Lyke}}, \citenamefont {{de la Macorra}},
  \citenamefont {{Martini}}, \citenamefont {{Masters}}, \citenamefont
  {{Mohammad}}, \citenamefont {{Moon}}, \citenamefont {{Mueller}},
  \citenamefont {{Mu{\~n}oz-Guti{\'e}rrez}}, \citenamefont {{Myers}},
  \citenamefont {{Nadathur}}, \citenamefont {{Neveux}}, \citenamefont
  {{Newman}}, \citenamefont {{Noterdaeme}}, \citenamefont {{Oravetz}},
  \citenamefont {{Oravetz}}, \citenamefont {{Palanque-Delabrouille}},
  \citenamefont {{Pan}}, \citenamefont {{Paviot}}, \citenamefont {{Percival}},
  \citenamefont {{P{\'e}rez-R{\`a}fols}}, \citenamefont {{Petitjean}},
  \citenamefont {{Pieri}}, \citenamefont {{Prakash}}, \citenamefont
  {{Raichoor}}, \citenamefont {{Ravoux}}, \citenamefont {{Rezaie}},
  \citenamefont {{Rich}}, \citenamefont {{Ross}}, \citenamefont {{Rossi}},
  \citenamefont {{Ruggeri}}, \citenamefont {{Ruhlmann-Kleider}}, \citenamefont
  {{S{\'a}nchez}}, \citenamefont {{S{\'a}nchez}}, \citenamefont
  {{S{\'a}nchez-Gallego}}, \citenamefont {{Sayres}}, \citenamefont
  {{Schneider}}, \citenamefont {{Seo}}, \citenamefont {{Shafieloo}},
  \citenamefont {{Slosar}}, \citenamefont {{Smith}}, \citenamefont {{Stermer}},
  \citenamefont {{Tamone}}, \citenamefont {{Tinker}}, \citenamefont
  {{Tojeiro}}, \citenamefont {{Vargas-Maga{\~n}a}}, \citenamefont {{Variu}},
  \citenamefont {{Wang}}, \citenamefont {{Weaver}}, \citenamefont {{Weijmans}},
  \citenamefont {{Y{\`e}che}}, \citenamefont {{Zarrouk}}, \citenamefont
  {{Zhao}}, \citenamefont {{Zhao}},\ and\ \citenamefont {{Zheng}}}]{Alam:2021}%
  \BibitemOpen
  \bibfield  {author} {\bibinfo {author} {\bibfnamefont {S.}~\bibnamefont
  {{Alam}}}, \bibinfo {author} {\bibfnamefont {M.}~\bibnamefont {{Aubert}}},
  \bibinfo {author} {\bibfnamefont {S.}~\bibnamefont {{Avila}}}, \bibinfo
  {author} {\bibfnamefont {C.}~\bibnamefont {{Balland}}}, \bibinfo {author}
  {\bibfnamefont {J.~E.}\ \bibnamefont {{Bautista}}}, \bibinfo {author}
  {\bibfnamefont {M.~A.}\ \bibnamefont {{Bershady}}}, \bibinfo {author}
  {\bibfnamefont {D.}~\bibnamefont {{Bizyaev}}}, \bibinfo {author}
  {\bibfnamefont {M.~R.}\ \bibnamefont {{Blanton}}}, \bibinfo {author}
  {\bibfnamefont {A.~S.}\ \bibnamefont {{Bolton}}}, \bibinfo {author}
  {\bibfnamefont {J.}~\bibnamefont {{Bovy}}}, \bibinfo {author} {\bibfnamefont
  {J.}~\bibnamefont {{Brinkmann}}}, \bibinfo {author} {\bibfnamefont {J.~R.}\
  \bibnamefont {{Brownstein}}}, \bibinfo {author} {\bibfnamefont
  {E.}~\bibnamefont {{Burtin}}}, \bibinfo {author} {\bibfnamefont
  {S.}~\bibnamefont {{Chabanier}}}, \bibinfo {author} {\bibfnamefont {M.~J.}\
  \bibnamefont {{Chapman}}}, \bibnamefont {and~others},\ }\href
  {https://doi.org/10.1103/PhysRevD.103.083533} {\bibfield  {journal} {\bibinfo
   {journal} {\prd}\ }\textbf {\bibinfo {volume} {103}},\ \bibinfo {eid}
  {083533} (\bibinfo {year} {2021})},\ \Eprint
  {https://arxiv.org/abs/2007.08991} {arXiv:2007.08991 [astro-ph.CO]}
  \BibitemShut {NoStop}%
\bibitem [{\citenamefont {{Abbott}}\ \emph
  {et~al.}(2022{\natexlab{a}})\citenamefont {{Abbott}}, \citenamefont
  {{Aguena}}, \citenamefont {{Alarcon}}, \citenamefont {{Allam}}, \citenamefont
  {{Alves}}, \citenamefont {{Amon}}, \citenamefont {{Andrade-Oliveira}},
  \citenamefont {{Annis}}, \citenamefont {{Avila}}, \citenamefont {{Bacon}},
  \citenamefont {{Baxter}}, \citenamefont {{Bechtol}}, \citenamefont
  {{Becker}}, \citenamefont {{Bernstein}}, \citenamefont {{Bhargava}},
  \citenamefont {{Birrer}}, \citenamefont {{Blazek}}, \citenamefont
  {{Brandao-Souza}}, \citenamefont {{Bridle}}, \citenamefont {{Brooks}},
  \citenamefont {{Buckley-Geer}}, \citenamefont {{Burke}}, \citenamefont
  {{Camacho}}, \citenamefont {{Campos}}, \citenamefont {{Carnero Rosell}},
  \citenamefont {{Carrasco Kind}}, \citenamefont {{Carretero}}, \citenamefont
  {{Castander}}, \citenamefont {{Cawthon}}, \citenamefont {{Chang}},
  \citenamefont {{Chen}}, \citenamefont {{Chen}}, \citenamefont {{Choi}},
  \citenamefont {{Conselice}}, \citenamefont {{Cordero}}, \citenamefont
  {{Costanzi}}, \citenamefont {{Crocce}}, \citenamefont {{da Costa}},
  \citenamefont {{da Silva Pereira}}, \citenamefont {{Davis}}, \citenamefont
  {{Davis}}, \citenamefont {{De Vicente}}, \citenamefont {{DeRose}},
  \citenamefont {{Desai}}, \citenamefont {{Di Valentino}}, \citenamefont
  {{Diehl}}, \citenamefont {{Dietrich}}, \citenamefont {{Dodelson}},
  \citenamefont {{Doel}}, \citenamefont {{Doux}}, \citenamefont
  {{Drlica-Wagner}}, \citenamefont {{Eckert}}, \citenamefont {{Eifler}},
  \citenamefont {{Elsner}}, \citenamefont {{Elvin-Poole}}, \citenamefont
  {{Everett}}, \citenamefont {{Evrard}}, \citenamefont {{Fang}}, \citenamefont
  {{Farahi}}, \citenamefont {{Fernandez}}, \citenamefont {{Ferrero}},
  \citenamefont {{Fert{\'e}}}, \citenamefont {{Fosalba}}, \citenamefont
  {{Friedrich}}, \citenamefont {{Frieman}}, \citenamefont
  {{Garc{\'\i}a-Bellido}}, \citenamefont {{Gatti}}, \citenamefont
  {{Gaztanaga}}, \citenamefont {{Gerdes}}, \citenamefont {{Giannantonio}},
  \citenamefont {{Giannini}}, \citenamefont {{Gruen}}, \citenamefont
  {{Gruendl}}, \citenamefont {{Gschwend}}, \citenamefont {{Gutierrez}},
  \citenamefont {{Harrison}}, \citenamefont {{Hartley}}, \citenamefont
  {{Herner}}, \citenamefont {{Hinton}}, \citenamefont {{Hollowood}},
  \citenamefont {{Honscheid}}, \citenamefont {{Hoyle}}, \citenamefont {{Huff}},
  \citenamefont {{Huterer}}, \citenamefont {{Jain}}, \citenamefont {{James}},
  \citenamefont {{Jarvis}}, \citenamefont {{Jeffrey}}, \citenamefont
  {{Jeltema}}, \citenamefont {{Kovacs}}, \citenamefont {{Krause}},
  \citenamefont {{Kron}}, \citenamefont {{Kuehn}}, \citenamefont
  {{Kuropatkin}}, \citenamefont {{Lahav}}, \citenamefont {{Leget}},
  \citenamefont {{Lemos}}, \citenamefont {{Liddle}}, \citenamefont {{Lidman}},
  \citenamefont {{Lima}}, \citenamefont {{Lin}}, \citenamefont {{MacCrann}},
  \citenamefont {{Maia}}, \citenamefont {{Marshall}}, \citenamefont
  {{Martini}}, \citenamefont {{McCullough}}, \citenamefont {{Melchior}},
  \citenamefont {{Mena-Fern{\'a}ndez}}, \citenamefont {{Menanteau}},
  \citenamefont {{Miquel}}, \citenamefont {{Mohr}}, \citenamefont {{Morgan}},
  \citenamefont {{Muir}}, \citenamefont {{Myles}}, \citenamefont {{Nadathur}},
  \citenamefont {{Navarro-Alsina}}, \citenamefont {{Nichol}}, \citenamefont
  {{Ogando}}, \citenamefont {{Omori}}, \citenamefont {{Palmese}}, \citenamefont
  {{Pandey}}, \citenamefont {{Park}}, \citenamefont {{Paz-Chinch{\'o}n}},
  \citenamefont {{Petravick}}, \citenamefont {{Pieres}}, \citenamefont {{Plazas
  Malag{\'o}n}}, \citenamefont {{Porredon}}, \citenamefont {{Prat}},
  \citenamefont {{Raveri}}, \citenamefont {{Rodriguez-Monroy}}, \citenamefont
  {{Rollins}}, \citenamefont {{Romer}}, \citenamefont {{Roodman}},
  \citenamefont {{Rosenfeld}}, \citenamefont {{Ross}}, \citenamefont
  {{Rykoff}}, \citenamefont {{Samuroff}}, \citenamefont {{S{\'a}nchez}},
  \citenamefont {{Sanchez}}, \citenamefont {{Sanchez}}, \citenamefont {{Sanchez
  Cid}}, \citenamefont {{Scarpine}}, \citenamefont {{Schubnell}}, \citenamefont
  {{Scolnic}}, \citenamefont {{Secco}}, \citenamefont {{Serrano}},
  \citenamefont {{Sevilla-Noarbe}}, \citenamefont {{Sheldon}}, \citenamefont
  {{Shin}}, \citenamefont {{Smith}}, \citenamefont {{Soares-Santos}},
  \citenamefont {{Suchyta}}, \citenamefont {{Swanson}}, \citenamefont
  {{Tabbutt}}, \citenamefont {{Tarle}}, \citenamefont {{Thomas}}, \citenamefont
  {{To}}, \citenamefont {{Troja}}, \citenamefont {{Troxel}}, \citenamefont
  {{Tucker}}, \citenamefont {{Tutusaus}}, \citenamefont {{Varga}},
  \citenamefont {{Walker}}, \citenamefont {{Weaverdyck}}, \citenamefont
  {{Wechsler}}, \citenamefont {{Weller}}, \citenamefont {{Yanny}},
  \citenamefont {{Yin}}, \citenamefont {{Zhang}}, \citenamefont {{Zuntz}},\
  and\ \citenamefont {{DES Collaboration}}}]{Abbott:2022}%
  \BibitemOpen
  \bibfield  {author} {\bibinfo {author} {\bibfnamefont {T.~M.~C.}\
  \bibnamefont {{Abbott}}}, \bibinfo {author} {\bibfnamefont {M.}~\bibnamefont
  {{Aguena}}}, \bibinfo {author} {\bibfnamefont {A.}~\bibnamefont {{Alarcon}}},
  \bibinfo {author} {\bibfnamefont {S.}~\bibnamefont {{Allam}}}, \bibinfo
  {author} {\bibfnamefont {O.}~\bibnamefont {{Alves}}}, \bibinfo {author}
  {\bibfnamefont {A.}~\bibnamefont {{Amon}}}, \bibinfo {author} {\bibfnamefont
  {F.}~\bibnamefont {{Andrade-Oliveira}}}, \bibinfo {author} {\bibfnamefont
  {J.}~\bibnamefont {{Annis}}}, \bibinfo {author} {\bibfnamefont
  {S.}~\bibnamefont {{Avila}}}, \bibinfo {author} {\bibfnamefont
  {D.}~\bibnamefont {{Bacon}}}, \bibinfo {author} {\bibfnamefont
  {E.}~\bibnamefont {{Baxter}}}, \bibinfo {author} {\bibfnamefont
  {K.}~\bibnamefont {{Bechtol}}}, \bibinfo {author} {\bibfnamefont {M.~R.}\
  \bibnamefont {{Becker}}}, \bibinfo {author} {\bibfnamefont {G.~M.}\
  \bibnamefont {{Bernstein}}}, \bibinfo {author} {\bibfnamefont
  {S.}~\bibnamefont {{Bhargava}}}, \bibnamefont {and~others},\ }\href
  {https://doi.org/10.1103/PhysRevD.105.023520} {\bibfield  {journal} {\bibinfo
   {journal} {\prd}\ }\textbf {\bibinfo {volume} {105}},\ \bibinfo {eid}
  {023520} (\bibinfo {year} {2022}{\natexlab{a}})},\ \Eprint
  {https://arxiv.org/abs/2105.13549} {arXiv:2105.13549 [astro-ph.CO]}
  \BibitemShut {NoStop}%
\bibitem [{\citenamefont {{Abbott}}\ \emph
  {et~al.}(2022{\natexlab{b}})\citenamefont {{Abbott}}, \citenamefont
  {{Aguena}}, \citenamefont {{Allam}}, \citenamefont {{Amon}}, \citenamefont
  {{Andrade-Oliveira}}, \citenamefont {{Asorey}}, \citenamefont {{Avila}},
  \citenamefont {{Bernstein}}, \citenamefont {{Bertin}}, \citenamefont
  {{Brandao-Souza}}, \citenamefont {{Brooks}}, \citenamefont {{Burke}},
  \citenamefont {{Calcino}}, \citenamefont {{Camacho}}, \citenamefont {{Carnero
  Rosell}}, \citenamefont {{Carollo}}, \citenamefont {{Carrasco Kind}},
  \citenamefont {{Carretero}}, \citenamefont {{Castander}}, \citenamefont
  {{Cawthon}}, \citenamefont {{Chan}}, \citenamefont {{Choi}}, \citenamefont
  {{Conselice}}, \citenamefont {{Costanzi}}, \citenamefont {{Crocce}},
  \citenamefont {{da Costa}}, \citenamefont {{Pereira}}, \citenamefont
  {{Davis}}, \citenamefont {{De Vicente}}, \citenamefont {{Desai}},
  \citenamefont {{Diehl}}, \citenamefont {{Doel}}, \citenamefont {{Eckert}},
  \citenamefont {{Elvin-Poole}}, \citenamefont {{Everett}}, \citenamefont
  {{Evrard}}, \citenamefont {{Fang}}, \citenamefont {{Ferrero}}, \citenamefont
  {{Fert{\'e}}}, \citenamefont {{Flaugher}}, \citenamefont {{Fosalba}},
  \citenamefont {{Garc{\'\i}a-Bellido}}, \citenamefont {{Gaztanaga}},
  \citenamefont {{Gerdes}}, \citenamefont {{Giannantonio}}, \citenamefont
  {{Glazebrook}}, \citenamefont {{Gomes}}, \citenamefont {{Gruen}},
  \citenamefont {{Gruendl}}, \citenamefont {{Gschwend}}, \citenamefont
  {{Gutierrez}}, \citenamefont {{Hinton}}, \citenamefont {{Hollowood}},
  \citenamefont {{Honscheid}}, \citenamefont {{Huterer}}, \citenamefont
  {{Jain}}, \citenamefont {{James}}, \citenamefont {{Jeltema}}, \citenamefont
  {{Kokron}}, \citenamefont {{Krause}}, \citenamefont {{Kuehn}}, \citenamefont
  {{Lahav}}, \citenamefont {{Lewis}}, \citenamefont {{Lidman}}, \citenamefont
  {{Lima}}, \citenamefont {{Lin}}, \citenamefont {{Maia}}, \citenamefont
  {{Malik}}, \citenamefont {{Martini}}, \citenamefont {{Melchior}},
  \citenamefont {{Mena-Fern{\'a}ndez}}, \citenamefont {{Menanteau}},
  \citenamefont {{Miquel}}, \citenamefont {{Mohr}}, \citenamefont {{Morgan}},
  \citenamefont {{Muir}}, \citenamefont {{Myles}}, \citenamefont
  {{M{\"o}ller}}, \citenamefont {{Palmese}}, \citenamefont
  {{Paz-Chinch{\'o}n}}, \citenamefont {{Percival}}, \citenamefont {{Pieres}},
  \citenamefont {{Plazas Malag{\'o}n}}, \citenamefont {{Porredon}},
  \citenamefont {{Prat}}, \citenamefont {{Reil}}, \citenamefont
  {{Rodriguez-Monroy}}, \citenamefont {{Romer}}, \citenamefont {{Roodman}},
  \citenamefont {{Rosenfeld}}, \citenamefont {{Ross}}, \citenamefont
  {{Sanchez}}, \citenamefont {{Sanchez Cid}}, \citenamefont {{Scarpine}},
  \citenamefont {{Serrano}}, \citenamefont {{Sevilla-Noarbe}}, \citenamefont
  {{Sheldon}}, \citenamefont {{Smith}}, \citenamefont {{Soares-Santos}},
  \citenamefont {{Suchyta}}, \citenamefont {{Swanson}}, \citenamefont
  {{Tarle}}, \citenamefont {{Thomas}}, \citenamefont {{To}}, \citenamefont
  {{Troxel}}, \citenamefont {{Tucker}}, \citenamefont {{Tucker}}, \citenamefont
  {{Tutusaus}}, \citenamefont {{Uddin}}, \citenamefont {{Varga}}, \citenamefont
  {{Weller}}, \citenamefont {{Wilkinson}},\ and\ \citenamefont {{DES
  Collaboration}}}]{DES:2022}%
  \BibitemOpen
  \bibfield  {author} {\bibinfo {author} {\bibfnamefont {T.~M.~C.}\
  \bibnamefont {{Abbott}}}, \bibinfo {author} {\bibfnamefont {M.}~\bibnamefont
  {{Aguena}}}, \bibinfo {author} {\bibfnamefont {S.}~\bibnamefont {{Allam}}},
  \bibinfo {author} {\bibfnamefont {A.}~\bibnamefont {{Amon}}}, \bibinfo
  {author} {\bibfnamefont {F.}~\bibnamefont {{Andrade-Oliveira}}}, \bibinfo
  {author} {\bibfnamefont {J.}~\bibnamefont {{Asorey}}}, \bibinfo {author}
  {\bibfnamefont {S.}~\bibnamefont {{Avila}}}, \bibinfo {author} {\bibfnamefont
  {G.~M.}\ \bibnamefont {{Bernstein}}}, \bibinfo {author} {\bibfnamefont
  {E.}~\bibnamefont {{Bertin}}}, \bibinfo {author} {\bibfnamefont
  {A.}~\bibnamefont {{Brandao-Souza}}}, \bibinfo {author} {\bibfnamefont
  {D.}~\bibnamefont {{Brooks}}}, \bibinfo {author} {\bibfnamefont {D.~L.}\
  \bibnamefont {{Burke}}}, \bibinfo {author} {\bibfnamefont {J.}~\bibnamefont
  {{Calcino}}}, \bibinfo {author} {\bibfnamefont {H.}~\bibnamefont
  {{Camacho}}}, \bibinfo {author} {\bibfnamefont {A.}~\bibnamefont {{Carnero
  Rosell}}}, \bibnamefont {and~others},\ }\href
  {https://doi.org/10.1103/PhysRevD.105.043512} {\bibfield  {journal} {\bibinfo
   {journal} {\prd}\ }\textbf {\bibinfo {volume} {105}},\ \bibinfo {eid}
  {043512} (\bibinfo {year} {2022}{\natexlab{b}})},\ \Eprint
  {https://arxiv.org/abs/2107.04646} {arXiv:2107.04646 [astro-ph.CO]}
  \BibitemShut {NoStop}%
\bibitem [{\citenamefont {{Nadathur}}\ \emph {et~al.}(2020)\citenamefont
  {{Nadathur}}, \citenamefont {{Percival}}, \citenamefont {{Beutler}},\ and\
  \citenamefont {{Winther}}}]{Nadathur:2020}%
  \BibitemOpen
  \bibfield  {author} {\bibinfo {author} {\bibfnamefont {S.}~\bibnamefont
  {{Nadathur}}}, \bibinfo {author} {\bibfnamefont {W.~J.}\ \bibnamefont
  {{Percival}}}, \bibinfo {author} {\bibfnamefont {F.}~\bibnamefont
  {{Beutler}}},\ \bibnamefont {and}\ \bibinfo {author} {\bibfnamefont {H.~A.}\
  \bibnamefont {{Winther}}},\ }\href
  {https://doi.org/10.1103/PhysRevLett.124.221301} {\bibfield  {journal}
  {\bibinfo  {journal} {\prl}\ }\textbf {\bibinfo {volume} {124}},\ \bibinfo
  {eid} {221301} (\bibinfo {year} {2020})},\ \Eprint
  {https://arxiv.org/abs/2001.11044} {arXiv:2001.11044 [astro-ph.CO]}
  \BibitemShut {NoStop}%
\bibitem [{\citenamefont {{Brieden}}\ \emph {et~al.}(2022)\citenamefont
  {{Brieden}}, \citenamefont {{Gil-Mar{\'\i}n}},\ and\ \citenamefont
  {{Verde}}}]{Brieden:2022}%
  \BibitemOpen
  \bibfield  {author} {\bibinfo {author} {\bibfnamefont {S.}~\bibnamefont
  {{Brieden}}}, \bibinfo {author} {\bibfnamefont {H.}~\bibnamefont
  {{Gil-Mar{\'\i}n}}},\ \bibnamefont {and}\ \bibinfo {author} {\bibfnamefont
  {L.}~\bibnamefont {{Verde}}},\ }\href
  {https://doi.org/10.1088/1475-7516/2022/08/024} {\bibfield  {journal}
  {\bibinfo  {journal} {\jcap}\ }\textbf {\bibinfo {volume} {2022}},\ \bibinfo
  {eid} {024} (\bibinfo {year} {2022})},\ \Eprint
  {https://arxiv.org/abs/2204.11868} {arXiv:2204.11868 [astro-ph.CO]}
  \BibitemShut {NoStop}%
\bibitem [{\citenamefont {{Eisenstein}}\ \emph {et~al.}(2005)\citenamefont
  {{Eisenstein}}, \citenamefont {{Zehavi}}, \citenamefont {{Hogg}},
  \citenamefont {{Scoccimarro}}, \citenamefont {{Blanton}}, \citenamefont
  {{Nichol}}, \citenamefont {{Scranton}}, \citenamefont {{Seo}}, \citenamefont
  {{Tegmark}}, \citenamefont {{Zheng}}, \citenamefont {{Anderson}},
  \citenamefont {{Annis}}, \citenamefont {{Bahcall}}, \citenamefont
  {{Brinkmann}}, \citenamefont {{Burles}}, \citenamefont {{Castand er}},
  \citenamefont {{Connolly}}, \citenamefont {{Csabai}}, \citenamefont {{Doi}},
  \citenamefont {{Fukugita}}, \citenamefont {{Frieman}}, \citenamefont
  {{Glazebrook}}, \citenamefont {{Gunn}}, \citenamefont {{Hendry}},
  \citenamefont {{Hennessy}}, \citenamefont {{Ivezi{\'c}}}, \citenamefont
  {{Kent}}, \citenamefont {{Knapp}}, \citenamefont {{Lin}}, \citenamefont
  {{Loh}}, \citenamefont {{Lupton}}, \citenamefont {{Margon}}, \citenamefont
  {{McKay}}, \citenamefont {{Meiksin}}, \citenamefont {{Munn}}, \citenamefont
  {{Pope}}, \citenamefont {{Richmond}}, \citenamefont {{Schlegel}},
  \citenamefont {{Schneider}}, \citenamefont {{Shimasaku}}, \citenamefont
  {{Stoughton}}, \citenamefont {{Strauss}}, \citenamefont {{SubbaRao}},
  \citenamefont {{Szalay}}, \citenamefont {{Szapudi}}, \citenamefont
  {{Tucker}}, \citenamefont {{Yanny}},\ and\ \citenamefont
  {{York}}}]{Eisenstein:2005}%
  \BibitemOpen
  \bibfield  {author} {\bibinfo {author} {\bibfnamefont {D.~J.}\ \bibnamefont
  {{Eisenstein}}}, \bibinfo {author} {\bibfnamefont {I.}~\bibnamefont
  {{Zehavi}}}, \bibinfo {author} {\bibfnamefont {D.~W.}\ \bibnamefont
  {{Hogg}}}, \bibinfo {author} {\bibfnamefont {R.}~\bibnamefont
  {{Scoccimarro}}}, \bibinfo {author} {\bibfnamefont {M.~R.}\ \bibnamefont
  {{Blanton}}}, \bibinfo {author} {\bibfnamefont {R.~C.}\ \bibnamefont
  {{Nichol}}}, \bibinfo {author} {\bibfnamefont {R.}~\bibnamefont
  {{Scranton}}}, \bibinfo {author} {\bibfnamefont {H.-J.}\ \bibnamefont
  {{Seo}}}, \bibinfo {author} {\bibfnamefont {M.}~\bibnamefont {{Tegmark}}},
  \bibinfo {author} {\bibfnamefont {Z.}~\bibnamefont {{Zheng}}}, \bibinfo
  {author} {\bibfnamefont {S.~F.}\ \bibnamefont {{Anderson}}}, \bibinfo
  {author} {\bibfnamefont {J.}~\bibnamefont {{Annis}}}, \bibinfo {author}
  {\bibfnamefont {N.}~\bibnamefont {{Bahcall}}}, \bibinfo {author}
  {\bibfnamefont {J.}~\bibnamefont {{Brinkmann}}}, \bibinfo {author}
  {\bibfnamefont {S.}~\bibnamefont {{Burles}}}, \bibnamefont {and~others},\
  }\href {https://doi.org/10.1086/466512} {\bibfield  {journal} {\bibinfo
  {journal} {\apj}\ }\textbf {\bibinfo {volume} {633}},\ \bibinfo {pages} {560}
  (\bibinfo {year} {2005})},\ \Eprint {https://arxiv.org/abs/astro-ph/0501171}
  {arXiv:astro-ph/0501171 [astro-ph]} \BibitemShut {NoStop}%
\bibitem [{\citenamefont {{Cole}}\ \emph {et~al.}(2005)\citenamefont {{Cole}},
  \citenamefont {{Percival}}, \citenamefont {{Peacock}}, \citenamefont
  {{Norberg}}, \citenamefont {{Baugh}}, \citenamefont {{Frenk}}, \citenamefont
  {{Baldry}}, \citenamefont {{Bland-Hawthorn}}, \citenamefont {{Bridges}},
  \citenamefont {{Cannon}}, \citenamefont {{Colless}}, \citenamefont
  {{Collins}}, \citenamefont {{Couch}}, \citenamefont {{Cross}}, \citenamefont
  {{Dalton}}, \citenamefont {{Eke}}, \citenamefont {{De Propris}},
  \citenamefont {{Driver}}, \citenamefont {{Efstathiou}}, \citenamefont
  {{Ellis}}, \citenamefont {{Glazebrook}}, \citenamefont {{Jackson}},
  \citenamefont {{Jenkins}}, \citenamefont {{Lahav}}, \citenamefont {{Lewis}},
  \citenamefont {{Lumsden}}, \citenamefont {{Maddox}}, \citenamefont
  {{Madgwick}}, \citenamefont {{Peterson}}, \citenamefont {{Sutherland}},\ and\
  \citenamefont {{Taylor}}}]{Cole:2005}%
  \BibitemOpen
  \bibfield  {author} {\bibinfo {author} {\bibfnamefont {S.}~\bibnamefont
  {{Cole}}}, \bibinfo {author} {\bibfnamefont {W.~J.}\ \bibnamefont
  {{Percival}}}, \bibinfo {author} {\bibfnamefont {J.~A.}\ \bibnamefont
  {{Peacock}}}, \bibinfo {author} {\bibfnamefont {P.}~\bibnamefont
  {{Norberg}}}, \bibinfo {author} {\bibfnamefont {C.~M.}\ \bibnamefont
  {{Baugh}}}, \bibinfo {author} {\bibfnamefont {C.~S.}\ \bibnamefont
  {{Frenk}}}, \bibinfo {author} {\bibfnamefont {I.}~\bibnamefont {{Baldry}}},
  \bibinfo {author} {\bibfnamefont {J.}~\bibnamefont {{Bland-Hawthorn}}},
  \bibinfo {author} {\bibfnamefont {T.}~\bibnamefont {{Bridges}}}, \bibinfo
  {author} {\bibfnamefont {R.}~\bibnamefont {{Cannon}}}, \bibinfo {author}
  {\bibfnamefont {M.}~\bibnamefont {{Colless}}}, \bibinfo {author}
  {\bibfnamefont {C.}~\bibnamefont {{Collins}}}, \bibinfo {author}
  {\bibfnamefont {W.}~\bibnamefont {{Couch}}}, \bibinfo {author} {\bibfnamefont
  {N.~J.~G.}\ \bibnamefont {{Cross}}}, \bibinfo {author} {\bibfnamefont
  {G.}~\bibnamefont {{Dalton}}}, \bibnamefont {and~others},\ }\href
  {https://doi.org/10.1111/j.1365-2966.2005.09318.x} {\bibfield  {journal}
  {\bibinfo  {journal} {\mnras}\ }\textbf {\bibinfo {volume} {362}},\ \bibinfo
  {pages} {505} (\bibinfo {year} {2005})},\ \Eprint
  {https://arxiv.org/abs/astro-ph/0501174} {arXiv:astro-ph/0501174 [astro-ph]}
  \BibitemShut {NoStop}%
\bibitem [{\citenamefont {{Satpathy}}\ \emph {et~al.}(2017)\citenamefont
  {{Satpathy}}, \citenamefont {{Alam}}, \citenamefont {{Ho}}, \citenamefont
  {{White}}, \citenamefont {{Bahcall}}, \citenamefont {{Beutler}},
  \citenamefont {{Brownstein}}, \citenamefont {{Chuang}}, \citenamefont
  {{Eisenstein}}, \citenamefont {{Grieb}}, \citenamefont {{Kitaura}},
  \citenamefont {{Olmstead}}, \citenamefont {{Percival}}, \citenamefont
  {{Salazar-Albornoz}}, \citenamefont {{S{\'a}nchez}}, \citenamefont {{Seo}},
  \citenamefont {{Thomas}}, \citenamefont {{Tinker}},\ and\ \citenamefont
  {{Tojeiro}}}]{Satpathy:2017}%
  \BibitemOpen
  \bibfield  {author} {\bibinfo {author} {\bibfnamefont {S.}~\bibnamefont
  {{Satpathy}}}, \bibinfo {author} {\bibfnamefont {S.}~\bibnamefont {{Alam}}},
  \bibinfo {author} {\bibfnamefont {S.}~\bibnamefont {{Ho}}}, \bibinfo {author}
  {\bibfnamefont {M.}~\bibnamefont {{White}}}, \bibinfo {author} {\bibfnamefont
  {N.~A.}\ \bibnamefont {{Bahcall}}}, \bibinfo {author} {\bibfnamefont
  {F.}~\bibnamefont {{Beutler}}}, \bibinfo {author} {\bibfnamefont {J.~R.}\
  \bibnamefont {{Brownstein}}}, \bibinfo {author} {\bibfnamefont {C.-H.}\
  \bibnamefont {{Chuang}}}, \bibinfo {author} {\bibfnamefont {D.~J.}\
  \bibnamefont {{Eisenstein}}}, \bibinfo {author} {\bibfnamefont {J.~N.}\
  \bibnamefont {{Grieb}}}, \bibinfo {author} {\bibfnamefont {F.}~\bibnamefont
  {{Kitaura}}}, \bibinfo {author} {\bibfnamefont {M.~D.}\ \bibnamefont
  {{Olmstead}}}, \bibinfo {author} {\bibfnamefont {W.~J.}\ \bibnamefont
  {{Percival}}}, \bibinfo {author} {\bibfnamefont {S.}~\bibnamefont
  {{Salazar-Albornoz}}}, \bibinfo {author} {\bibfnamefont {A.~G.}\ \bibnamefont
  {{S{\'a}nchez}}}, \bibnamefont {and~others},\ }\href
  {https://doi.org/10.1093/mnras/stx883} {\bibfield  {journal} {\bibinfo
  {journal} {\mnras}\ }\textbf {\bibinfo {volume} {469}},\ \bibinfo {pages}
  {1369} (\bibinfo {year} {2017})},\ \Eprint {https://arxiv.org/abs/1607.03148}
  {arXiv:1607.03148 [astro-ph.CO]} \BibitemShut {NoStop}%
\bibitem [{\citenamefont {{Beutler}}\ \emph {et~al.}(2017)\citenamefont
  {{Beutler}}, \citenamefont {{Seo}}, \citenamefont {{Saito}}, \citenamefont
  {{Chuang}}, \citenamefont {{Cuesta}}, \citenamefont {{Eisenstein}},
  \citenamefont {{Gil-Mar{\'\i}n}}, \citenamefont {{Grieb}}, \citenamefont
  {{Hand}}, \citenamefont {{Kitaura}}, \citenamefont {{Modi}}, \citenamefont
  {{Nichol}}, \citenamefont {{Olmstead}}, \citenamefont {{Percival}},
  \citenamefont {{Prada}}, \citenamefont {{S{\'a}nchez}}, \citenamefont
  {{Rodriguez-Torres}}, \citenamefont {{Ross}}, \citenamefont {{Ross}},
  \citenamefont {{Schneider}}, \citenamefont {{Tinker}}, \citenamefont
  {{Tojeiro}},\ and\ \citenamefont {{Vargas-Maga{\~n}a}}}]{Beutler:2017}%
  \BibitemOpen
  \bibfield  {author} {\bibinfo {author} {\bibfnamefont {F.}~\bibnamefont
  {{Beutler}}}, \bibinfo {author} {\bibfnamefont {H.-J.}\ \bibnamefont
  {{Seo}}}, \bibinfo {author} {\bibfnamefont {S.}~\bibnamefont {{Saito}}},
  \bibinfo {author} {\bibfnamefont {C.-H.}\ \bibnamefont {{Chuang}}}, \bibinfo
  {author} {\bibfnamefont {A.~J.}\ \bibnamefont {{Cuesta}}}, \bibinfo {author}
  {\bibfnamefont {D.~J.}\ \bibnamefont {{Eisenstein}}}, \bibinfo {author}
  {\bibfnamefont {H.}~\bibnamefont {{Gil-Mar{\'\i}n}}}, \bibinfo {author}
  {\bibfnamefont {J.~N.}\ \bibnamefont {{Grieb}}}, \bibinfo {author}
  {\bibfnamefont {N.}~\bibnamefont {{Hand}}}, \bibinfo {author} {\bibfnamefont
  {F.-S.}\ \bibnamefont {{Kitaura}}}, \bibinfo {author} {\bibfnamefont
  {C.}~\bibnamefont {{Modi}}}, \bibinfo {author} {\bibfnamefont {R.~C.}\
  \bibnamefont {{Nichol}}}, \bibinfo {author} {\bibfnamefont {M.~D.}\
  \bibnamefont {{Olmstead}}}, \bibinfo {author} {\bibfnamefont {W.~J.}\
  \bibnamefont {{Percival}}}, \bibinfo {author} {\bibfnamefont
  {F.}~\bibnamefont {{Prada}}}, \bibnamefont {and~others},\ }\href
  {https://doi.org/10.1093/mnras/stw3298} {\bibfield  {journal} {\bibinfo
  {journal} {\mnras}\ }\textbf {\bibinfo {volume} {466}},\ \bibinfo {pages}
  {2242} (\bibinfo {year} {2017})},\ \Eprint {https://arxiv.org/abs/1607.03150}
  {arXiv:1607.03150 [astro-ph.CO]} \BibitemShut {NoStop}%
\bibitem [{\citenamefont {{Alcock}}\ and\ \citenamefont
  {{Paczynski}}(1979)}]{Alcock:1979}%
  \BibitemOpen
  \bibfield  {author} {\bibinfo {author} {\bibfnamefont {C.}~\bibnamefont
  {{Alcock}}}\ \bibnamefont {and}\ \bibinfo {author} {\bibfnamefont
  {B.}~\bibnamefont {{Paczynski}}},\ }\href {https://doi.org/10.1038/281358a0}
  {\bibfield  {journal} {\bibinfo  {journal} {\nat}\ }\textbf {\bibinfo
  {volume} {281}},\ \bibinfo {pages} {358} (\bibinfo {year}
  {1979})}\BibitemShut {NoStop}%
\bibitem [{\citenamefont {{Nadathur}}\ \emph {et~al.}(2019)\citenamefont
  {{Nadathur}}, \citenamefont {{Carter}}, \citenamefont {{Percival}},
  \citenamefont {{Winther}},\ and\ \citenamefont {{Bautista}}}]{Nadathur:2019}%
  \BibitemOpen
  \bibfield  {author} {\bibinfo {author} {\bibfnamefont {S.}~\bibnamefont
  {{Nadathur}}}, \bibinfo {author} {\bibfnamefont {P.~M.}\ \bibnamefont
  {{Carter}}}, \bibinfo {author} {\bibfnamefont {W.~J.}\ \bibnamefont
  {{Percival}}}, \bibinfo {author} {\bibfnamefont {H.~A.}\ \bibnamefont
  {{Winther}}},\ \bibnamefont {and}\ \bibinfo {author} {\bibfnamefont {J.~E.}\
  \bibnamefont {{Bautista}}},\ }\href
  {https://doi.org/10.1103/PhysRevD.100.023504} {\bibfield  {journal} {\bibinfo
   {journal} {\prd}\ }\textbf {\bibinfo {volume} {100}},\ \bibinfo {eid}
  {023504} (\bibinfo {year} {2019})},\ \Eprint
  {https://arxiv.org/abs/1904.01030} {arXiv:1904.01030 [astro-ph.CO]}
  \BibitemShut {NoStop}%
\bibitem [{\citenamefont {{du Mas des Bourboux}}\ \emph
  {et~al.}(2020)\citenamefont {{du Mas des Bourboux}}, \citenamefont {{Rich}},
  \citenamefont {{Font-Ribera}}, \citenamefont {{de Sainte Agathe}},
  \citenamefont {{Farr}}, \citenamefont {{Etourneau}}, \citenamefont {{Le
  Goff}}, \citenamefont {{Cuceu}}, \citenamefont {{Balland}}, \citenamefont
  {{Bautista}}, \citenamefont {{Blomqvist}}, \citenamefont {{Brinkmann}},
  \citenamefont {{Brownstein}}, \citenamefont {{Chabanier}}, \citenamefont
  {{Chaussidon}}, \citenamefont {{Dawson}}, \citenamefont
  {{Gonz{\'a}lez-Morales}}, \citenamefont {{Guy}}, \citenamefont {{Lyke}},
  \citenamefont {{de la Macorra}}, \citenamefont {{Mueller}}, \citenamefont
  {{Myers}}, \citenamefont {{Nitschelm}}, \citenamefont {{Mu{\~n}oz
  Guti{\'e}rrez}}, \citenamefont {{Palanque-Delabrouille}}, \citenamefont
  {{Parker}}, \citenamefont {{Percival}}, \citenamefont
  {{P{\'e}rez-R{\`a}fols}}, \citenamefont {{Petitjean}}, \citenamefont
  {{Pieri}}, \citenamefont {{Ravoux}}, \citenamefont {{Rossi}}, \citenamefont
  {{Schneider}}, \citenamefont {{Seo}}, \citenamefont {{Slosar}}, \citenamefont
  {{Stermer}}, \citenamefont {{Vivek}}, \citenamefont {{Y{\`e}che}},\ and\
  \citenamefont {{Youles}}}]{duMasdesBourboux:2020}%
  \BibitemOpen
  \bibfield  {author} {\bibinfo {author} {\bibfnamefont {H.}~\bibnamefont {{du
  Mas des Bourboux}}}, \bibinfo {author} {\bibfnamefont {J.}~\bibnamefont
  {{Rich}}}, \bibinfo {author} {\bibfnamefont {A.}~\bibnamefont
  {{Font-Ribera}}}, \bibinfo {author} {\bibfnamefont {V.}~\bibnamefont {{de
  Sainte Agathe}}}, \bibinfo {author} {\bibfnamefont {J.}~\bibnamefont
  {{Farr}}}, \bibinfo {author} {\bibfnamefont {T.}~\bibnamefont {{Etourneau}}},
  \bibinfo {author} {\bibfnamefont {J.-M.}\ \bibnamefont {{Le Goff}}}, \bibinfo
  {author} {\bibfnamefont {A.}~\bibnamefont {{Cuceu}}}, \bibinfo {author}
  {\bibfnamefont {C.}~\bibnamefont {{Balland}}}, \bibinfo {author}
  {\bibfnamefont {J.~E.}\ \bibnamefont {{Bautista}}}, \bibinfo {author}
  {\bibfnamefont {M.}~\bibnamefont {{Blomqvist}}}, \bibinfo {author}
  {\bibfnamefont {J.}~\bibnamefont {{Brinkmann}}}, \bibinfo {author}
  {\bibfnamefont {J.~R.}\ \bibnamefont {{Brownstein}}}, \bibinfo {author}
  {\bibfnamefont {S.}~\bibnamefont {{Chabanier}}}, \bibinfo {author}
  {\bibfnamefont {E.}~\bibnamefont {{Chaussidon}}}, \bibnamefont {and~others},\
  }\href {https://doi.org/10.3847/1538-4357/abb085} {\bibfield  {journal}
  {\bibinfo  {journal} {\apj}\ }\textbf {\bibinfo {volume} {901}},\ \bibinfo
  {eid} {153} (\bibinfo {year} {2020})},\ \Eprint
  {https://arxiv.org/abs/2007.08995} {arXiv:2007.08995 [astro-ph.CO]}
  \BibitemShut {NoStop}%
\bibitem [{\citenamefont {{Ahumada}}\ \emph {et~al.}(2020)\citenamefont
  {{Ahumada}}, \citenamefont {{Prieto}}, \citenamefont {{Almeida}},
  \citenamefont {{Anders}}, \citenamefont {{Anderson}}, \citenamefont
  {{Andrews}}, \citenamefont {{Anguiano}}, \citenamefont {{Arcodia}},
  \citenamefont {{Armengaud}}, \citenamefont {{Aubert}}, \citenamefont
  {{Avila}}, \citenamefont {{Avila-Reese}}, \citenamefont {{Badenes}},
  \citenamefont {{Balland}}, \citenamefont {{Barger}}, \citenamefont
  {{Barrera-Ballesteros}}, \citenamefont {{Basu}}, \citenamefont {{Bautista}},
  \citenamefont {{Beaton}}, \citenamefont {{Beers}}, \citenamefont
  {{Benavides}}, \citenamefont {{Bender}}, \citenamefont {{Bernardi}},
  \citenamefont {{Bershady}}, \citenamefont {{Beutler}}, \citenamefont
  {{Bidin}}, \citenamefont {{Bird}}, \citenamefont {{Bizyaev}}, \citenamefont
  {{Blanc}}, \citenamefont {{Blanton}}, \citenamefont {{Boquien}},
  \citenamefont {{Borissova}}, \citenamefont {{Bovy}}, \citenamefont
  {{Brandt}}, \citenamefont {{Brinkmann}}, \citenamefont {{Brownstein}},
  \citenamefont {{Bundy}}, \citenamefont {{Bureau}}, \citenamefont
  {{Burgasser}}, \citenamefont {{Burtin}}, \citenamefont {{Cano-D{\'\i}az}},
  \citenamefont {{Capasso}}, \citenamefont {{Cappellari}}, \citenamefont
  {{Carrera}}, \citenamefont {{Chabanier}}, \citenamefont {{Chaplin}},
  \citenamefont {{Chapman}}, \citenamefont {{Cherinka}}, \citenamefont
  {{Chiappini}}, \citenamefont {{Doohyun Choi}}, \citenamefont {{Chojnowski}},
  \citenamefont {{Chung}}, \citenamefont {{Clerc}}, \citenamefont {{Coffey}},
  \citenamefont {{Comerford}}, \citenamefont {{Comparat}}, \citenamefont {{da
  Costa}}, \citenamefont {{Cousinou}}, \citenamefont {{Covey}}, \citenamefont
  {{Crane}}, \citenamefont {{Cunha}}, \citenamefont {{Ilha}}, \citenamefont
  {{Dai}}, \citenamefont {{Damsted}}, \citenamefont {{Darling}}, \citenamefont
  {{Davidson}}, \citenamefont {{Davies}}, \citenamefont {{Dawson}},
  \citenamefont {{De}}, \citenamefont {{de la Macorra}}, \citenamefont {{De
  Lee}}, \citenamefont {{Queiroz}}, \citenamefont {{Deconto Machado}},
  \citenamefont {{de la Torre}}, \citenamefont {{Dell'Agli}}, \citenamefont
  {{du Mas des Bourboux}}, \citenamefont {{Diamond-Stanic}}, \citenamefont
  {{Dillon}}, \citenamefont {{Donor}}, \citenamefont {{Drory}}, \citenamefont
  {{Duckworth}}, \citenamefont {{Dwelly}}, \citenamefont {{Ebelke}},
  \citenamefont {{Eftekharzadeh}}, \citenamefont {{Davis Eigenbrot}},
  \citenamefont {{Elsworth}}, \citenamefont {{Eracleous}}, \citenamefont
  {{Erfanianfar}}, \citenamefont {{Escoffier}}, \citenamefont {{Fan}},
  \citenamefont {{Farr}}, \citenamefont {{Fern{\'a}ndez-Trincado}},
  \citenamefont {{Feuillet}}, \citenamefont {{Finoguenov}}, \citenamefont
  {{Fofie}}, \citenamefont {{Fraser-McKelvie}}, \citenamefont {{Frinchaboy}},
  \citenamefont {{Fromenteau}}, \citenamefont {{Fu}}, \citenamefont
  {{Galbany}}, \citenamefont {{Garcia}}, \citenamefont
  {{Garc{\'\i}a-Hern{\'a}ndez}}, \citenamefont {{Oehmichen}}, \citenamefont
  {{Ge}}, \citenamefont {{Maia}}, \citenamefont {{Geisler}}, \citenamefont
  {{Gelfand}}, \citenamefont {{Goddy}}, \citenamefont {{Gonzalez-Perez}},
  \citenamefont {{Grabowski}}, \citenamefont {{Green}}, \citenamefont
  {{Grier}}, \citenamefont {{Guo}}, \citenamefont {{Guy}}, \citenamefont
  {{Harding}}, \citenamefont {{Hasselquist}}, \citenamefont {{Hawken}},
  \citenamefont {{Hayes}}, \citenamefont {{Hearty}}, \citenamefont {{Hekker}},
  \citenamefont {{Hogg}}, \citenamefont {{Holtzman}}, \citenamefont {{Horta}},
  \citenamefont {{Hou}}, \citenamefont {{Hsieh}}, \citenamefont {{Huber}},
  \citenamefont {{Hunt}}, \citenamefont {{Chitham}}, \citenamefont {{Imig}},
  \citenamefont {{Jaber}}, \citenamefont {{Angel}}, \citenamefont {{Johnson}},
  \citenamefont {{Jones}}, \citenamefont {{J{\"o}nsson}}, \citenamefont
  {{Jullo}}, \citenamefont {{Kim}}, \citenamefont {{Kinemuchi}}, \citenamefont
  {{Kirkpatrick}}, \citenamefont {{Kite}}, \citenamefont {{Klaene}},
  \citenamefont {{Kneib}}, \citenamefont {{Kollmeier}}, \citenamefont {{Kong}},
  \citenamefont {{Kounkel}}, \citenamefont {{Krishnarao}}, \citenamefont
  {{Lacerna}}, \citenamefont {{Lan}}, \citenamefont {{Lane}}, \citenamefont
  {{Law}}, \citenamefont {{Le Goff}}, \citenamefont {{Leung}}, \citenamefont
  {{Lewis}}, \citenamefont {{Li}}, \citenamefont {{Lian}}, \citenamefont
  {{Lin}}, \citenamefont {{Long}}, \citenamefont {{Longa-Pe{\~n}a}},
  \citenamefont {{Lundgren}}, \citenamefont {{Lyke}}, \citenamefont {{Ted
  Mackereth}}, \citenamefont {{MacLeod}}, \citenamefont {{Majewski}},
  \citenamefont {{Manchado}}, \citenamefont {{Maraston}}, \citenamefont
  {{Martini}}, \citenamefont {{Masseron}}, \citenamefont {{Masters}},
  \citenamefont {{Mathur}}, \citenamefont {{McDermid}}, \citenamefont
  {{Merloni}}, \citenamefont {{Merrifield}}, \citenamefont
  {{M{\'e}sz{\'a}ros}}, \citenamefont {{Miglio}}, \citenamefont {{Minniti}},
  \citenamefont {{Minsley}}, \citenamefont {{Miyaji}}, \citenamefont
  {{Mohammad}}, \citenamefont {{Mosser}}, \citenamefont {{Mueller}},
  \citenamefont {{Muna}}, \citenamefont {{Mu{\~n}oz-Guti{\'e}rrez}},
  \citenamefont {{Myers}}, \citenamefont {{Nadathur}}, \citenamefont {{Nair}},
  \citenamefont {{Nandra}}, \citenamefont {{do Nascimento}}, \citenamefont
  {{Nevin}}, \citenamefont {{Newman}}, \citenamefont {{Nidever}}, \citenamefont
  {{Nitschelm}}, \citenamefont {{Noterdaeme}}, \citenamefont {{O'Connell}},
  \citenamefont {{Olmstead}}, \citenamefont {{Oravetz}}, \citenamefont
  {{Oravetz}}, \citenamefont {{Osorio}}, \citenamefont {{Pace}}, \citenamefont
  {{Padilla}}, \citenamefont {{Palanque-Delabrouille}}, \citenamefont
  {{Palicio}}, \citenamefont {{Pan}}, \citenamefont {{Pan}}, \citenamefont
  {{Parker}}, \citenamefont {{Paviot}}, \citenamefont {{Peirani}},
  \citenamefont {{Ram{\'r}ez}}, \citenamefont {{Penny}}, \citenamefont
  {{Percival}}, \citenamefont {{Perez-Fournon}}, \citenamefont
  {{P{\'e}rez-R{\`a}fols}}, \citenamefont {{Petitjean}}, \citenamefont
  {{Pieri}}, \citenamefont {{Pinsonneault}}, \citenamefont {{Poovelil}},
  \citenamefont {{Povick}}, \citenamefont {{Prakash}}, \citenamefont
  {{Price-Whelan}}, \citenamefont {{Raddick}}, \citenamefont {{Raichoor}},
  \citenamefont {{Ray}}, \citenamefont {{Rembold}}, \citenamefont {{Rezaie}},
  \citenamefont {{Riffel}}, \citenamefont {{Riffel}}, \citenamefont {{Rix}},
  \citenamefont {{Robin}}, \citenamefont {{Roman-Lopes}}, \citenamefont
  {{Rom{\'a}n-Z{\'u}{\~n}iga}}, \citenamefont {{Rose}}, \citenamefont {{Ross}},
  \citenamefont {{Rossi}}, \citenamefont {{Rowlands}}, \citenamefont {{Rubin}},
  \citenamefont {{Salvato}}, \citenamefont {{S{\'a}nchez}}, \citenamefont
  {{S{\'a}nchez-Menguiano}}, \citenamefont {{S{\'a}nchez-Gallego}},
  \citenamefont {{Sayres}}, \citenamefont {{Schaefer}}, \citenamefont
  {{Schiavon}}, \citenamefont {{Schimoia}}, \citenamefont {{Schlafly}},
  \citenamefont {{Schlegel}}, \citenamefont {{Schneider}}, \citenamefont
  {{Schultheis}}, \citenamefont {{Schwope}}, \citenamefont {{Seo}},
  \citenamefont {{Serenelli}}, \citenamefont {{Shafieloo}}, \citenamefont
  {{Shamsi}}, \citenamefont {{Shao}}, \citenamefont {{Shen}}, \citenamefont
  {{Shetrone}}, \citenamefont {{Shirley}}, \citenamefont {{Aguirre}},
  \citenamefont {{Simon}}, \citenamefont {{Skrutskie}}, \citenamefont
  {{Slosar}}, \citenamefont {{Smethurst}}, \citenamefont {{Sobeck}},
  \citenamefont {{Sodi}}, \citenamefont {{Souto}}, \citenamefont {{Stark}},
  \citenamefont {{Stassun}}, \citenamefont {{Steinmetz}}, \citenamefont
  {{Stello}}, \citenamefont {{Stermer}}, \citenamefont {{Storchi-Bergmann}},
  \citenamefont {{Streblyanska}}, \citenamefont {{Stringfellow}}, \citenamefont
  {{Stutz}}, \citenamefont {{Su{\'a}rez}}, \citenamefont {{Sun}}, \citenamefont
  {{Taghizadeh-Popp}}, \citenamefont {{Talbot}}, \citenamefont {{Tayar}},
  \citenamefont {{Thakar}}, \citenamefont {{Theriault}}, \citenamefont
  {{Thomas}}, \citenamefont {{Thomas}}, \citenamefont {{Tinker}}, \citenamefont
  {{Tojeiro}}, \citenamefont {{Toledo}}, \citenamefont {{Tremonti}},
  \citenamefont {{Troup}}, \citenamefont {{Tuttle}}, \citenamefont
  {{Unda-Sanzana}}, \citenamefont {{Valentini}}, \citenamefont
  {{Vargas-Gonz{\'a}lez}}, \citenamefont {{Vargas-Maga{\~n}a}}, \citenamefont
  {{V{\'a}zquez-Mata}}, \citenamefont {{Vivek}}, \citenamefont {{Wake}},
  \citenamefont {{Wang}}, \citenamefont {{Weaver}}, \citenamefont {{Weijmans}},
  \citenamefont {{Wild}}, \citenamefont {{Wilson}}, \citenamefont {{Wilson}},
  \citenamefont {{Wolthuis}}, \citenamefont {{Wood-Vasey}}, \citenamefont
  {{Yan}}, \citenamefont {{Yang}}, \citenamefont {{Y{\`e}che}}, \citenamefont
  {{Zamora}}, \citenamefont {{Zarrouk}}, \citenamefont {{Zasowski}},
  \citenamefont {{Zhang}}, \citenamefont {{Zhao}}, \citenamefont {{Zhao}},
  \citenamefont {{Zheng}}, \citenamefont {{Zheng}}, \citenamefont {{Zhu}},\
  and\ \citenamefont {{Zou}}}]{Ahumada:2020}%
  \BibitemOpen
  \bibfield  {author} {\bibinfo {author} {\bibfnamefont {R.}~\bibnamefont
  {{Ahumada}}}, \bibinfo {author} {\bibfnamefont {C.~A.}\ \bibnamefont
  {{Prieto}}}, \bibinfo {author} {\bibfnamefont {A.}~\bibnamefont {{Almeida}}},
  \bibinfo {author} {\bibfnamefont {F.}~\bibnamefont {{Anders}}}, \bibinfo
  {author} {\bibfnamefont {S.~F.}\ \bibnamefont {{Anderson}}}, \bibinfo
  {author} {\bibfnamefont {B.~H.}\ \bibnamefont {{Andrews}}}, \bibinfo {author}
  {\bibfnamefont {B.}~\bibnamefont {{Anguiano}}}, \bibinfo {author}
  {\bibfnamefont {R.}~\bibnamefont {{Arcodia}}}, \bibinfo {author}
  {\bibfnamefont {E.}~\bibnamefont {{Armengaud}}}, \bibinfo {author}
  {\bibfnamefont {M.}~\bibnamefont {{Aubert}}}, \bibinfo {author}
  {\bibfnamefont {S.}~\bibnamefont {{Avila}}}, \bibinfo {author} {\bibfnamefont
  {V.}~\bibnamefont {{Avila-Reese}}}, \bibinfo {author} {\bibfnamefont
  {C.}~\bibnamefont {{Badenes}}}, \bibinfo {author} {\bibfnamefont
  {C.}~\bibnamefont {{Balland}}}, \bibinfo {author} {\bibfnamefont
  {K.}~\bibnamefont {{Barger}}}, \bibnamefont {and~others},\ }\href
  {https://doi.org/10.3847/1538-4365/ab929e} {\bibfield  {journal} {\bibinfo
  {journal} {\apjs}\ }\textbf {\bibinfo {volume} {249}},\ \bibinfo {eid} {3}
  (\bibinfo {year} {2020})},\ \Eprint {https://arxiv.org/abs/1912.02905}
  {arXiv:1912.02905 [astro-ph.GA]} \BibitemShut {NoStop}%
\bibitem [{\citenamefont {{Eisenstein}}\ \emph {et~al.}(2011)\citenamefont
  {{Eisenstein}}, \citenamefont {{Weinberg}}, \citenamefont {{Agol}},
  \citenamefont {{Aihara}}, \citenamefont {{Allende Prieto}}, \citenamefont
  {{Anderson}}, \citenamefont {{Arns}}, \citenamefont {{Aubourg}},
  \citenamefont {{Bailey}}, \citenamefont {{Balbinot}}, \citenamefont
  {{Barkhouser}}, \citenamefont {{Beers}}, \citenamefont {{Berlind}},
  \citenamefont {{Bickerton}}, \citenamefont {{Bizyaev}}, \citenamefont
  {{Blanton}}, \citenamefont {{Bochanski}}, \citenamefont {{Bolton}},
  \citenamefont {{Bosman}}, \citenamefont {{Bovy}}, \citenamefont {{Brandt}},
  \citenamefont {{Breslauer}}, \citenamefont {{Brewington}}, \citenamefont
  {{Brinkmann}}, \citenamefont {{Brown}}, \citenamefont {{Brownstein}},
  \citenamefont {{Burger}}, \citenamefont {{Busca}}, \citenamefont
  {{Campbell}}, \citenamefont {{Cargile}}, \citenamefont {{Carithers}},
  \citenamefont {{Carlberg}}, \citenamefont {{Carr}}, \citenamefont {{Chang}},
  \citenamefont {{Chen}}, \citenamefont {{Chiappini}}, \citenamefont
  {{Comparat}}, \citenamefont {{Connolly}}, \citenamefont {{Cortes}},
  \citenamefont {{Croft}}, \citenamefont {{Cunha}}, \citenamefont {{da Costa}},
  \citenamefont {{Davenport}}, \citenamefont {{Dawson}}, \citenamefont {{De
  Lee}}, \citenamefont {{Porto de Mello}}, \citenamefont {{de Simoni}},
  \citenamefont {{Dean}}, \citenamefont {{Dhital}}, \citenamefont {{Ealet}},
  \citenamefont {{Ebelke}}, \citenamefont {{Edmondson}}, \citenamefont
  {{Eiting}}, \citenamefont {{Escoffier}}, \citenamefont {{Esposito}},
  \citenamefont {{Evans}}, \citenamefont {{Fan}}, \citenamefont {{Femen{\'\i}a
  Castell{\'a}}}, \citenamefont {{Dutra Ferreira}}, \citenamefont
  {{Fitzgerald}}, \citenamefont {{Fleming}}, \citenamefont {{Font-Ribera}},
  \citenamefont {{Ford}}, \citenamefont {{Frinchaboy}}, \citenamefont
  {{Garc{\'\i}a P{\'e}rez}}, \citenamefont {{Gaudi}}, \citenamefont {{Ge}},
  \citenamefont {{Ghezzi}}, \citenamefont {{Gillespie}}, \citenamefont
  {{Gilmore}}, \citenamefont {{Girardi}}, \citenamefont {{Gott}}, \citenamefont
  {{Gould}}, \citenamefont {{Grebel}}, \citenamefont {{Gunn}}, \citenamefont
  {{Hamilton}}, \citenamefont {{Harding}}, \citenamefont {{Harris}},
  \citenamefont {{Hawley}}, \citenamefont {{Hearty}}, \citenamefont
  {{Hennawi}}, \citenamefont {{Gonz{\'a}lez Hern{\'a}ndez}}, \citenamefont
  {{Ho}}, \citenamefont {{Hogg}}, \citenamefont {{Holtzman}}, \citenamefont
  {{Honscheid}}, \citenamefont {{Inada}}, \citenamefont {{Ivans}},
  \citenamefont {{Jiang}}, \citenamefont {{Jiang}}, \citenamefont {{Johnson}},
  \citenamefont {{Jordan}}, \citenamefont {{Jordan}}, \citenamefont
  {{Kauffmann}}, \citenamefont {{Kazin}}, \citenamefont {{Kirkby}},
  \citenamefont {{Klaene}}, \citenamefont {{Knapp}}, \citenamefont {{Kneib}},
  \citenamefont {{Kochanek}}, \citenamefont {{Koesterke}}, \citenamefont
  {{Kollmeier}}, \citenamefont {{Kron}}, \citenamefont {{Lampeitl}},
  \citenamefont {{Lang}}, \citenamefont {{Lawler}}, \citenamefont {{Le Goff}},
  \citenamefont {{Lee}}, \citenamefont {{Lee}}, \citenamefont {{Leisenring}},
  \citenamefont {{Lin}}, \citenamefont {{Liu}}, \citenamefont {{Long}},
  \citenamefont {{Loomis}}, \citenamefont {{Lucatello}}, \citenamefont
  {{Lundgren}}, \citenamefont {{Lupton}}, \citenamefont {{Ma}}, \citenamefont
  {{Ma}}, \citenamefont {{MacDonald}}, \citenamefont {{Mack}}, \citenamefont
  {{Mahadevan}}, \citenamefont {{Maia}}, \citenamefont {{Majewski}},
  \citenamefont {{Makler}}, \citenamefont {{Malanushenko}}, \citenamefont
  {{Malanushenko}}, \citenamefont {{Mandelbaum}}, \citenamefont {{Maraston}},
  \citenamefont {{Margala}}, \citenamefont {{Maseman}}, \citenamefont
  {{Masters}}, \citenamefont {{McBride}}, \citenamefont {{McDonald}},
  \citenamefont {{McGreer}}, \citenamefont {{McMahon}}, \citenamefont {{Mena
  Requejo}}, \citenamefont {{M{\'e}nard}}, \citenamefont
  {{Miralda-Escud{\'e}}}, \citenamefont {{Morrison}}, \citenamefont
  {{Mullally}}, \citenamefont {{Muna}}, \citenamefont {{Murayama}},
  \citenamefont {{Myers}}, \citenamefont {{Naugle}}, \citenamefont {{Neto}},
  \citenamefont {{Nguyen}}, \citenamefont {{Nichol}}, \citenamefont
  {{Nidever}}, \citenamefont {{O'Connell}}, \citenamefont {{Ogando}},
  \citenamefont {{Olmstead}}, \citenamefont {{Oravetz}}, \citenamefont
  {{Padmanabhan}}, \citenamefont {{Paegert}}, \citenamefont
  {{Palanque-Delabrouille}}, \citenamefont {{Pan}}, \citenamefont {{Pandey}},
  \citenamefont {{Parejko}}, \citenamefont {{P{\^a}ris}}, \citenamefont
  {{Pellegrini}}, \citenamefont {{Pepper}}, \citenamefont {{Percival}},
  \citenamefont {{Petitjean}}, \citenamefont {{Pfaffenberger}}, \citenamefont
  {{Pforr}}, \citenamefont {{Phleps}}, \citenamefont {{Pichon}}, \citenamefont
  {{Pieri}}, \citenamefont {{Prada}}, \citenamefont {{Price-Whelan}},
  \citenamefont {{Raddick}}, \citenamefont {{Ramos}}, \citenamefont {{Reid}},
  \citenamefont {{Reyle}}, \citenamefont {{Rich}}, \citenamefont {{Richards}},
  \citenamefont {{Rieke}}, \citenamefont {{Rieke}}, \citenamefont {{Rix}},
  \citenamefont {{Robin}}, \citenamefont {{Rocha-Pinto}}, \citenamefont
  {{Rockosi}}, \citenamefont {{Roe}}, \citenamefont {{Rollinde}}, \citenamefont
  {{Ross}}, \citenamefont {{Ross}}, \citenamefont {{Rossetto}}, \citenamefont
  {{S{\'a}nchez}}, \citenamefont {{Santiago}}, \citenamefont {{Sayres}},
  \citenamefont {{Schiavon}}, \citenamefont {{Schlegel}}, \citenamefont
  {{Schlesinger}}, \citenamefont {{Schmidt}}, \citenamefont {{Schneider}},
  \citenamefont {{Sellgren}}, \citenamefont {{Shelden}}, \citenamefont
  {{Sheldon}}, \citenamefont {{Shetrone}}, \citenamefont {{Shu}}, \citenamefont
  {{Silverman}}, \citenamefont {{Simmerer}}, \citenamefont {{Simmons}},
  \citenamefont {{Sivarani}}, \citenamefont {{Skrutskie}}, \citenamefont
  {{Slosar}}, \citenamefont {{Smee}}, \citenamefont {{Smith}}, \citenamefont
  {{Snedden}}, \citenamefont {{Stassun}}, \citenamefont {{Steele}},
  \citenamefont {{Steinmetz}}, \citenamefont {{Stockett}}, \citenamefont
  {{Stollberg}}, \citenamefont {{Strauss}}, \citenamefont {{Szalay}},
  \citenamefont {{Tanaka}}, \citenamefont {{Thakar}}, \citenamefont {{Thomas}},
  \citenamefont {{Tinker}}, \citenamefont {{Tofflemire}}, \citenamefont
  {{Tojeiro}}, \citenamefont {{Tremonti}}, \citenamefont {{Vargas Maga{\~n}a}},
  \citenamefont {{Verde}}, \citenamefont {{Vogt}}, \citenamefont {{Wake}},
  \citenamefont {{Wan}}, \citenamefont {{Wang}}, \citenamefont {{Weaver}},
  \citenamefont {{White}}, \citenamefont {{White}}, \citenamefont {{Wilson}},
  \citenamefont {{Wisniewski}}, \citenamefont {{Wood-Vasey}}, \citenamefont
  {{Yanny}}, \citenamefont {{Yasuda}}, \citenamefont {{Y{\`e}che}},
  \citenamefont {{York}}, \citenamefont {{Young}}, \citenamefont {{Zasowski}},
  \citenamefont {{Zehavi}},\ and\ \citenamefont {{Zhao}}}]{Eisenstein:2011}%
  \BibitemOpen
  \bibfield  {author} {\bibinfo {author} {\bibfnamefont {D.~J.}\ \bibnamefont
  {{Eisenstein}}}, \bibinfo {author} {\bibfnamefont {D.~H.}\ \bibnamefont
  {{Weinberg}}}, \bibinfo {author} {\bibfnamefont {E.}~\bibnamefont {{Agol}}},
  \bibinfo {author} {\bibfnamefont {H.}~\bibnamefont {{Aihara}}}, \bibinfo
  {author} {\bibfnamefont {C.}~\bibnamefont {{Allende Prieto}}}, \bibinfo
  {author} {\bibfnamefont {S.~F.}\ \bibnamefont {{Anderson}}}, \bibinfo
  {author} {\bibfnamefont {J.~A.}\ \bibnamefont {{Arns}}}, \bibinfo {author}
  {\bibfnamefont {{\'E}.}~\bibnamefont {{Aubourg}}}, \bibinfo {author}
  {\bibfnamefont {S.}~\bibnamefont {{Bailey}}}, \bibinfo {author}
  {\bibfnamefont {E.}~\bibnamefont {{Balbinot}}}, \bibinfo {author}
  {\bibfnamefont {R.}~\bibnamefont {{Barkhouser}}}, \bibinfo {author}
  {\bibfnamefont {T.~C.}\ \bibnamefont {{Beers}}}, \bibinfo {author}
  {\bibfnamefont {A.~A.}\ \bibnamefont {{Berlind}}}, \bibinfo {author}
  {\bibfnamefont {S.~J.}\ \bibnamefont {{Bickerton}}}, \bibinfo {author}
  {\bibfnamefont {D.}~\bibnamefont {{Bizyaev}}}, \bibnamefont {and~others},\
  }\href {https://doi.org/10.1088/0004-6256/142/3/72} {\bibfield  {journal}
  {\bibinfo  {journal} {\aj}\ }\textbf {\bibinfo {volume} {142}},\ \bibinfo
  {eid} {72} (\bibinfo {year} {2011})},\ \Eprint
  {https://arxiv.org/abs/1101.1529} {arXiv:1101.1529 [astro-ph.IM]}
  \BibitemShut {NoStop}%
\bibitem [{\citenamefont {{Dawson}}\ \emph {et~al.}(2016)\citenamefont
  {{Dawson}}, \citenamefont {{Kneib}}, \citenamefont {{Percival}},
  \citenamefont {{Alam}}, \citenamefont {{Albareti}}, \citenamefont
  {{Anderson}}, \citenamefont {{Armengaud}}, \citenamefont {{Aubourg}},
  \citenamefont {{Bailey}}, \citenamefont {{Bautista}}, \citenamefont
  {{Berlind}}, \citenamefont {{Bershady}}, \citenamefont {{Beutler}},
  \citenamefont {{Bizyaev}}, \citenamefont {{Blanton}}, \citenamefont
  {{Blomqvist}}, \citenamefont {{Bolton}}, \citenamefont {{Bovy}},
  \citenamefont {{Brandt}}, \citenamefont {{Brinkmann}}, \citenamefont
  {{Brownstein}}, \citenamefont {{Burtin}}, \citenamefont {{Busca}},
  \citenamefont {{Cai}}, \citenamefont {{Chuang}}, \citenamefont {{Clerc}},
  \citenamefont {{Comparat}}, \citenamefont {{Cope}}, \citenamefont {{Croft}},
  \citenamefont {{Cruz-Gonzalez}}, \citenamefont {{da Costa}}, \citenamefont
  {{Cousinou}}, \citenamefont {{Darling}}, \citenamefont {{de la Macorra}},
  \citenamefont {{de la Torre}}, \citenamefont {{Delubac}}, \citenamefont {{du
  Mas des Bourboux}}, \citenamefont {{Dwelly}}, \citenamefont {{Ealet}},
  \citenamefont {{Eisenstein}}, \citenamefont {{Eracleous}}, \citenamefont
  {{Escoffier}}, \citenamefont {{Fan}}, \citenamefont {{Finoguenov}},
  \citenamefont {{Font-Ribera}}, \citenamefont {{Frinchaboy}}, \citenamefont
  {{Gaulme}}, \citenamefont {{Georgakakis}}, \citenamefont {{Green}},
  \citenamefont {{Guo}}, \citenamefont {{Guy}}, \citenamefont {{Ho}},
  \citenamefont {{Holder}}, \citenamefont {{Huehnerhoff}}, \citenamefont
  {{Hutchinson}}, \citenamefont {{Jing}}, \citenamefont {{Jullo}},
  \citenamefont {{Kamble}}, \citenamefont {{Kinemuchi}}, \citenamefont
  {{Kirkby}}, \citenamefont {{Kitaura}}, \citenamefont {{Klaene}},
  \citenamefont {{Laher}}, \citenamefont {{Lang}}, \citenamefont {{Laurent}},
  \citenamefont {{Le Goff}}, \citenamefont {{Li}}, \citenamefont {{Liang}},
  \citenamefont {{Lima}}, \citenamefont {{Lin}}, \citenamefont {{Lin}},
  \citenamefont {{Lin}}, \citenamefont {{Long}}, \citenamefont {{Lundgren}},
  \citenamefont {{MacDonald}}, \citenamefont {{Geimba Maia}}, \citenamefont
  {{Malanushenko}}, \citenamefont {{Malanushenko}}, \citenamefont
  {{Mariappan}}, \citenamefont {{McBride}}, \citenamefont {{McGreer}},
  \citenamefont {{M{\'e}nard}}, \citenamefont {{Merloni}}, \citenamefont
  {{Meza}}, \citenamefont {{Montero-Dorta}}, \citenamefont {{Muna}},
  \citenamefont {{Myers}}, \citenamefont {{Nandra}}, \citenamefont {{Naugle}},
  \citenamefont {{Newman}}, \citenamefont {{Noterdaeme}}, \citenamefont
  {{Nugent}}, \citenamefont {{Ogando}}, \citenamefont {{Olmstead}},
  \citenamefont {{Oravetz}}, \citenamefont {{Oravetz}}, \citenamefont
  {{Padmanabhan}}, \citenamefont {{Palanque-Delabrouille}}, \citenamefont
  {{Pan}}, \citenamefont {{Parejko}}, \citenamefont {{P{\^a}ris}},
  \citenamefont {{Peacock}}, \citenamefont {{Petitjean}}, \citenamefont
  {{Pieri}}, \citenamefont {{Pisani}}, \citenamefont {{Prada}}, \citenamefont
  {{Prakash}}, \citenamefont {{Raichoor}}, \citenamefont {{Reid}},
  \citenamefont {{Rich}}, \citenamefont {{Ridl}}, \citenamefont
  {{Rodriguez-Torres}}, \citenamefont {{Carnero Rosell}}, \citenamefont
  {{Ross}}, \citenamefont {{Rossi}}, \citenamefont {{Ruan}}, \citenamefont
  {{Salvato}}, \citenamefont {{Sayres}}, \citenamefont {{Schneider}},
  \citenamefont {{Schlegel}}, \citenamefont {{Seljak}}, \citenamefont {{Seo}},
  \citenamefont {{Sesar}}, \citenamefont {{Shandera}}, \citenamefont {{Shu}},
  \citenamefont {{Slosar}}, \citenamefont {{Sobreira}}, \citenamefont
  {{Streblyanska}}, \citenamefont {{Suzuki}}, \citenamefont {{Taylor}},
  \citenamefont {{Tao}}, \citenamefont {{Tinker}}, \citenamefont {{Tojeiro}},
  \citenamefont {{Vargas-Maga{\~n}a}}, \citenamefont {{Wang}}, \citenamefont
  {{Weaver}}, \citenamefont {{Weinberg}}, \citenamefont {{White}},
  \citenamefont {{Wood-Vasey}}, \citenamefont {{Yeche}}, \citenamefont
  {{Zhai}}, \citenamefont {{Zhao}}, \citenamefont {{Zhao}}, \citenamefont
  {{Zheng}}, \citenamefont {{Ben Zhu}},\ and\ \citenamefont
  {{Zou}}}]{Dawson:2016}%
  \BibitemOpen
  \bibfield  {author} {\bibinfo {author} {\bibfnamefont {K.~S.}\ \bibnamefont
  {{Dawson}}}, \bibinfo {author} {\bibfnamefont {J.-P.}\ \bibnamefont
  {{Kneib}}}, \bibinfo {author} {\bibfnamefont {W.~J.}\ \bibnamefont
  {{Percival}}}, \bibinfo {author} {\bibfnamefont {S.}~\bibnamefont {{Alam}}},
  \bibinfo {author} {\bibfnamefont {F.~D.}\ \bibnamefont {{Albareti}}},
  \bibinfo {author} {\bibfnamefont {S.~F.}\ \bibnamefont {{Anderson}}},
  \bibinfo {author} {\bibfnamefont {E.}~\bibnamefont {{Armengaud}}}, \bibinfo
  {author} {\bibfnamefont {{\'E}.}~\bibnamefont {{Aubourg}}}, \bibinfo {author}
  {\bibfnamefont {S.}~\bibnamefont {{Bailey}}}, \bibinfo {author}
  {\bibfnamefont {J.~E.}\ \bibnamefont {{Bautista}}}, \bibinfo {author}
  {\bibfnamefont {A.~A.}\ \bibnamefont {{Berlind}}}, \bibinfo {author}
  {\bibfnamefont {M.~A.}\ \bibnamefont {{Bershady}}}, \bibinfo {author}
  {\bibfnamefont {F.}~\bibnamefont {{Beutler}}}, \bibinfo {author}
  {\bibfnamefont {D.}~\bibnamefont {{Bizyaev}}}, \bibinfo {author}
  {\bibfnamefont {M.~R.}\ \bibnamefont {{Blanton}}}, \bibnamefont
  {and~others},\ }\href {https://doi.org/10.3847/0004-6256/151/2/44} {\bibfield
   {journal} {\bibinfo  {journal} {\aj}\ }\textbf {\bibinfo {volume} {151}},\
  \bibinfo {eid} {44} (\bibinfo {year} {2016})},\ \Eprint
  {https://arxiv.org/abs/1508.04473} {arXiv:1508.04473 [astro-ph.CO]}
  \BibitemShut {NoStop}%
\bibitem [{\citenamefont {{Lyke}}\ \emph {et~al.}(2020)\citenamefont {{Lyke}},
  \citenamefont {{Higley}}, \citenamefont {{McLane}}, \citenamefont
  {{Schurhammer}}, \citenamefont {{Myers}}, \citenamefont {{Ross}},
  \citenamefont {{Dawson}}, \citenamefont {{Chabanier}}, \citenamefont
  {{Martini}}, \citenamefont {{Busca}}, \citenamefont {{Mas des Bourboux}},
  \citenamefont {{Salvato}}, \citenamefont {{Streblyanska}}, \citenamefont
  {{Zarrouk}}, \citenamefont {{Burtin}}, \citenamefont {{Anderson}},
  \citenamefont {{Bautista}}, \citenamefont {{Bizyaev}}, \citenamefont
  {{Brandt}}, \citenamefont {{Brinkmann}}, \citenamefont {{Brownstein}},
  \citenamefont {{Comparat}}, \citenamefont {{Green}}, \citenamefont {{de la
  Macorra}}, \citenamefont {{Mu{\~n}oz Guti{\'e}rrez}}, \citenamefont {{Hou}},
  \citenamefont {{Newman}}, \citenamefont {{Palanque-Delabrouille}},
  \citenamefont {{P{\^a}ris}}, \citenamefont {{Percival}}, \citenamefont
  {{Petitjean}}, \citenamefont {{Rich}}, \citenamefont {{Rossi}}, \citenamefont
  {{Schneider}}, \citenamefont {{Smith}}, \citenamefont {{Vivek}},\ and\
  \citenamefont {{Weaver}}}]{Lyke:2020}%
  \BibitemOpen
  \bibfield  {author} {\bibinfo {author} {\bibfnamefont {B.~W.}\ \bibnamefont
  {{Lyke}}}, \bibinfo {author} {\bibfnamefont {A.~N.}\ \bibnamefont
  {{Higley}}}, \bibinfo {author} {\bibfnamefont {J.~N.}\ \bibnamefont
  {{McLane}}}, \bibinfo {author} {\bibfnamefont {D.~P.}\ \bibnamefont
  {{Schurhammer}}}, \bibinfo {author} {\bibfnamefont {A.~D.}\ \bibnamefont
  {{Myers}}}, \bibinfo {author} {\bibfnamefont {A.~J.}\ \bibnamefont {{Ross}}},
  \bibinfo {author} {\bibfnamefont {K.}~\bibnamefont {{Dawson}}}, \bibinfo
  {author} {\bibfnamefont {S.}~\bibnamefont {{Chabanier}}}, \bibinfo {author}
  {\bibfnamefont {P.}~\bibnamefont {{Martini}}}, \bibinfo {author}
  {\bibfnamefont {N.~G.}\ \bibnamefont {{Busca}}}, \bibinfo {author}
  {\bibfnamefont {H.~d.}\ \bibnamefont {{Mas des Bourboux}}}, \bibinfo {author}
  {\bibfnamefont {M.}~\bibnamefont {{Salvato}}}, \bibinfo {author}
  {\bibfnamefont {A.}~\bibnamefont {{Streblyanska}}}, \bibinfo {author}
  {\bibfnamefont {P.}~\bibnamefont {{Zarrouk}}}, \bibinfo {author}
  {\bibfnamefont {E.}~\bibnamefont {{Burtin}}}, \bibnamefont {and~others},\
  }\href {https://doi.org/10.3847/1538-4365/aba623} {\bibfield  {journal}
  {\bibinfo  {journal} {\apjs}\ }\textbf {\bibinfo {volume} {250}},\ \bibinfo
  {eid} {8} (\bibinfo {year} {2020})},\ \Eprint
  {https://arxiv.org/abs/2007.09001} {arXiv:2007.09001 [astro-ph.GA]}
  \BibitemShut {NoStop}%
\bibitem [{\citenamefont {Lewis}\ \emph {et~al.}(2000)\citenamefont {Lewis},
  \citenamefont {Challinor},\ and\ \citenamefont {Lasenby}}]{Lewis:1999}%
  \BibitemOpen
  \bibfield  {author} {\bibinfo {author} {\bibfnamefont {A.}~\bibnamefont
  {Lewis}}, \bibinfo {author} {\bibfnamefont {A.}~\bibnamefont {Challinor}},\
  \bibnamefont {and}\ \bibinfo {author} {\bibfnamefont {A.}~\bibnamefont
  {Lasenby}},\ }\href {https://doi.org/10.1086/309179} {\bibfield  {journal}
  {\bibinfo  {journal} {\apj}\ }\textbf {\bibinfo {volume} {538}},\ \bibinfo
  {pages} {473} (\bibinfo {year} {2000})},\ \Eprint
  {https://arxiv.org/abs/astro-ph/9911177} {arXiv:astro-ph/9911177 [astro-ph]}
  \BibitemShut {NoStop}%
\bibitem [{\citenamefont {{Kirkby}}\ \emph {et~al.}(2013)\citenamefont
  {{Kirkby}}, \citenamefont {{Margala}}, \citenamefont {{Slosar}},
  \citenamefont {{Bailey}}, \citenamefont {{Busca}}, \citenamefont {{Delubac}},
  \citenamefont {{Rich}}, \citenamefont {{Bautista}}, \citenamefont
  {{Blomqvist}}, \citenamefont {{Brownstein}}, \citenamefont {{Carithers}},
  \citenamefont {{Croft}}, \citenamefont {{Dawson}}, \citenamefont
  {{Font-Ribera}}, \citenamefont {{Miralda-Escud{\'e}}}, \citenamefont
  {{Myers}}, \citenamefont {{Nichol}}, \citenamefont {{Palanque-Delabrouille}},
  \citenamefont {{P{\^a}ris}}, \citenamefont {{Petitjean}}, \citenamefont
  {{Rossi}}, \citenamefont {{Schlegel}}, \citenamefont {{Schneider}},
  \citenamefont {{Viel}}, \citenamefont {{Weinberg}},\ and\ \citenamefont
  {{Y{\`e}che}}}]{Kirkby:2013}%
  \BibitemOpen
  \bibfield  {author} {\bibinfo {author} {\bibfnamefont {D.}~\bibnamefont
  {{Kirkby}}}, \bibinfo {author} {\bibfnamefont {D.}~\bibnamefont {{Margala}}},
  \bibinfo {author} {\bibfnamefont {A.}~\bibnamefont {{Slosar}}}, \bibinfo
  {author} {\bibfnamefont {S.}~\bibnamefont {{Bailey}}}, \bibinfo {author}
  {\bibfnamefont {N.~G.}\ \bibnamefont {{Busca}}}, \bibinfo {author}
  {\bibfnamefont {T.}~\bibnamefont {{Delubac}}}, \bibinfo {author}
  {\bibfnamefont {J.}~\bibnamefont {{Rich}}}, \bibinfo {author} {\bibfnamefont
  {J.~E.}\ \bibnamefont {{Bautista}}}, \bibinfo {author} {\bibfnamefont
  {M.}~\bibnamefont {{Blomqvist}}}, \bibinfo {author} {\bibfnamefont {J.~R.}\
  \bibnamefont {{Brownstein}}}, \bibinfo {author} {\bibfnamefont
  {B.}~\bibnamefont {{Carithers}}}, \bibinfo {author} {\bibfnamefont
  {R.~A.~C.}\ \bibnamefont {{Croft}}}, \bibinfo {author} {\bibfnamefont
  {K.~S.}\ \bibnamefont {{Dawson}}}, \bibinfo {author} {\bibfnamefont
  {A.}~\bibnamefont {{Font-Ribera}}}, \bibinfo {author} {\bibfnamefont
  {J.}~\bibnamefont {{Miralda-Escud{\'e}}}}, \bibnamefont {and~others},\ }\href
  {https://doi.org/10.1088/1475-7516/2013/03/024} {\bibfield  {journal}
  {\bibinfo  {journal} {\jcap}\ }\textbf {\bibinfo {volume} {2013}},\ \bibinfo
  {eid} {024} (\bibinfo {year} {2013})},\ \Eprint
  {https://arxiv.org/abs/1301.3456} {arXiv:1301.3456 [astro-ph.CO]}
  \BibitemShut {NoStop}%
\bibitem [{\citenamefont {{Cuceu}}\ \emph {et~al.}(2023)\citenamefont
  {{Cuceu}}, \citenamefont {{Font-Ribera}}, \citenamefont {{Martini}},
  \citenamefont {{Joachimi}}, \citenamefont {{Nadathur}}, \citenamefont
  {{Rich}}, \citenamefont {{Gonz{\'a}lez-Morales}}, \citenamefont {{du Mas des
  Bourboux}},\ and\ \citenamefont {{Farr}}}]{Cuceu:2022}%
  \BibitemOpen
  \bibfield  {author} {\bibinfo {author} {\bibfnamefont {A.}~\bibnamefont
  {{Cuceu}}}, \bibinfo {author} {\bibfnamefont {A.}~\bibnamefont
  {{Font-Ribera}}}, \bibinfo {author} {\bibfnamefont {P.}~\bibnamefont
  {{Martini}}}, \bibinfo {author} {\bibfnamefont {B.}~\bibnamefont
  {{Joachimi}}}, \bibinfo {author} {\bibfnamefont {S.}~\bibnamefont
  {{Nadathur}}}, \bibinfo {author} {\bibfnamefont {J.}~\bibnamefont {{Rich}}},
  \bibinfo {author} {\bibfnamefont {A.~X.}\ \bibnamefont
  {{Gonz{\'a}lez-Morales}}}, \bibinfo {author} {\bibfnamefont {H.}~\bibnamefont
  {{du Mas des Bourboux}}},\ \bibnamefont {and}\ \bibinfo {author}
  {\bibfnamefont {J.}~\bibnamefont {{Farr}}},\ }\href
  {https://doi.org/10.1093/mnras/stad1546} {\bibfield  {journal} {\bibinfo
  {journal} {\mnras}\ }\textbf {\bibinfo {volume} {523}},\ \bibinfo {pages}
  {3773} (\bibinfo {year} {2023})},\ \Eprint {https://arxiv.org/abs/2209.12931}
  {arXiv:2209.12931 [astro-ph.CO]} \BibitemShut {NoStop}%
\bibitem [{Note1()}]{Note1}%
  \BibitemOpen
  \bibinfo {note} {\protect \url
  {https://github.com/andreicuceu/vega}}\BibitemShut {NoStop}%
\bibitem [{\citenamefont {{G{\'o}rski}}\ \emph {et~al.}(2005)\citenamefont
  {{G{\'o}rski}}, \citenamefont {{Hivon}}, \citenamefont {{Banday}},
  \citenamefont {{Wandelt}}, \citenamefont {{Hansen}}, \citenamefont
  {{Reinecke}},\ and\ \citenamefont {{Bartelmann}}}]{Gorski:2005}%
  \BibitemOpen
  \bibfield  {author} {\bibinfo {author} {\bibfnamefont {K.~M.}\ \bibnamefont
  {{G{\'o}rski}}}, \bibinfo {author} {\bibfnamefont {E.}~\bibnamefont
  {{Hivon}}}, \bibinfo {author} {\bibfnamefont {A.~J.}\ \bibnamefont
  {{Banday}}}, \bibinfo {author} {\bibfnamefont {B.~D.}\ \bibnamefont
  {{Wandelt}}}, \bibinfo {author} {\bibfnamefont {F.~K.}\ \bibnamefont
  {{Hansen}}}, \bibinfo {author} {\bibfnamefont {M.}~\bibnamefont
  {{Reinecke}}},\ \bibnamefont {and}\ \bibinfo {author} {\bibfnamefont
  {M.}~\bibnamefont {{Bartelmann}}},\ }\href {https://doi.org/10.1086/427976}
  {\bibfield  {journal} {\bibinfo  {journal} {\apj}\ }\textbf {\bibinfo
  {volume} {622}},\ \bibinfo {pages} {759} (\bibinfo {year} {2005})},\ \Eprint
  {https://arxiv.org/abs/astro-ph/0409513} {arXiv:astro-ph/0409513 [astro-ph]}
  \BibitemShut {NoStop}%
\bibitem [{\citenamefont {{Bautista}}\ \emph {et~al.}(2017)\citenamefont
  {{Bautista}}, \citenamefont {{Busca}}, \citenamefont {{Guy}}, \citenamefont
  {{Rich}}, \citenamefont {{Blomqvist}}, \citenamefont {{du Mas des Bourboux}},
  \citenamefont {{Pieri}}, \citenamefont {{Font-Ribera}}, \citenamefont
  {{Bailey}}, \citenamefont {{Delubac}}, \citenamefont {{Kirkby}},
  \citenamefont {{Le Goff}}, \citenamefont {{Margala}}, \citenamefont
  {{Slosar}}, \citenamefont {{Vazquez}}, \citenamefont {{Brownstein}},
  \citenamefont {{Dawson}}, \citenamefont {{Eisenstein}}, \citenamefont
  {{Miralda-Escud{\'e}}}, \citenamefont {{Noterdaeme}}, \citenamefont
  {{Palanque-Delabrouille}}, \citenamefont {{P{\^a}ris}}, \citenamefont
  {{Petitjean}}, \citenamefont {{Ross}}, \citenamefont {{Schneider}},
  \citenamefont {{Weinberg}},\ and\ \citenamefont
  {{Y{\`e}che}}}]{Bautista:2017}%
  \BibitemOpen
  \bibfield  {author} {\bibinfo {author} {\bibfnamefont {J.~E.}\ \bibnamefont
  {{Bautista}}}, \bibinfo {author} {\bibfnamefont {N.~G.}\ \bibnamefont
  {{Busca}}}, \bibinfo {author} {\bibfnamefont {J.}~\bibnamefont {{Guy}}},
  \bibinfo {author} {\bibfnamefont {J.}~\bibnamefont {{Rich}}}, \bibinfo
  {author} {\bibfnamefont {M.}~\bibnamefont {{Blomqvist}}}, \bibinfo {author}
  {\bibfnamefont {H.}~\bibnamefont {{du Mas des Bourboux}}}, \bibinfo {author}
  {\bibfnamefont {M.~M.}\ \bibnamefont {{Pieri}}}, \bibinfo {author}
  {\bibfnamefont {A.}~\bibnamefont {{Font-Ribera}}}, \bibinfo {author}
  {\bibfnamefont {S.}~\bibnamefont {{Bailey}}}, \bibinfo {author}
  {\bibfnamefont {T.}~\bibnamefont {{Delubac}}}, \bibinfo {author}
  {\bibfnamefont {D.}~\bibnamefont {{Kirkby}}}, \bibinfo {author}
  {\bibfnamefont {J.-M.}\ \bibnamefont {{Le Goff}}}, \bibinfo {author}
  {\bibfnamefont {D.}~\bibnamefont {{Margala}}}, \bibinfo {author}
  {\bibfnamefont {A.}~\bibnamefont {{Slosar}}}, \bibinfo {author}
  {\bibfnamefont {J.~A.}\ \bibnamefont {{Vazquez}}}, \bibnamefont
  {and~others},\ }\href {https://doi.org/10.1051/0004-6361/201730533}
  {\bibfield  {journal} {\bibinfo  {journal} {\aap}\ }\textbf {\bibinfo
  {volume} {603}},\ \bibinfo {eid} {A12} (\bibinfo {year} {2017})},\ \Eprint
  {https://arxiv.org/abs/1702.00176} {arXiv:1702.00176 [astro-ph.CO]}
  \BibitemShut {NoStop}%
\bibitem [{\citenamefont {{du Mas des Bourboux}}\ \emph
  {et~al.}(2017)\citenamefont {{du Mas des Bourboux}}, \citenamefont {{Le
  Goff}}, \citenamefont {{Blomqvist}}, \citenamefont {{Busca}}, \citenamefont
  {{Guy}}, \citenamefont {{Rich}}, \citenamefont {{Y{\`e}che}}, \citenamefont
  {{Bautista}}, \citenamefont {{Burtin}}, \citenamefont {{Dawson}},
  \citenamefont {{Eisenstein}}, \citenamefont {{Font-Ribera}}, \citenamefont
  {{Kirkby}}, \citenamefont {{Miralda-Escud{\'e}}}, \citenamefont
  {{Noterdaeme}}, \citenamefont {{Palanque-Delabrouille}}, \citenamefont
  {{P{\^a}ris}}, \citenamefont {{Petitjean}}, \citenamefont
  {{P{\'e}rez-R{\`a}fols}}, \citenamefont {{Pieri}}, \citenamefont {{Ross}},
  \citenamefont {{Schlegel}}, \citenamefont {{Schneider}}, \citenamefont
  {{Slosar}}, \citenamefont {{Weinberg}},\ and\ \citenamefont
  {{Zarrouk}}}]{duMasdesBourboux:2017}%
  \BibitemOpen
  \bibfield  {author} {\bibinfo {author} {\bibfnamefont {H.}~\bibnamefont {{du
  Mas des Bourboux}}}, \bibinfo {author} {\bibfnamefont {J.-M.}\ \bibnamefont
  {{Le Goff}}}, \bibinfo {author} {\bibfnamefont {M.}~\bibnamefont
  {{Blomqvist}}}, \bibinfo {author} {\bibfnamefont {N.~G.}\ \bibnamefont
  {{Busca}}}, \bibinfo {author} {\bibfnamefont {J.}~\bibnamefont {{Guy}}},
  \bibinfo {author} {\bibfnamefont {J.}~\bibnamefont {{Rich}}}, \bibinfo
  {author} {\bibfnamefont {C.}~\bibnamefont {{Y{\`e}che}}}, \bibinfo {author}
  {\bibfnamefont {J.~E.}\ \bibnamefont {{Bautista}}}, \bibinfo {author}
  {\bibfnamefont {{\'E}.}~\bibnamefont {{Burtin}}}, \bibinfo {author}
  {\bibfnamefont {K.~S.}\ \bibnamefont {{Dawson}}}, \bibinfo {author}
  {\bibfnamefont {D.~J.}\ \bibnamefont {{Eisenstein}}}, \bibinfo {author}
  {\bibfnamefont {A.}~\bibnamefont {{Font-Ribera}}}, \bibinfo {author}
  {\bibfnamefont {D.}~\bibnamefont {{Kirkby}}}, \bibinfo {author}
  {\bibfnamefont {J.}~\bibnamefont {{Miralda-Escud{\'e}}}}, \bibinfo {author}
  {\bibfnamefont {P.}~\bibnamefont {{Noterdaeme}}}, \bibnamefont {and~others},\
  }\href {https://doi.org/10.1051/0004-6361/201731731} {\bibfield  {journal}
  {\bibinfo  {journal} {\aap}\ }\textbf {\bibinfo {volume} {608}},\ \bibinfo
  {eid} {A130} (\bibinfo {year} {2017})},\ \Eprint
  {https://arxiv.org/abs/1708.02225} {arXiv:1708.02225 [astro-ph.CO]}
  \BibitemShut {NoStop}%
\bibitem [{\citenamefont {{du Mas des Bourboux}}\ \emph
  {et~al.}(2021)\citenamefont {{du Mas des Bourboux}}, \citenamefont {{Rich}},
  \citenamefont {{Font-Ribera}}, \citenamefont {{de Sainte Agathe}},
  \citenamefont {{Farr}}, \citenamefont {{Etourneau}}, \citenamefont {{Le
  Goff}}, \citenamefont {{Cuceu}}, \citenamefont {{Balland}}, \citenamefont
  {{Bautista}}, \citenamefont {{Blomqvist}}, \citenamefont {{Brinkmann}},
  \citenamefont {{Brownstein}}, \citenamefont {{Chabanier}}, \citenamefont
  {{Chaussidon}}, \citenamefont {{Dawson}}, \citenamefont
  {{Gonz{\'a}lez-Morales}}, \citenamefont {{Guy}}, \citenamefont {{Lyke}},
  \citenamefont {{de la Macorra}}, \citenamefont {{Mueller}}, \citenamefont
  {{Myers}}, \citenamefont {{Nitschelm}}, \citenamefont {{Mu{\~n}oz
  Guti{\'e}rrez}}, \citenamefont {{Palanque-Delabrouille}}, \citenamefont
  {{Parker}}, \citenamefont {{Percival}}, \citenamefont
  {{P{\'e}rez-R{\`a}fols}}, \citenamefont {{Petitjean}}, \citenamefont
  {{Pieri}}, \citenamefont {{Ravoux}}, \citenamefont {{Rossi}}, \citenamefont
  {{Schneider}}, \citenamefont {{Seo}}, \citenamefont {{Slosar}}, \citenamefont
  {{Stermer}}, \citenamefont {{Vivek}}, \citenamefont {{Y{\`e}che}},\ and\
  \citenamefont {{Youles}}}]{Bourboux:2021}%
  \BibitemOpen
  \bibfield  {author} {\bibinfo {author} {\bibfnamefont {H.}~\bibnamefont {{du
  Mas des Bourboux}}}, \bibinfo {author} {\bibfnamefont {J.}~\bibnamefont
  {{Rich}}}, \bibinfo {author} {\bibfnamefont {A.}~\bibnamefont
  {{Font-Ribera}}}, \bibinfo {author} {\bibfnamefont {V.}~\bibnamefont {{de
  Sainte Agathe}}}, \bibinfo {author} {\bibfnamefont {J.}~\bibnamefont
  {{Farr}}}, \bibinfo {author} {\bibfnamefont {T.}~\bibnamefont {{Etourneau}}},
  \bibinfo {author} {\bibfnamefont {J.-M.}\ \bibnamefont {{Le Goff}}}, \bibinfo
  {author} {\bibfnamefont {A.}~\bibnamefont {{Cuceu}}}, \bibinfo {author}
  {\bibfnamefont {C.}~\bibnamefont {{Balland}}}, \bibinfo {author}
  {\bibfnamefont {J.~E.}\ \bibnamefont {{Bautista}}}, \bibinfo {author}
  {\bibfnamefont {M.}~\bibnamefont {{Blomqvist}}}, \bibinfo {author}
  {\bibfnamefont {J.}~\bibnamefont {{Brinkmann}}}, \bibinfo {author}
  {\bibfnamefont {J.~R.}\ \bibnamefont {{Brownstein}}}, \bibinfo {author}
  {\bibfnamefont {S.}~\bibnamefont {{Chabanier}}}, \bibinfo {author}
  {\bibfnamefont {E.}~\bibnamefont {{Chaussidon}}}, \bibnamefont {and~others},\
  }\href@noop {} {\bibinfo {title} {{picca: Package for Igm
  Cosmological-Correlations Analyses}}},\ \bibinfo {howpublished} {Astrophysics
  Source Code Library, record ascl:2106.018} (\bibinfo {year} {2021}),\ \Eprint
  {https://arxiv.org/abs/2106.018} {ascl:2106.018} \BibitemShut {NoStop}%
\bibitem [{\citenamefont {{Ram{\'\i}rez-P{\'e}rez}}\ \emph
  {et~al.}(2022)\citenamefont {{Ram{\'\i}rez-P{\'e}rez}}, \citenamefont
  {{Sanchez}}, \citenamefont {{Alonso}},\ and\ \citenamefont
  {{Font-Ribera}}}]{Ramirez:2022}%
  \BibitemOpen
  \bibfield  {author} {\bibinfo {author} {\bibfnamefont {C.}~\bibnamefont
  {{Ram{\'\i}rez-P{\'e}rez}}}, \bibinfo {author} {\bibfnamefont
  {J.}~\bibnamefont {{Sanchez}}}, \bibinfo {author} {\bibfnamefont
  {D.}~\bibnamefont {{Alonso}}},\ \bibnamefont {and}\ \bibinfo {author}
  {\bibfnamefont {A.}~\bibnamefont {{Font-Ribera}}},\ }\href
  {https://doi.org/10.1088/1475-7516/2022/05/002} {\bibfield  {journal}
  {\bibinfo  {journal} {\jcap}\ }\textbf {\bibinfo {volume} {2022}},\ \bibinfo
  {eid} {002} (\bibinfo {year} {2022})},\ \Eprint
  {https://arxiv.org/abs/2111.05069} {arXiv:2111.05069 [astro-ph.CO]}
  \BibitemShut {NoStop}%
\bibitem [{\citenamefont {{Farr}}\ \emph {et~al.}(2020)\citenamefont {{Farr}},
  \citenamefont {{Font-Ribera}}, \citenamefont {{du Mas des Bourboux}},
  \citenamefont {{Mu{\~n}oz-Guti{\'e}rrez}}, \citenamefont {{S{\'a}nchez}},
  \citenamefont {{Pontzen}}, \citenamefont {{Xochitl Gonz{\'a}lez-Morales}},
  \citenamefont {{Alonso}}, \citenamefont {{Brooks}}, \citenamefont {{Doel}},
  \citenamefont {{Etourneau}}, \citenamefont {{Guy}}, \citenamefont {{Le
  Goff}}, \citenamefont {{de la Macorra}}, \citenamefont
  {{Palanque-Delabrouille}}, \citenamefont {{P{\'e}rez-R{\`a}fols}},
  \citenamefont {{Rich}}, \citenamefont {{Slosar}}, \citenamefont {{Tarle}},
  \citenamefont {{Yutong}},\ and\ \citenamefont {{Zhang}}}]{Farr:2020}%
  \BibitemOpen
  \bibfield  {author} {\bibinfo {author} {\bibfnamefont {J.}~\bibnamefont
  {{Farr}}}, \bibinfo {author} {\bibfnamefont {A.}~\bibnamefont
  {{Font-Ribera}}}, \bibinfo {author} {\bibfnamefont {H.}~\bibnamefont {{du Mas
  des Bourboux}}}, \bibinfo {author} {\bibfnamefont {A.}~\bibnamefont
  {{Mu{\~n}oz-Guti{\'e}rrez}}}, \bibinfo {author} {\bibfnamefont {F.~J.}\
  \bibnamefont {{S{\'a}nchez}}}, \bibinfo {author} {\bibfnamefont
  {A.}~\bibnamefont {{Pontzen}}}, \bibinfo {author} {\bibfnamefont
  {A.}~\bibnamefont {{Xochitl Gonz{\'a}lez-Morales}}}, \bibinfo {author}
  {\bibfnamefont {D.}~\bibnamefont {{Alonso}}}, \bibinfo {author}
  {\bibfnamefont {D.}~\bibnamefont {{Brooks}}}, \bibinfo {author}
  {\bibfnamefont {P.}~\bibnamefont {{Doel}}}, \bibinfo {author} {\bibfnamefont
  {T.}~\bibnamefont {{Etourneau}}}, \bibinfo {author} {\bibfnamefont
  {J.}~\bibnamefont {{Guy}}}, \bibinfo {author} {\bibfnamefont {J.-M.}\
  \bibnamefont {{Le Goff}}}, \bibinfo {author} {\bibfnamefont {A.}~\bibnamefont
  {{de la Macorra}}}, \bibinfo {author} {\bibfnamefont {N.}~\bibnamefont
  {{Palanque-Delabrouille}}}, \bibnamefont {and~others},\ }\href
  {https://doi.org/10.1088/1475-7516/2020/03/068} {\bibfield  {journal}
  {\bibinfo  {journal} {\jcap}\ }\textbf {\bibinfo {volume} {2020}},\ \bibinfo
  {eid} {068} (\bibinfo {year} {2020})},\ \Eprint
  {https://arxiv.org/abs/1912.02763} {arXiv:1912.02763 [astro-ph.CO]}
  \BibitemShut {NoStop}%
\bibitem [{\citenamefont {{de Sainte Agathe}}\ \emph
  {et~al.}(2019)\citenamefont {{de Sainte Agathe}}, \citenamefont {{Balland}},
  \citenamefont {{du Mas des Bourboux}}, \citenamefont {{Busca}}, \citenamefont
  {{Blomqvist}}, \citenamefont {{Guy}}, \citenamefont {{Rich}}, \citenamefont
  {{Font-Ribera}}, \citenamefont {{Pieri}}, \citenamefont {{Bautista}},
  \citenamefont {{Dawson}}, \citenamefont {{Le Goff}}, \citenamefont {{de la
  Macorra}}, \citenamefont {{Palanque-Delabrouille}}, \citenamefont
  {{Percival}}, \citenamefont {{P{\'e}rez-R{\`a}fols}}, \citenamefont
  {{Schneider}}, \citenamefont {{Slosar}},\ and\ \citenamefont
  {{Y{\`e}che}}}]{deSainteAgathe:2019}%
  \BibitemOpen
  \bibfield  {author} {\bibinfo {author} {\bibfnamefont {V.}~\bibnamefont {{de
  Sainte Agathe}}}, \bibinfo {author} {\bibfnamefont {C.}~\bibnamefont
  {{Balland}}}, \bibinfo {author} {\bibfnamefont {H.}~\bibnamefont {{du Mas des
  Bourboux}}}, \bibinfo {author} {\bibfnamefont {N.~G.}\ \bibnamefont
  {{Busca}}}, \bibinfo {author} {\bibfnamefont {M.}~\bibnamefont
  {{Blomqvist}}}, \bibinfo {author} {\bibfnamefont {J.}~\bibnamefont {{Guy}}},
  \bibinfo {author} {\bibfnamefont {J.}~\bibnamefont {{Rich}}}, \bibinfo
  {author} {\bibfnamefont {A.}~\bibnamefont {{Font-Ribera}}}, \bibinfo {author}
  {\bibfnamefont {M.~M.}\ \bibnamefont {{Pieri}}}, \bibinfo {author}
  {\bibfnamefont {J.~E.}\ \bibnamefont {{Bautista}}}, \bibinfo {author}
  {\bibfnamefont {K.}~\bibnamefont {{Dawson}}}, \bibinfo {author}
  {\bibfnamefont {J.-M.}\ \bibnamefont {{Le Goff}}}, \bibinfo {author}
  {\bibfnamefont {A.}~\bibnamefont {{de la Macorra}}}, \bibinfo {author}
  {\bibfnamefont {N.}~\bibnamefont {{Palanque-Delabrouille}}}, \bibinfo
  {author} {\bibfnamefont {W.~J.}\ \bibnamefont {{Percival}}}, \bibnamefont
  {and~others},\ }\href {https://doi.org/10.1051/0004-6361/201935638}
  {\bibfield  {journal} {\bibinfo  {journal} {\aap}\ }\textbf {\bibinfo
  {volume} {629}},\ \bibinfo {eid} {A85} (\bibinfo {year} {2019})},\ \Eprint
  {https://arxiv.org/abs/1904.03400} {arXiv:1904.03400 [astro-ph.CO]}
  \BibitemShut {NoStop}%
\bibitem [{\citenamefont {{Arinyo-i-Prats}}\ \emph {et~al.}(2015)\citenamefont
  {{Arinyo-i-Prats}}, \citenamefont {{Miralda-Escud{\'e}}}, \citenamefont
  {{Viel}},\ and\ \citenamefont {{Cen}}}]{Arinyo:2015}%
  \BibitemOpen
  \bibfield  {author} {\bibinfo {author} {\bibfnamefont {A.}~\bibnamefont
  {{Arinyo-i-Prats}}}, \bibinfo {author} {\bibfnamefont {J.}~\bibnamefont
  {{Miralda-Escud{\'e}}}}, \bibinfo {author} {\bibfnamefont {M.}~\bibnamefont
  {{Viel}}},\ \bibnamefont {and}\ \bibinfo {author} {\bibfnamefont
  {R.}~\bibnamefont {{Cen}}},\ }\href
  {https://doi.org/10.1088/1475-7516/2015/12/017} {\bibfield  {journal}
  {\bibinfo  {journal} {\jcap}\ }\textbf {\bibinfo {volume} {2015}},\ \bibinfo
  {eid} {017} (\bibinfo {year} {2015})},\ \Eprint
  {https://arxiv.org/abs/1506.04519} {arXiv:1506.04519 [astro-ph.CO]}
  \BibitemShut {NoStop}%
\bibitem [{\citenamefont {{Givans}}\ \emph {et~al.}(2022)\citenamefont
  {{Givans}}, \citenamefont {{Font-Ribera}}, \citenamefont {{Slosar}},
  \citenamefont {{Seeyave}}, \citenamefont {{Pedersen}}, \citenamefont
  {{Rogers}}, \citenamefont {{Garny}}, \citenamefont {{Blas}},\ and\
  \citenamefont {{Ir{\v{s}}i{\v{c}}}}}]{Givans:2022}%
  \BibitemOpen
  \bibfield  {author} {\bibinfo {author} {\bibfnamefont {J.~J.}\ \bibnamefont
  {{Givans}}}, \bibinfo {author} {\bibfnamefont {A.}~\bibnamefont
  {{Font-Ribera}}}, \bibinfo {author} {\bibfnamefont {A.}~\bibnamefont
  {{Slosar}}}, \bibinfo {author} {\bibfnamefont {L.}~\bibnamefont {{Seeyave}}},
  \bibinfo {author} {\bibfnamefont {C.}~\bibnamefont {{Pedersen}}}, \bibinfo
  {author} {\bibfnamefont {K.~K.}\ \bibnamefont {{Rogers}}}, \bibinfo {author}
  {\bibfnamefont {M.}~\bibnamefont {{Garny}}}, \bibinfo {author} {\bibfnamefont
  {D.}~\bibnamefont {{Blas}}},\ \bibnamefont {and}\ \bibinfo {author}
  {\bibfnamefont {V.}~\bibnamefont {{Ir{\v{s}}i{\v{c}}}}},\ }\href
  {https://doi.org/10.1088/1475-7516/2022/09/070} {\bibfield  {journal}
  {\bibinfo  {journal} {\jcap}\ }\textbf {\bibinfo {volume} {2022}},\ \bibinfo
  {eid} {070} (\bibinfo {year} {2022})},\ \Eprint
  {https://arxiv.org/abs/2205.00962} {arXiv:2205.00962 [astro-ph.CO]}
  \BibitemShut {NoStop}%
\bibitem [{\citenamefont {{Percival}}\ and\ \citenamefont
  {{White}}(2009)}]{Percival:2009}%
  \BibitemOpen
  \bibfield  {author} {\bibinfo {author} {\bibfnamefont {W.~J.}\ \bibnamefont
  {{Percival}}}\ \bibnamefont {and}\ \bibinfo {author} {\bibfnamefont
  {M.}~\bibnamefont {{White}}},\ }\href
  {https://doi.org/10.1111/j.1365-2966.2008.14211.x} {\bibfield  {journal}
  {\bibinfo  {journal} {\mnras}\ }\textbf {\bibinfo {volume} {393}},\ \bibinfo
  {pages} {297} (\bibinfo {year} {2009})},\ \Eprint
  {https://arxiv.org/abs/0808.0003} {arXiv:0808.0003 [astro-ph]} \BibitemShut
  {NoStop}%
\bibitem [{\citenamefont {{Cuceu}}\ \emph {et~al.}(2021)\citenamefont
  {{Cuceu}}, \citenamefont {{Font-Ribera}}, \citenamefont {{Joachimi}},\ and\
  \citenamefont {{Nadathur}}}]{Cuceu:2021}%
  \BibitemOpen
  \bibfield  {author} {\bibinfo {author} {\bibfnamefont {A.}~\bibnamefont
  {{Cuceu}}}, \bibinfo {author} {\bibfnamefont {A.}~\bibnamefont
  {{Font-Ribera}}}, \bibinfo {author} {\bibfnamefont {B.}~\bibnamefont
  {{Joachimi}}},\ \bibnamefont {and}\ \bibinfo {author} {\bibfnamefont
  {S.}~\bibnamefont {{Nadathur}}},\ }\href
  {https://doi.org/10.1093/mnras/stab1999} {\bibfield  {journal} {\bibinfo
  {journal} {\mnras}\ }\textbf {\bibinfo {volume} {506}},\ \bibinfo {pages}
  {5439} (\bibinfo {year} {2021})},\ \Eprint {https://arxiv.org/abs/2103.14075}
  {arXiv:2103.14075 [astro-ph.CO]} \BibitemShut {NoStop}%
\bibitem [{\citenamefont {{Gontcho A Gontcho}}\ \emph
  {et~al.}(2014)\citenamefont {{Gontcho A Gontcho}}, \citenamefont
  {{Miralda-Escud{\'e}}},\ and\ \citenamefont {{Busca}}}]{Gontcho:2014}%
  \BibitemOpen
  \bibfield  {author} {\bibinfo {author} {\bibfnamefont {S.}~\bibnamefont
  {{Gontcho A Gontcho}}}, \bibinfo {author} {\bibfnamefont {J.}~\bibnamefont
  {{Miralda-Escud{\'e}}}},\ \bibnamefont {and}\ \bibinfo {author}
  {\bibfnamefont {N.~G.}\ \bibnamefont {{Busca}}},\ }\href
  {https://doi.org/10.1093/mnras/stu860} {\bibfield  {journal} {\bibinfo
  {journal} {\mnras}\ }\textbf {\bibinfo {volume} {442}},\ \bibinfo {pages}
  {187} (\bibinfo {year} {2014})},\ \Eprint {https://arxiv.org/abs/1404.7425}
  {arXiv:1404.7425 [astro-ph.CO]} \BibitemShut {NoStop}%
\bibitem [{\citenamefont {{Rogers}}\ \emph {et~al.}(2018)\citenamefont
  {{Rogers}}, \citenamefont {{Bird}}, \citenamefont {{Peiris}}, \citenamefont
  {{Pontzen}}, \citenamefont {{Font-Ribera}},\ and\ \citenamefont
  {{Leistedt}}}]{Rogers:2018}%
  \BibitemOpen
  \bibfield  {author} {\bibinfo {author} {\bibfnamefont {K.~K.}\ \bibnamefont
  {{Rogers}}}, \bibinfo {author} {\bibfnamefont {S.}~\bibnamefont {{Bird}}},
  \bibinfo {author} {\bibfnamefont {H.~V.}\ \bibnamefont {{Peiris}}}, \bibinfo
  {author} {\bibfnamefont {A.}~\bibnamefont {{Pontzen}}}, \bibinfo {author}
  {\bibfnamefont {A.}~\bibnamefont {{Font-Ribera}}},\ \bibnamefont {and}\
  \bibinfo {author} {\bibfnamefont {B.}~\bibnamefont {{Leistedt}}},\ }\href
  {https://doi.org/10.1093/mnras/sty603} {\bibfield  {journal} {\bibinfo
  {journal} {\mnras}\ }\textbf {\bibinfo {volume} {476}},\ \bibinfo {pages}
  {3716} (\bibinfo {year} {2018})},\ \Eprint {https://arxiv.org/abs/1711.06275}
  {arXiv:1711.06275 [astro-ph.CO]} \BibitemShut {NoStop}%
\bibitem [{\citenamefont {{Font-Ribera}}\ \emph {et~al.}(2013)\citenamefont
  {{Font-Ribera}}, \citenamefont {{Arnau}}, \citenamefont
  {{Miralda-Escud{\'e}}}, \citenamefont {{Rollinde}}, \citenamefont
  {{Brinkmann}}, \citenamefont {{Brownstein}}, \citenamefont {{Lee}},
  \citenamefont {{Myers}}, \citenamefont {{Palanque-Delabrouille}},
  \citenamefont {{P{\^a}ris}}, \citenamefont {{Petitjean}}, \citenamefont
  {{Rich}}, \citenamefont {{Ross}}, \citenamefont {{Schneider}},\ and\
  \citenamefont {{White}}}]{FontRibera:2013}%
  \BibitemOpen
  \bibfield  {author} {\bibinfo {author} {\bibfnamefont {A.}~\bibnamefont
  {{Font-Ribera}}}, \bibinfo {author} {\bibfnamefont {E.}~\bibnamefont
  {{Arnau}}}, \bibinfo {author} {\bibfnamefont {J.}~\bibnamefont
  {{Miralda-Escud{\'e}}}}, \bibinfo {author} {\bibfnamefont {E.}~\bibnamefont
  {{Rollinde}}}, \bibinfo {author} {\bibfnamefont {J.}~\bibnamefont
  {{Brinkmann}}}, \bibinfo {author} {\bibfnamefont {J.~R.}\ \bibnamefont
  {{Brownstein}}}, \bibinfo {author} {\bibfnamefont {K.-G.}\ \bibnamefont
  {{Lee}}}, \bibinfo {author} {\bibfnamefont {A.~D.}\ \bibnamefont {{Myers}}},
  \bibinfo {author} {\bibfnamefont {N.}~\bibnamefont
  {{Palanque-Delabrouille}}}, \bibinfo {author} {\bibfnamefont
  {I.}~\bibnamefont {{P{\^a}ris}}}, \bibinfo {author} {\bibfnamefont
  {P.}~\bibnamefont {{Petitjean}}}, \bibinfo {author} {\bibfnamefont
  {J.}~\bibnamefont {{Rich}}}, \bibinfo {author} {\bibfnamefont {N.~P.}\
  \bibnamefont {{Ross}}}, \bibinfo {author} {\bibfnamefont {D.~P.}\
  \bibnamefont {{Schneider}}},\ \bibnamefont {and}\ \bibinfo {author}
  {\bibfnamefont {M.}~\bibnamefont {{White}}},\ }\href
  {https://doi.org/10.1088/1475-7516/2013/05/018} {\bibfield  {journal}
  {\bibinfo  {journal} {\jcap}\ }\textbf {\bibinfo {volume} {2013}},\ \bibinfo
  {eid} {018} (\bibinfo {year} {2013})},\ \Eprint
  {https://arxiv.org/abs/1303.1937} {arXiv:1303.1937 [astro-ph.CO]}
  \BibitemShut {NoStop}%
\bibitem [{\citenamefont {{Youles}}\ \emph {et~al.}(2022)\citenamefont
  {{Youles}}, \citenamefont {{Bautista}}, \citenamefont {{Font-Ribera}},
  \citenamefont {{Bacon}}, \citenamefont {{Rich}}, \citenamefont {{Brooks}},
  \citenamefont {{Davis}}, \citenamefont {{Dawson}}, \citenamefont {{de la
  Macorra}}, \citenamefont {{Dhungana}}, \citenamefont {{Doel}}, \citenamefont
  {{Fanning}}, \citenamefont {{Gazta{\~n}aga}}, \citenamefont {{Gontcho A
  Gontcho}}, \citenamefont {{Gonzalez-Morales}}, \citenamefont {{Guy}},
  \citenamefont {{Honscheid}}, \citenamefont {{Ir{\v{s}}i{\v{c}}}},
  \citenamefont {{Kehoe}}, \citenamefont {{Kirkby}}, \citenamefont {{Kisner}},
  \citenamefont {{Landriau}}, \citenamefont {{Le Guillou}}, \citenamefont
  {{Levi}}, \citenamefont {{Martini}}, \citenamefont
  {{Mu{\~n}oz-Guti{\'e}rrez}}, \citenamefont {{Palanque-Delabrouille}},
  \citenamefont {{P{\'e}rez-R{\`a}fols}}, \citenamefont {{Poppett}},
  \citenamefont {{Ram{\'\i}rez-P{\'e}rez}}, \citenamefont {{Schubnell}},
  \citenamefont {{Tarl{\'e}}},\ and\ \citenamefont {{Walther}}}]{Youles:2022}%
  \BibitemOpen
  \bibfield  {author} {\bibinfo {author} {\bibfnamefont {S.}~\bibnamefont
  {{Youles}}}, \bibinfo {author} {\bibfnamefont {J.~E.}\ \bibnamefont
  {{Bautista}}}, \bibinfo {author} {\bibfnamefont {A.}~\bibnamefont
  {{Font-Ribera}}}, \bibinfo {author} {\bibfnamefont {D.}~\bibnamefont
  {{Bacon}}}, \bibinfo {author} {\bibfnamefont {J.}~\bibnamefont {{Rich}}},
  \bibinfo {author} {\bibfnamefont {D.}~\bibnamefont {{Brooks}}}, \bibinfo
  {author} {\bibfnamefont {T.~M.}\ \bibnamefont {{Davis}}}, \bibinfo {author}
  {\bibfnamefont {K.}~\bibnamefont {{Dawson}}}, \bibinfo {author}
  {\bibfnamefont {A.}~\bibnamefont {{de la Macorra}}}, \bibinfo {author}
  {\bibfnamefont {G.}~\bibnamefont {{Dhungana}}}, \bibinfo {author}
  {\bibfnamefont {P.}~\bibnamefont {{Doel}}}, \bibinfo {author} {\bibfnamefont
  {K.}~\bibnamefont {{Fanning}}}, \bibinfo {author} {\bibfnamefont
  {E.}~\bibnamefont {{Gazta{\~n}aga}}}, \bibinfo {author} {\bibfnamefont
  {S.}~\bibnamefont {{Gontcho A Gontcho}}}, \bibinfo {author} {\bibfnamefont
  {A.~X.}\ \bibnamefont {{Gonzalez-Morales}}}, \bibnamefont {and~others},\
  }\href {https://doi.org/10.1093/mnras/stac2102} {\bibfield  {journal}
  {\bibinfo  {journal} {\mnras}\ }\textbf {\bibinfo {volume} {516}},\ \bibinfo
  {pages} {421} (\bibinfo {year} {2022})},\ \Eprint
  {https://arxiv.org/abs/2205.06648} {arXiv:2205.06648 [astro-ph.CO]}
  \BibitemShut {NoStop}%
\bibitem [{\citenamefont {{DESI Collaboration}}\ \emph
  {et~al.}(2022)\citenamefont {{DESI Collaboration}}, \citenamefont
  {{Abareshi}}, \citenamefont {{Aguilar}}, \citenamefont {{Ahlen}},
  \citenamefont {{Alam}}, \citenamefont {{Alexander}}, \citenamefont
  {{Alfarsy}}, \citenamefont {{Allen}}, \citenamefont {{Allende Prieto}},
  \citenamefont {{Alves}}, \citenamefont {{Ameel}}, \citenamefont
  {{Armengaud}}, \citenamefont {{Asorey}}, \citenamefont {{Aviles}},
  \citenamefont {{Bailey}}, \citenamefont {{Balaguera-Antol{\'\i}nez}},
  \citenamefont {{Ballester}}, \citenamefont {{Baltay}}, \citenamefont
  {{Bault}}, \citenamefont {{Beltran}}, \citenamefont {{Benavides}},
  \citenamefont {{BenZvi}}, \citenamefont {{Berti}}, \citenamefont {{Besuner}},
  \citenamefont {{Beutler}}, \citenamefont {{Bianchi}}, \citenamefont
  {{Blake}}, \citenamefont {{Blanc}}, \citenamefont {{Blum}}, \citenamefont
  {{Bolton}}, \citenamefont {{Bose}}, \citenamefont {{Bramall}}, \citenamefont
  {{Brieden}}, \citenamefont {{Brodzeller}}, \citenamefont {{Brooks}},
  \citenamefont {{Brownewell}}, \citenamefont {{Buckley-Geer}}, \citenamefont
  {{Cahn}}, \citenamefont {{Cai}}, \citenamefont {{Canning}}, \citenamefont
  {{Capasso}}, \citenamefont {{Carnero Rosell}}, \citenamefont {{Carton}},
  \citenamefont {{Casas}}, \citenamefont {{Castander}}, \citenamefont
  {{Cervantes-Cota}}, \citenamefont {{Chabanier}}, \citenamefont
  {{Chaussidon}}, \citenamefont {{Chuang}}, \citenamefont {{Circosta}},
  \citenamefont {{Cole}}, \citenamefont {{Cooper}}, \citenamefont {{da Costa}},
  \citenamefont {{Cousinou}}, \citenamefont {{Cuceu}}, \citenamefont {{Davis}},
  \citenamefont {{Dawson}}, \citenamefont {{de la Cruz-Noriega}}, \citenamefont
  {{de la Macorra}}, \citenamefont {{de Mattia}}, \citenamefont {{Della
  Costa}}, \citenamefont {{Demmer}}, \citenamefont {{Derwent}}, \citenamefont
  {{Dey}}, \citenamefont {{Dey}}, \citenamefont {{Dhungana}}, \citenamefont
  {{Ding}}, \citenamefont {{Dobson}}, \citenamefont {{Doel}}, \citenamefont
  {{Donald-McCann}}, \citenamefont {{Donaldson}}, \citenamefont {{Douglass}},
  \citenamefont {{Duan}}, \citenamefont {{Dunlop}}, \citenamefont
  {{Edelstein}}, \citenamefont {{Eftekharzadeh}}, \citenamefont {{Eisenstein}},
  \citenamefont {{Enriquez-Vargas}}, \citenamefont {{Escoffier}}, \citenamefont
  {{Evatt}}, \citenamefont {{Fagrelius}}, \citenamefont {{Fan}}, \citenamefont
  {{Fanning}}, \citenamefont {{Fawcett}}, \citenamefont {{Ferraro}},
  \citenamefont {{Ereza}}, \citenamefont {{Flaugher}}, \citenamefont
  {{Font-Ribera}}, \citenamefont {{Forero-Romero}}, \citenamefont {{Frenk}},
  \citenamefont {{Fromenteau}}, \citenamefont {{G{\"a}nsicke}}, \citenamefont
  {{Garcia-Quintero}}, \citenamefont {{Garrison}}, \citenamefont
  {{Gazta{\~n}aga}}, \citenamefont {{Gerardi}}, \citenamefont
  {{Gil-Mar{\'\i}n}}, \citenamefont {{Gontcho a Gontcho}}, \citenamefont
  {{Gonzalez-Morales}}, \citenamefont {{Gonzalez-de-Rivera}}, \citenamefont
  {{Gonzalez-Perez}}, \citenamefont {{Gordon}}, \citenamefont {{Graur}},
  \citenamefont {{Green}}, \citenamefont {{Grove}}, \citenamefont {{Gruen}},
  \citenamefont {{Gutierrez}}, \citenamefont {{Guy}}, \citenamefont {{Hahn}},
  \citenamefont {{Harris}}, \citenamefont {{Herrera}}, \citenamefont
  {{Herrera-Alcantar}}, \citenamefont {{Honscheid}}, \citenamefont {{Howlett}},
  \citenamefont {{Huterer}}, \citenamefont {{Ir{\v{s}}i{\v{c}}}}, \citenamefont
  {{Ishak}}, \citenamefont {{Jelinsky}}, \citenamefont {{Jiang}}, \citenamefont
  {{Jimenez}}, \citenamefont {{Jing}}, \citenamefont {{Joyce}}, \citenamefont
  {{Jullo}}, \citenamefont {{Juneau}}, \citenamefont {{Kara{\c{c}}ayl{\i}}},
  \citenamefont {{Karamanis}}, \citenamefont {{Karcher}}, \citenamefont
  {{Karim}}, \citenamefont {{Kehoe}}, \citenamefont {{Kent}}, \citenamefont
  {{Kirkby}}, \citenamefont {{Kisner}}, \citenamefont {{Kitaura}},
  \citenamefont {{Koposov}}, \citenamefont {{Kov{\'a}cs}}, \citenamefont
  {{Kremin}}, \citenamefont {{Krolewski}}, \citenamefont {{L'Huillier}},
  \citenamefont {{Lahav}}, \citenamefont {{Lambert}}, \citenamefont {{Lamman}},
  \citenamefont {{Lan}}, \citenamefont {{Landriau}}, \citenamefont {{Lane}},
  \citenamefont {{Lang}}, \citenamefont {{Lange}}, \citenamefont {{Lasker}},
  \citenamefont {{Le Guillou}}, \citenamefont {{Leauthaud}}, \citenamefont {{Le
  Van Suu}}, \citenamefont {{Levi}}, \citenamefont {{Li}}, \citenamefont
  {{Magneville}}, \citenamefont {{Manera}}, \citenamefont {{Manser}},
  \citenamefont {{Marshall}}, \citenamefont {{Martini}}, \citenamefont
  {{McCollam}}, \citenamefont {{McDonald}}, \citenamefont {{Meisner}},
  \citenamefont {{Mena-Fern{\'a}ndez}}, \citenamefont {{Meneses-Rizo}},
  \citenamefont {{Mezcua}}, \citenamefont {{Miller}}, \citenamefont {{Miquel}},
  \citenamefont {{Montero-Camacho}}, \citenamefont {{Moon}}, \citenamefont
  {{Moustakas}}, \citenamefont {{Mueller}}, \citenamefont
  {{Mu{\~n}oz-Guti{\'e}rrez}}, \citenamefont {{Myers}}, \citenamefont
  {{Nadathur}}, \citenamefont {{Najita}}, \citenamefont {{Napolitano}},
  \citenamefont {{Neilsen}}, \citenamefont {{Newman}}, \citenamefont {{Nie}},
  \citenamefont {{Ning}}, \citenamefont {{Niz}}, \citenamefont {{Norberg}},
  \citenamefont {{Noriega}}, \citenamefont {{O'Brien}}, \citenamefont
  {{Obuljen}}, \citenamefont {{Palanque-Delabrouille}}, \citenamefont
  {{Palmese}}, \citenamefont {{Zhiwei}}, \citenamefont {{Pappalardo}},
  \citenamefont {{PENG}}, \citenamefont {{Percival}}, \citenamefont
  {{Perruchot}}, \citenamefont {{Pogge}}, \citenamefont {{Poppett}},
  \citenamefont {{Porredon}}, \citenamefont {{Prada}}, \citenamefont
  {{Prochaska}}, \citenamefont {{Pucha}}, \citenamefont
  {{P{\'e}rez-Fern{\'a}ndez}}, \citenamefont {{P{\'e}rez-R{\`a}fols}},
  \citenamefont {{Rabinowitz}}, \citenamefont {{Raichoor}}, \citenamefont
  {{Ramirez-Solano}}, \citenamefont {{Ram{\'\i}rez-P{\'e}rez}}, \citenamefont
  {{Ravoux}}, \citenamefont {{Reil}}, \citenamefont {{Rezaie}}, \citenamefont
  {{Rocher}}, \citenamefont {{Rockosi}}, \citenamefont {{Roe}}, \citenamefont
  {{Roodman}}, \citenamefont {{Ross}}, \citenamefont {{Rossi}}, \citenamefont
  {{Ruggeri}}, \citenamefont {{Ruhlmann-Kleider}}, \citenamefont {{Sabiu}},
  \citenamefont {{Safonova}}, \citenamefont {{Said}}, \citenamefont
  {{Saintonge}}, \citenamefont {{Salas Catonga}}, \citenamefont {{Samushia}},
  \citenamefont {{Sanchez}}, \citenamefont {{Saulder}}, \citenamefont
  {{Schaan}}, \citenamefont {{Schlafly}}, \citenamefont {{Schlegel}},
  \citenamefont {{Schmoll}}, \citenamefont {{Scholte}}, \citenamefont
  {{Schubnell}}, \citenamefont {{Secroun}}, \citenamefont {{Seo}},
  \citenamefont {{Serrano}}, \citenamefont {{Sharples}}, \citenamefont
  {{Sholl}}, \citenamefont {{Silber}}, \citenamefont {{Silva}}, \citenamefont
  {{Sirk}}, \citenamefont {{Siudek}}, \citenamefont {{Smith}}, \citenamefont
  {{Sprayberry}}, \citenamefont {{Staten}}, \citenamefont {{Stupak}},
  \citenamefont {{Tan}}, \citenamefont {{Tarl{\'e}}}, \citenamefont {{Tie}},
  \citenamefont {{Tojeiro}}, \citenamefont {{Ure{\~n}a-L{\'o}pez}},
  \citenamefont {{Valdes}}, \citenamefont {{Valenzuela}}, \citenamefont
  {{Valluri}}, \citenamefont {{Vargas-Maga{\~n}a}}, \citenamefont {{Verde}},
  \citenamefont {{Walther}}, \citenamefont {{Wang}}, \citenamefont {{Wang}},
  \citenamefont {{Weaver}}, \citenamefont {{Weaverdyck}}, \citenamefont
  {{Wechsler}}, \citenamefont {{Wilson}}, \citenamefont {{Yang}}, \citenamefont
  {{Yu}}, \citenamefont {{Yuan}}, \citenamefont {{Y{\`e}che}}, \citenamefont
  {{Zhang}}, \citenamefont {{Zhang}}, \citenamefont {{Zhao}}, \citenamefont
  {{Zhou}}, \citenamefont {{Zhou}}, \citenamefont {{Zou}}, \citenamefont
  {{Zou}}, \citenamefont {{Zou}}, \citenamefont {{Zu}},\ and\ \citenamefont
  {{DESI Collaboration}}}]{Abareshi:2022}%
  \BibitemOpen
  \bibfield  {author} {\bibinfo {author} {\bibnamefont {{DESI Collaboration}}},
  \bibinfo {author} {\bibfnamefont {B.}~\bibnamefont {{Abareshi}}}, \bibinfo
  {author} {\bibfnamefont {J.}~\bibnamefont {{Aguilar}}}, \bibinfo {author}
  {\bibfnamefont {S.}~\bibnamefont {{Ahlen}}}, \bibinfo {author} {\bibfnamefont
  {S.}~\bibnamefont {{Alam}}}, \bibinfo {author} {\bibfnamefont {D.~M.}\
  \bibnamefont {{Alexander}}}, \bibinfo {author} {\bibfnamefont
  {R.}~\bibnamefont {{Alfarsy}}}, \bibinfo {author} {\bibfnamefont
  {L.}~\bibnamefont {{Allen}}}, \bibinfo {author} {\bibfnamefont
  {C.}~\bibnamefont {{Allende Prieto}}}, \bibinfo {author} {\bibfnamefont
  {O.}~\bibnamefont {{Alves}}}, \bibinfo {author} {\bibfnamefont
  {J.}~\bibnamefont {{Ameel}}}, \bibinfo {author} {\bibfnamefont
  {E.}~\bibnamefont {{Armengaud}}}, \bibinfo {author} {\bibfnamefont
  {J.}~\bibnamefont {{Asorey}}}, \bibinfo {author} {\bibfnamefont
  {A.}~\bibnamefont {{Aviles}}}, \bibinfo {author} {\bibfnamefont
  {S.}~\bibnamefont {{Bailey}}}, \bibnamefont {and~others},\ }\href
  {https://doi.org/10.3847/1538-3881/ac882b} {\bibfield  {journal} {\bibinfo
  {journal} {\aj}\ }\textbf {\bibinfo {volume} {164}},\ \bibinfo {eid} {207}
  (\bibinfo {year} {2022})},\ \Eprint {https://arxiv.org/abs/2205.10939}
  {arXiv:2205.10939 [astro-ph.IM]} \BibitemShut {NoStop}%
\bibitem [{\citenamefont {{Torrado}}\ and\ \citenamefont
  {{Lewis}}(2019)}]{Torrado:2019}%
  \BibitemOpen
  \bibfield  {author} {\bibinfo {author} {\bibfnamefont {J.}~\bibnamefont
  {{Torrado}}}\ \bibnamefont {and}\ \bibinfo {author} {\bibfnamefont
  {A.}~\bibnamefont {{Lewis}}},\ }\href@noop {} {\bibinfo {title} {{Cobaya:
  Bayesian analysis in cosmology}}},\ \bibinfo {howpublished} {Astrophysics
  Source Code Library, record ascl:1910.019} (\bibinfo {year} {2019}),\ \Eprint
  {https://arxiv.org/abs/1910.019} {ascl:1910.019} \BibitemShut {NoStop}%
\bibitem [{\citenamefont {{Torrado}}\ and\ \citenamefont
  {{Lewis}}(2021)}]{Torrado:2021}%
  \BibitemOpen
  \bibfield  {author} {\bibinfo {author} {\bibfnamefont {J.}~\bibnamefont
  {{Torrado}}}\ \bibnamefont {and}\ \bibinfo {author} {\bibfnamefont
  {A.}~\bibnamefont {{Lewis}}},\ }\href
  {https://doi.org/10.1088/1475-7516/2021/05/057} {\bibfield  {journal}
  {\bibinfo  {journal} {\jcap}\ }\textbf {\bibinfo {volume} {2021}},\ \bibinfo
  {eid} {057} (\bibinfo {year} {2021})},\ \Eprint
  {https://arxiv.org/abs/2005.05290} {arXiv:2005.05290 [astro-ph.IM]}
  \BibitemShut {NoStop}%
\bibitem [{\citenamefont {{Handley}}\ \emph
  {et~al.}(2015{\natexlab{a}})\citenamefont {{Handley}}, \citenamefont
  {{Hobson}},\ and\ \citenamefont {{Lasenby}}}]{Handley:2015a}%
  \BibitemOpen
  \bibfield  {author} {\bibinfo {author} {\bibfnamefont {W.~J.}\ \bibnamefont
  {{Handley}}}, \bibinfo {author} {\bibfnamefont {M.~P.}\ \bibnamefont
  {{Hobson}}},\ \bibnamefont {and}\ \bibinfo {author} {\bibfnamefont {A.~N.}\
  \bibnamefont {{Lasenby}}},\ }\href {https://doi.org/10.1093/mnrasl/slv047}
  {\bibfield  {journal} {\bibinfo  {journal} {\mnras}\ }\textbf {\bibinfo
  {volume} {450}},\ \bibinfo {pages} {L61} (\bibinfo {year}
  {2015}{\natexlab{a}})},\ \Eprint {https://arxiv.org/abs/1502.01856}
  {arXiv:1502.01856 [astro-ph.CO]} \BibitemShut {NoStop}%
\bibitem [{\citenamefont {{Handley}}\ \emph
  {et~al.}(2015{\natexlab{b}})\citenamefont {{Handley}}, \citenamefont
  {{Hobson}},\ and\ \citenamefont {{Lasenby}}}]{Handley:2015b}%
  \BibitemOpen
  \bibfield  {author} {\bibinfo {author} {\bibfnamefont {W.~J.}\ \bibnamefont
  {{Handley}}}, \bibinfo {author} {\bibfnamefont {M.~P.}\ \bibnamefont
  {{Hobson}}},\ \bibnamefont {and}\ \bibinfo {author} {\bibfnamefont {A.~N.}\
  \bibnamefont {{Lasenby}}},\ }\href {https://doi.org/10.1093/mnras/stv1911}
  {\bibfield  {journal} {\bibinfo  {journal} {\mnras}\ }\textbf {\bibinfo
  {volume} {453}},\ \bibinfo {pages} {4384} (\bibinfo {year}
  {2015}{\natexlab{b}})},\ \Eprint {https://arxiv.org/abs/1506.00171}
  {arXiv:1506.00171 [astro-ph.IM]} \BibitemShut {NoStop}%
\bibitem [{Note2()}]{Note2}%
  \BibitemOpen
  \bibinfo {note} {Note that the choice of isotropic scale parameter is an
  arbitrary one and not based on arguments about the optimal parameter to
  measure.}\BibitemShut {Stop}%
\bibitem [{Note3()}]{Note3}%
  \BibitemOpen
  \bibinfo {note} {Tension metrics were computed from the full posterior
  distributions by summing over probability densities.}\BibitemShut {Stop}%
\bibitem [{Note4()}]{Note4}%
  \BibitemOpen
  \bibinfo {note} {\protect \url
  {https://pla.esac.esa.int//##cosmology}}\BibitemShut {NoStop}%
\bibitem [{\citenamefont {{Aubourg}}\ \emph {et~al.}(2015)\citenamefont
  {{Aubourg}}, \citenamefont {{Bailey}}, \citenamefont {{Bautista}},
  \citenamefont {{Beutler}}, \citenamefont {{Bhardwaj}}, \citenamefont
  {{Bizyaev}}, \citenamefont {{Blanton}}, \citenamefont {{Blomqvist}},
  \citenamefont {{Bolton}}, \citenamefont {{Bovy}}, \citenamefont
  {{Brewington}}, \citenamefont {{Brinkmann}}, \citenamefont {{Brownstein}},
  \citenamefont {{Burden}}, \citenamefont {{Busca}}, \citenamefont
  {{Carithers}}, \citenamefont {{Chuang}}, \citenamefont {{Comparat}},
  \citenamefont {{Croft}}, \citenamefont {{Cuesta}}, \citenamefont {{Dawson}},
  \citenamefont {{Delubac}}, \citenamefont {{Eisenstein}}, \citenamefont
  {{Font-Ribera}}, \citenamefont {{Ge}}, \citenamefont {{Le Goff}},
  \citenamefont {{Gontcho}}, \citenamefont {{Gott}}, \citenamefont {{Gunn}},
  \citenamefont {{Guo}}, \citenamefont {{Guy}}, \citenamefont {{Hamilton}},
  \citenamefont {{Ho}}, \citenamefont {{Honscheid}}, \citenamefont {{Howlett}},
  \citenamefont {{Kirkby}}, \citenamefont {{Kitaura}}, \citenamefont {{Kneib}},
  \citenamefont {{Lee}}, \citenamefont {{Long}}, \citenamefont {{Lupton}},
  \citenamefont {{Maga{\~n}a}}, \citenamefont {{Malanushenko}}, \citenamefont
  {{Malanushenko}}, \citenamefont {{Manera}}, \citenamefont {{Maraston}},
  \citenamefont {{Margala}}, \citenamefont {{McBride}}, \citenamefont
  {{Miralda-Escud{\'e}}}, \citenamefont {{Myers}}, \citenamefont {{Nichol}},
  \citenamefont {{Noterdaeme}}, \citenamefont {{Nuza}}, \citenamefont
  {{Olmstead}}, \citenamefont {{Oravetz}}, \citenamefont {{P{\^a}ris}},
  \citenamefont {{Padmanabhan}}, \citenamefont {{Palanque-Delabrouille}},
  \citenamefont {{Pan}}, \citenamefont {{Pellejero-Ibanez}}, \citenamefont
  {{Percival}}, \citenamefont {{Petitjean}}, \citenamefont {{Pieri}},
  \citenamefont {{Prada}}, \citenamefont {{Reid}}, \citenamefont {{Rich}},
  \citenamefont {{Roe}}, \citenamefont {{Ross}}, \citenamefont {{Ross}},
  \citenamefont {{Rossi}}, \citenamefont {{Rubi{\~n}o-Mart{\'\i}n}},
  \citenamefont {{S{\'a}nchez}}, \citenamefont {{Samushia}}, \citenamefont
  {{G{\'e}nova-Santos}}, \citenamefont {{Sc{\'o}ccola}}, \citenamefont
  {{Schlegel}}, \citenamefont {{Schneider}}, \citenamefont {{Seo}},
  \citenamefont {{Sheldon}}, \citenamefont {{Simmons}}, \citenamefont
  {{Skibba}}, \citenamefont {{Slosar}}, \citenamefont {{Strauss}},
  \citenamefont {{Thomas}}, \citenamefont {{Tinker}}, \citenamefont
  {{Tojeiro}}, \citenamefont {{Vazquez}}, \citenamefont {{Viel}}, \citenamefont
  {{Wake}}, \citenamefont {{Weaver}}, \citenamefont {{Weinberg}}, \citenamefont
  {{Wood-Vasey}}, \citenamefont {{Y{\`e}che}}, \citenamefont {{Zehavi}},
  \citenamefont {{Zhao}},\ and\ \citenamefont {{BOSS
  Collaboration}}}]{Aubourg:2015}%
  \BibitemOpen
  \bibfield  {author} {\bibinfo {author} {\bibfnamefont {{\'E}.}~\bibnamefont
  {{Aubourg}}}, \bibinfo {author} {\bibfnamefont {S.}~\bibnamefont {{Bailey}}},
  \bibinfo {author} {\bibfnamefont {J.~E.}\ \bibnamefont {{Bautista}}},
  \bibinfo {author} {\bibfnamefont {F.}~\bibnamefont {{Beutler}}}, \bibinfo
  {author} {\bibfnamefont {V.}~\bibnamefont {{Bhardwaj}}}, \bibinfo {author}
  {\bibfnamefont {D.}~\bibnamefont {{Bizyaev}}}, \bibinfo {author}
  {\bibfnamefont {M.}~\bibnamefont {{Blanton}}}, \bibinfo {author}
  {\bibfnamefont {M.}~\bibnamefont {{Blomqvist}}}, \bibinfo {author}
  {\bibfnamefont {A.~S.}\ \bibnamefont {{Bolton}}}, \bibinfo {author}
  {\bibfnamefont {J.}~\bibnamefont {{Bovy}}}, \bibinfo {author} {\bibfnamefont
  {H.}~\bibnamefont {{Brewington}}}, \bibinfo {author} {\bibfnamefont
  {J.}~\bibnamefont {{Brinkmann}}}, \bibinfo {author} {\bibfnamefont {J.~R.}\
  \bibnamefont {{Brownstein}}}, \bibinfo {author} {\bibfnamefont
  {A.}~\bibnamefont {{Burden}}}, \bibinfo {author} {\bibfnamefont {N.~G.}\
  \bibnamefont {{Busca}}}, \bibnamefont {and~others},\ }\href
  {https://doi.org/10.1103/PhysRevD.92.123516} {\bibfield  {journal} {\bibinfo
  {journal} {\prd}\ }\textbf {\bibinfo {volume} {92}},\ \bibinfo {eid} {123516}
  (\bibinfo {year} {2015})},\ \Eprint {https://arxiv.org/abs/1411.1074}
  {arXiv:1411.1074 [astro-ph.CO]} \BibitemShut {NoStop}%
\bibitem [{\citenamefont {{Cooke}}\ \emph {et~al.}(2018)\citenamefont
  {{Cooke}}, \citenamefont {{Pettini}},\ and\ \citenamefont
  {{Steidel}}}]{Cooke:2018}%
  \BibitemOpen
  \bibfield  {author} {\bibinfo {author} {\bibfnamefont {R.~J.}\ \bibnamefont
  {{Cooke}}}, \bibinfo {author} {\bibfnamefont {M.}~\bibnamefont {{Pettini}}},\
  \bibnamefont {and}\ \bibinfo {author} {\bibfnamefont {C.~C.}\ \bibnamefont
  {{Steidel}}},\ }\href {https://doi.org/10.3847/1538-4357/aaab53} {\bibfield
  {journal} {\bibinfo  {journal} {\apj}\ }\textbf {\bibinfo {volume} {855}},\
  \bibinfo {eid} {102} (\bibinfo {year} {2018})},\ \Eprint
  {https://arxiv.org/abs/1710.11129} {arXiv:1710.11129 [astro-ph.CO]}
  \BibitemShut {NoStop}%
\bibitem [{\citenamefont {{Mossa}}\ \emph {et~al.}(2020)\citenamefont
  {{Mossa}}, \citenamefont {{St{\"o}ckel}}, \citenamefont {{Cavanna}},
  \citenamefont {{Ferraro}}, \citenamefont {{Aliotta}}, \citenamefont
  {{Barile}}, \citenamefont {{Bemmerer}}, \citenamefont {{Best}}, \citenamefont
  {{Boeltzig}}, \citenamefont {{Broggini}}, \citenamefont {{Bruno}},
  \citenamefont {{Caciolli}}, \citenamefont {{Chillery}}, \citenamefont
  {{Ciani}}, \citenamefont {{Corvisiero}}, \citenamefont {{Csedreki}},
  \citenamefont {{Davinson}}, \citenamefont {{Depalo}}, \citenamefont {{Di
  Leva}}, \citenamefont {{Elekes}}, \citenamefont {{Fiore}}, \citenamefont
  {{Formicola}}, \citenamefont {{F{\"u}l{\"o}p}}, \citenamefont {{Gervino}},
  \citenamefont {{Guglielmetti}}, \citenamefont {{Gustavino}}, \citenamefont
  {{Gy{\"u}rky}}, \citenamefont {{Imbriani}}, \citenamefont {{Junker}},
  \citenamefont {{Kievsky}}, \citenamefont {{Kochanek}}, \citenamefont
  {{Lugaro}}, \citenamefont {{Marcucci}}, \citenamefont {{Mangano}},
  \citenamefont {{Marigo}}, \citenamefont {{Masha}}, \citenamefont
  {{Menegazzo}}, \citenamefont {{Pantaleo}}, \citenamefont {{Paticchio}},
  \citenamefont {{Perrino}}, \citenamefont {{Piatti}}, \citenamefont
  {{Pisanti}}, \citenamefont {{Prati}}, \citenamefont {{Schiavulli}},
  \citenamefont {{Straniero}}, \citenamefont {{Sz{\"u}cs}}, \citenamefont
  {{Tak{\'a}cs}}, \citenamefont {{Trezzi}}, \citenamefont {{Viviani}},\ and\
  \citenamefont {{Zavatarelli}}}]{Mossa:2020}%
  \BibitemOpen
  \bibfield  {author} {\bibinfo {author} {\bibfnamefont {V.}~\bibnamefont
  {{Mossa}}}, \bibinfo {author} {\bibfnamefont {K.}~\bibnamefont
  {{St{\"o}ckel}}}, \bibinfo {author} {\bibfnamefont {F.}~\bibnamefont
  {{Cavanna}}}, \bibinfo {author} {\bibfnamefont {F.}~\bibnamefont
  {{Ferraro}}}, \bibinfo {author} {\bibfnamefont {M.}~\bibnamefont
  {{Aliotta}}}, \bibinfo {author} {\bibfnamefont {F.}~\bibnamefont {{Barile}}},
  \bibinfo {author} {\bibfnamefont {D.}~\bibnamefont {{Bemmerer}}}, \bibinfo
  {author} {\bibfnamefont {A.}~\bibnamefont {{Best}}}, \bibinfo {author}
  {\bibfnamefont {A.}~\bibnamefont {{Boeltzig}}}, \bibinfo {author}
  {\bibfnamefont {C.}~\bibnamefont {{Broggini}}}, \bibinfo {author}
  {\bibfnamefont {C.~G.}\ \bibnamefont {{Bruno}}}, \bibinfo {author}
  {\bibfnamefont {A.}~\bibnamefont {{Caciolli}}}, \bibinfo {author}
  {\bibfnamefont {T.}~\bibnamefont {{Chillery}}}, \bibinfo {author}
  {\bibfnamefont {G.~F.}\ \bibnamefont {{Ciani}}}, \bibinfo {author}
  {\bibfnamefont {P.}~\bibnamefont {{Corvisiero}}}, \bibnamefont {and~others},\
  }\href {https://doi.org/10.1038/s41586-020-2878-4} {\bibfield  {journal}
  {\bibinfo  {journal} {\nat}\ }\textbf {\bibinfo {volume} {587}},\ \bibinfo
  {pages} {210} (\bibinfo {year} {2020})}\BibitemShut {NoStop}%
\bibitem [{\citenamefont {{Gerardi}}\ \emph {et~al.}(2023)\citenamefont
  {{Gerardi}}, \citenamefont {{Cuceu}}, \citenamefont {{Font-Ribera}},
  \citenamefont {{Joachimi}},\ and\ \citenamefont {{Lemos}}}]{Gerardi:2022}%
  \BibitemOpen
  \bibfield  {author} {\bibinfo {author} {\bibfnamefont {F.}~\bibnamefont
  {{Gerardi}}}, \bibinfo {author} {\bibfnamefont {A.}~\bibnamefont {{Cuceu}}},
  \bibinfo {author} {\bibfnamefont {A.}~\bibnamefont {{Font-Ribera}}}, \bibinfo
  {author} {\bibfnamefont {B.}~\bibnamefont {{Joachimi}}},\ \bibnamefont {and}\
  \bibinfo {author} {\bibfnamefont {P.}~\bibnamefont {{Lemos}}},\ }\href
  {https://doi.org/10.1093/mnras/stac3257} {\bibfield  {journal} {\bibinfo
  {journal} {\mnras}\ }\textbf {\bibinfo {volume} {518}},\ \bibinfo {pages}
  {2567} (\bibinfo {year} {2023})},\ \Eprint {https://arxiv.org/abs/2209.11263}
  {arXiv:2209.11263 [astro-ph.CO]} \BibitemShut {NoStop}%
\bibitem [{\citenamefont {{Lewis}}(2019)}]{Lewis:2019}%
  \BibitemOpen
  \bibfield  {author} {\bibinfo {author} {\bibfnamefont {A.}~\bibnamefont
  {{Lewis}}},\ }\href@noop {} {\bibfield  {journal} {\bibinfo  {journal} {arXiv
  e-prints}\ ,\ \bibinfo {eid} {arXiv:1910.13970}} (\bibinfo {year} {2019})},\
  \Eprint {https://arxiv.org/abs/1910.13970} {arXiv:1910.13970 [astro-ph.IM]}
  \BibitemShut {NoStop}%
\bibitem [{\citenamefont {Harris}\ \emph {et~al.}(2020)\citenamefont {Harris},
  \citenamefont {Millman}, \citenamefont {van~der Walt}, \citenamefont
  {Gommers}, \citenamefont {Virtanen}, \citenamefont {Cournapeau},
  \citenamefont {Wieser}, \citenamefont {Taylor}, \citenamefont {Berg},
  \citenamefont {Smith}, \citenamefont {Kern}, \citenamefont {Picus},
  \citenamefont {Hoyer}, \citenamefont {van Kerkwijk}, \citenamefont {Brett},
  \citenamefont {Haldane}, \citenamefont {del R{\'{i}}o}, \citenamefont
  {Wiebe}, \citenamefont {Peterson}, \citenamefont {G{\'{e}}rard-Marchant},
  \citenamefont {Sheppard}, \citenamefont {Reddy}, \citenamefont {Weckesser},
  \citenamefont {Abbasi}, \citenamefont {Gohlke},\ and\ \citenamefont
  {Oliphant}}]{Harris:2020}%
  \BibitemOpen
  \bibfield  {author} {\bibinfo {author} {\bibfnamefont {C.~R.}\ \bibnamefont
  {Harris}}, \bibinfo {author} {\bibfnamefont {K.~J.}\ \bibnamefont {Millman}},
  \bibinfo {author} {\bibfnamefont {S.~J.}\ \bibnamefont {van~der Walt}},
  \bibinfo {author} {\bibfnamefont {R.}~\bibnamefont {Gommers}}, \bibinfo
  {author} {\bibfnamefont {P.}~\bibnamefont {Virtanen}}, \bibinfo {author}
  {\bibfnamefont {D.}~\bibnamefont {Cournapeau}}, \bibinfo {author}
  {\bibfnamefont {E.}~\bibnamefont {Wieser}}, \bibinfo {author} {\bibfnamefont
  {J.}~\bibnamefont {Taylor}}, \bibinfo {author} {\bibfnamefont
  {S.}~\bibnamefont {Berg}}, \bibinfo {author} {\bibfnamefont {N.~J.}\
  \bibnamefont {Smith}}, \bibinfo {author} {\bibfnamefont {R.}~\bibnamefont
  {Kern}}, \bibinfo {author} {\bibfnamefont {M.}~\bibnamefont {Picus}},
  \bibinfo {author} {\bibfnamefont {S.}~\bibnamefont {Hoyer}}, \bibinfo
  {author} {\bibfnamefont {M.~H.}\ \bibnamefont {van Kerkwijk}}, \bibinfo
  {author} {\bibfnamefont {M.}~\bibnamefont {Brett}}, \bibnamefont
  {and~others},\ }\href {https://doi.org/10.1038/s41586-020-2649-2} {\bibfield
  {journal} {\bibinfo  {journal} {Nature}\ }\textbf {\bibinfo {volume} {585}},\
  \bibinfo {pages} {357} (\bibinfo {year} {2020})}\BibitemShut {NoStop}%
\bibitem [{Note5()}]{Note5}%
  \BibitemOpen
  \bibinfo {note} {Https://healpix.sourceforge.io}\BibitemShut {NoStop}%
\bibitem [{Note6()}]{Note6}%
  \BibitemOpen
  \bibinfo {note} {Using the same weights that enter the computation of the
  correlations.}\BibitemShut {Stop}%
\end{thebibliography}%
\bibliographystyle{mod-apsrev4-2.bst}

\section{Data and Likelihood}

Our data consists of the \lyaf\ correlation functions measured by eBOSS from SDSS DR16 \cite{duMasdesBourboux:2020}. We use a set of four correlation functions which consists of two \lya\ auto-correlations, \lyalyalyalya\ and \lyalyalyalyb, and two \lya-QSO cross-correlations, \lyalyaqso\ and \lyalybqso, where \lya(\lya) represents \lya\ flux between the \lya\ and \lyb\ peaks, and \lya(\lyb) represents \lya\ flux blue-ward of the \lyb\ peak. The correlations are first computed independently in each HEALPix pixel \footnote{https://healpix.sourceforge.io} in the survey \cite{Gorski:2005,Bautista:2017,duMasdesBourboux:2017} using the \texttt{picca} package \cite{Bourboux:2021}. The size of a pixel on the sky corresponds to a $250\times250 \; (h^{-1}\mathrm{Mpc})^2$ patch at $z=2.33$, and there are roughly $880$ pixels (\texttt{nside} $= 16$) in the eBOSS footprint. The mean and covariance of this set of correlations give the measurement and uncertainty on the final correlation. The small correlations between samples are neglected. The accuracy of this method was verified using mock datasets through comparisons with the mock to mock variations, and also by comparing it with other methods of computing the covariance on the correlation functions \cite{duMasdesBourboux:2020}. 

The effective redshift of our measurement was computed by Ref. \cite{duMasdesBourboux:2020} through a weighted average of the mean redshift of all pixel pairs used to compute the correlation functions \footnote{Using the same weights that enter the computation of the correlations.}. For a detailed description of the \lyaf\ data from eBOSS, and the process of measuring correlation functions and their uncertainties, see Ref. \cite{duMasdesBourboux:2020}.

\begin{figure*}
\includegraphics[width=0.77\linewidth]{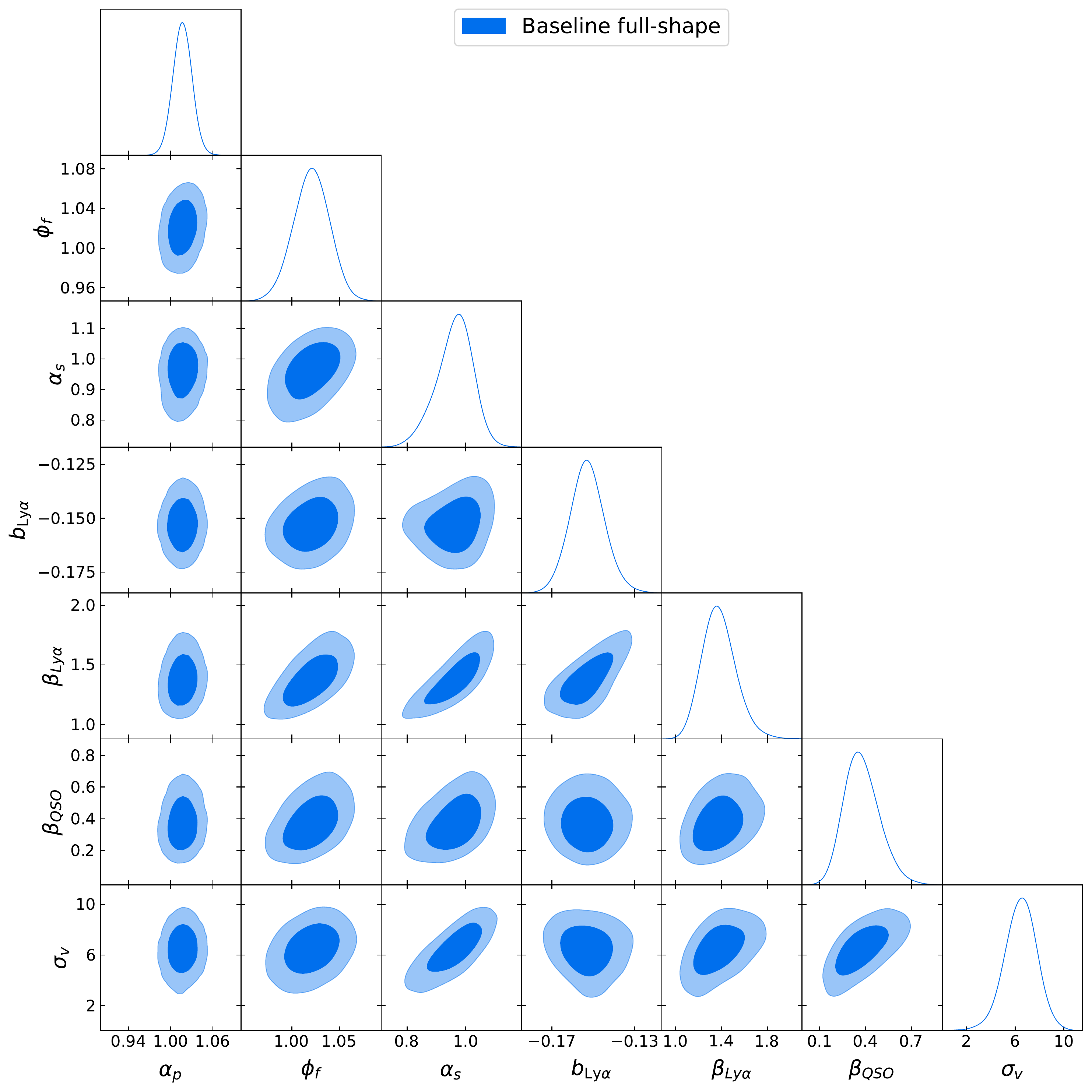}
\caption{Posterior distribution of the parameters of interest for cosmology, \alphap\ and \phif, along with the nuisance parameters which are correlated with \phif.}
\label{fig:posterior}
\end{figure*}

We compute a Gaussian likelihood without the cross-covariance between the different correlations which was found to be negligible by Ref. \cite{duMasdesBourboux:2020}. The free parameters in our model along with their priors are shown in \Cref{tab:priors}. For a detailed description of the model and parameters see Refs. \cite{Cuceu:2022} and \cite{duMasdesBourboux:2020}. In \Cref{fig:posterior}, we show the posterior distribution of \alphap\ and \phif, along with a selection of nuisance parameters based on their correlation with \phif. We note that there are no degeneracies between the two parameters of interest for cosmology and any of the nuisance parameters.

\begin{table}
    \centering
    \begin{tabular}{ | c | m{13em} | c | }
        \hline
         Parameters & Description & Prior \\
         \hline\hline
         $\phi$, $\alpha$ & Scale parameters & $U(0.01, 2.0)$ \\ [0.5ex]
         \hline
         $b_{\mathrm{Ly}\alpha}$ & \lya\ linear bias & $U(-2.0, 0.0)$ \\ [0.5ex]
         \hline
         $b_{\mathrm{QSO}}$ & QSO linear bias & $U(0.0, 10.0)$ \\ [0.5ex]
         \hline
         $\beta_{\mathrm{Ly}\alpha}$, $\beta_{\mathrm{QSO}}$ & RSD parameters of \lya\ and QSOs & $U(0.0, 5.0)$ \\ [0.5ex]
         \hline
         $b_{\mathrm{HCD}}$ & HCD linear bias & $U(-0.2, 0.0)$ \\ [0.5ex]
         \hline
         $\beta_{\mathrm{HCD}}$ & RSD parameter for HCDs & $\mathcal{N}(0.5, 0.2^2)$ \\ [0.5ex]
         \hline
         $\sigma_v$ [\hMpc] & Smoothing for redshift errors and QSO non-linear velocities & $U(0.0, 15.0)$ \\
         \hline
         $\Delta r_{||}$ [\hMpc] & Shift due to QSO redshift errors & $U(-3.0, 3.0)$ \\ [0.5ex]
         \hline
         $b_{X}$ & Linear bias of metal absorber& $U(-0.02, 0.0)$ \\ [0.5ex]
         \hline
         $\xi_0^\mathrm{TP}$ & Amplitude of transverse proximity effect & $U(0.0, 2.0)$ \\ [0.5ex]  
         \hline
         $A_\mathrm{sky}^{\mathrm{Ly}\alpha}$, $A_\mathrm{sky}^{\mathrm{Ly}\beta}$ & Gaussian amplitudes modelling contamination due to sky subtraction & $U(0.0, 0.1)$ \\ [0.5ex]  
         \hline
         $\sigma_\mathrm{sky}^{\mathrm{Ly}\alpha}$, $\sigma_\mathrm{sky}^{\mathrm{Ly}\beta}$ & Gaussian standard deviations modelling contamination due to sky subtraction & $U(0.0, 60.0)$ \\ [0.5ex]  
         \hline
    \end{tabular}
    \caption{List of the free parameters in our baseline model, along with their description and priors. $U(\text{min,max})$ represents a flat prior within that interval, while $\mathcal{N}(\mu, \sigma^2)$ represents a Gaussian prior. The metal lines we model include $\mathrm{SiII}(1190)$, $\mathrm{SiII}(1193)$, $\mathrm{SiIII}(1207)$, $\mathrm{SiII}(1260)$, and $\mathrm{CIV}$. See \cite{duMasdesBourboux:2020,Cuceu:2022} for detailed descriptions of these parameters.}
    \label{tab:priors}
\end{table}

\begin{figure}
\includegraphics[width=1\linewidth]{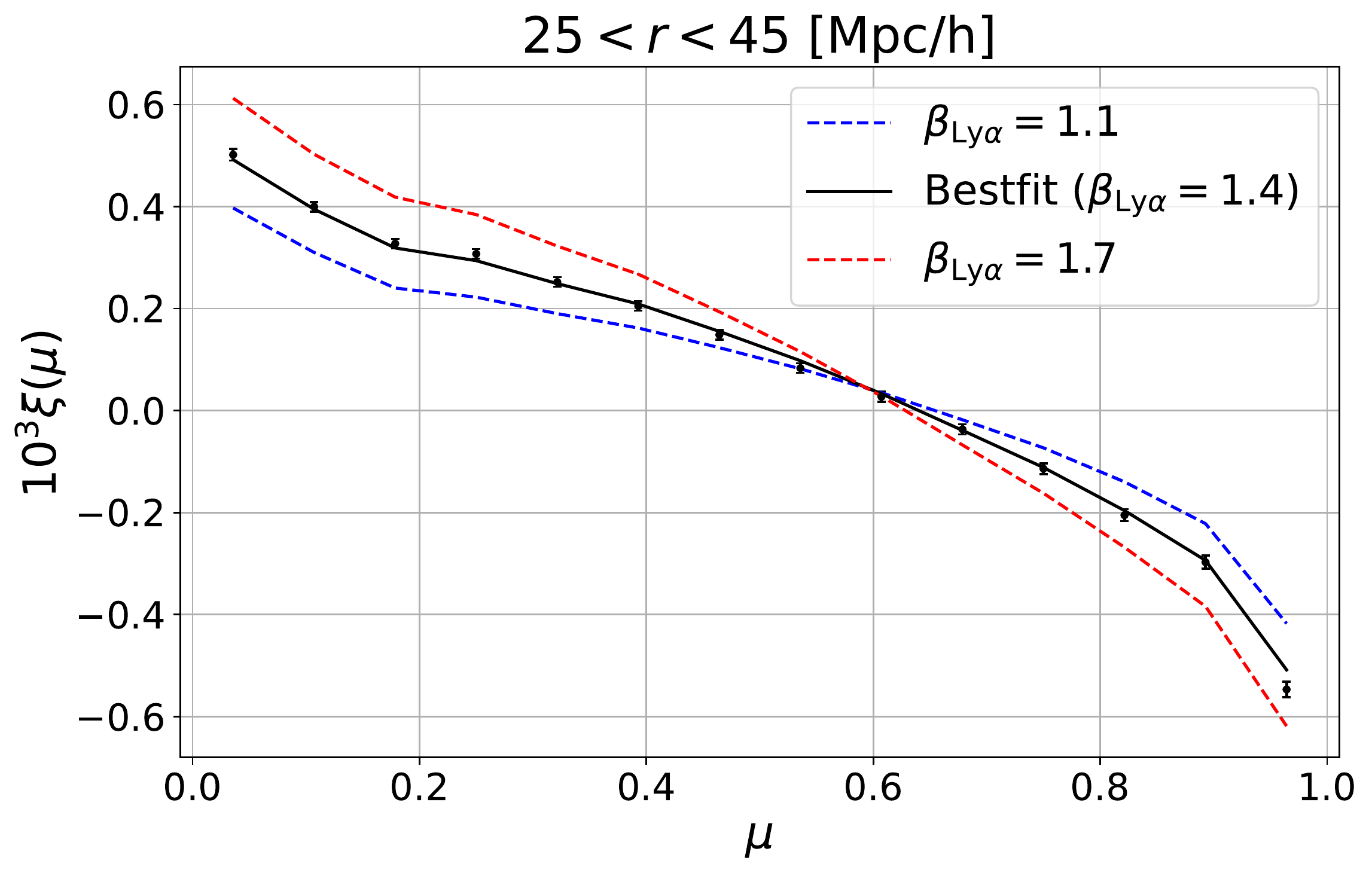}
\caption{The same as Figure 2 in the main text, but showing two models (red and blue) with different values of the redshift space distortion (RSD) parameter $\beta_{\mathrm{Ly}\alpha}$. This shows that RSD have a distinct impact on $\xi(\mu)$ when compared to AP (see main text), explaining why the two effects are not degenerate.}
\label{fig:shell_beta}
\end{figure}

To illustrate why the Alcock-Paczy\'nski (AP) effect is not degenerate with redshift space distortions (RSD), we show the impact of the RSD parameter $\beta_{\mathrm{Ly}\alpha}$ on a shell of the \lya\ auto-correlation function in \Cref{fig:shell_beta}. When compared with Figure 2 in the main text, which shows the impact of changing \phif, it is clear that RSD and the AP effect induce different $\mu$ dependencies in the correlation function. Therefore, the two effects can be disentangled, explaining our tight AP constraints.

\begin{figure*}
\includegraphics[width=0.91\linewidth]{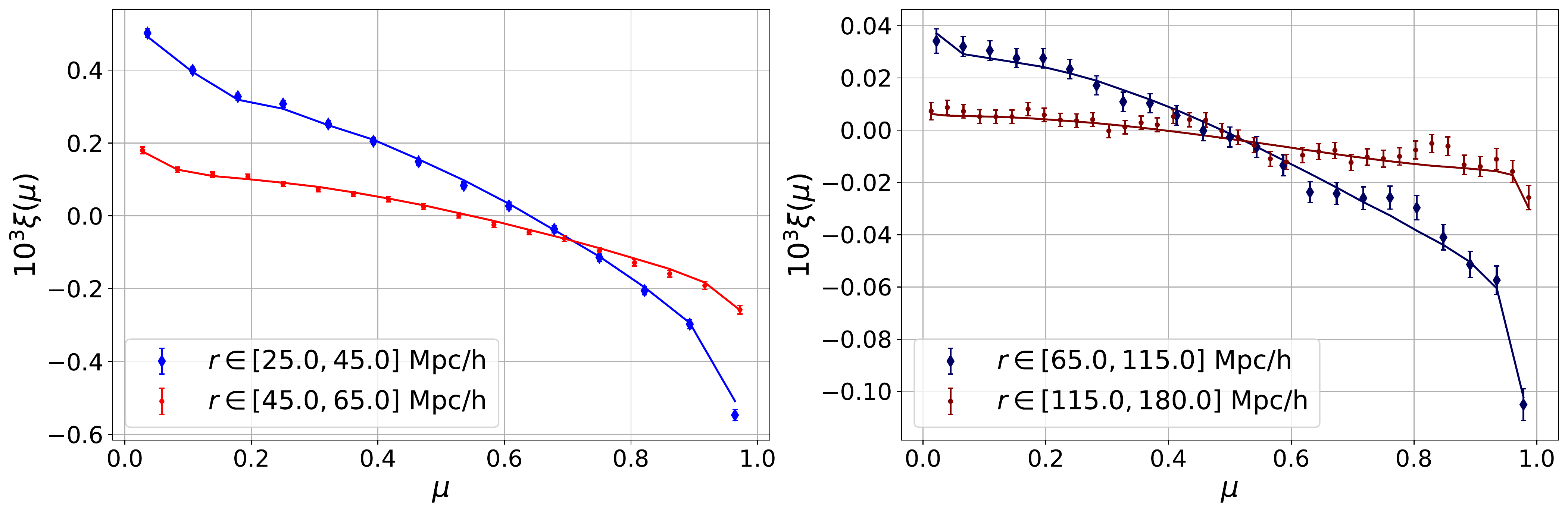}
\caption{\lyaf\ auto-correlation function (points with errorbars) compressed into shells in isotropic separation $r$, and shown as a function of $\mu$. This shows that the model fits $\xi(\mu)$ well, especially at small separations (left panel) where most of the AP information comes from.}
\label{fig:r_shells}
\end{figure*}

For completeness, in \Cref{fig:r_shells} we also compare the \lyalyalyalya\ measurement with the best fit model in shells of larger isotropic separation. This shows that the baseline model provides a good fit of the correlation as a function of $\mu$.

\section{Analysis validation}

In order to validate the full-shape Alcock-Paczy\'nski analysis, we performed a detailed study of systematic errors using a set of one hundred synthetic realizations of the eBOSS survey in a companion work \cite{Cuceu:2022}. We found that we can recover unbiased measurements of AP with our baseline model, when using a minimum scale $r_\mathrm{min}=25$\hMpc. In \cite{Cuceu:2022}, we also discussed the realism of the synthetic datasets, as well as which other effects currently not included in the mocks warrant further attention. Here, we briefly discuss how imperfections in the mock datasets could impact our analysis, while in the next section we perform a series of robustness tests meant to check our sensitivity to different modelling options that could not be tested with the synthetic datasets. See \cite{Cuceu:2022} for more details, and \cite{Ramirez:2022,Farr:2020} for a detailed description of the \lyaf\ synthetic datasets.

As described in the main text, the main sources of contamination for the \lyaf\ correlations are high column density (HCD) absorbers, metal absorbers, and the distortion due to quasar continuum fitting errors. Furthermore, quasar redshift errors are also a significant source of uncertainty for the \lya-QSO cross-correlation, which we discuss in the next section. Focusing on the first three effects, the distortion due to quasar continuum fitting is introduced by our analysis method, and therefore not directly affected by imperfections in the synthetic data. On the other hand, HCD and metal contamination are both astrophysical effects and imperfections in their modelling could impact our analysis.

On the HCD side, we measure an HCD bias parameter consistent with 0 (no HCD contamination) within the $1\sigma$ bounds when only fitting scales down to $r_\mathrm{min}=25$\hMpc. This is in contrast to BAO analyses which have been fitting down to $r_\mathrm{min}=10$\hMpc, and found strong evidence ($>10\sigma$) of HCD contamination \cite{Bautista:2017,deSainteAgathe:2019,duMasdesBourboux:2020}. Therefore, this indicates that at the level of eBOSS, unmasked HCDs only have a detectable effect on scales smaller than the smallest scale fitted in our analysis. In the next section we further test our sensitivity to HCD contamination by allowing the parameter that represents the typical scale of unmasked HCD systems to vary, and find it has no impact on our measurement.

The metal contamination introduces visible peaks in the correlation function along the line of sight (see Figure 1 in the main text). The most important ones for our analysis are at small scales, specifically the SiII(1190) and SiII(1193) lines that produce the blended peak at $\sim60$\hMpc, and the SiIII(1207) line that produces the peak at $\sim21$\hMpc\ \cite{Bautista:2017}. These peaks are due to the cross-correlations between the metals and \lya\ absorption. We model all these correlations using a linear theory model, similarly to the \lyaf\ model. However, we do not infer cosmology (through $\alpha$ and $\phi$) from these correlations. Rather, we simply marginalize over their effect by sampling a bias parameter for each of them. Nevertheless, due to the simplicity of the mocks, the performance of this approach, especially on small scales in \lya-Metal cross-correlations, could not be accurately tested. Therefore, we include two robustness tests in the next section to gauge our sensitivity to this effect. Finally, we note that using the synthetic datasets we found that even when not modelling this metal contamination (effectively ignoring their presence), we only recover a systematic bias of $\sim0.5\sigma$ at the level of eBOSS \cite{Cuceu:2022}.

\section{Robustness tests}

In order to test the robustness of our result, we perform a blind analysis in which we test a variety of different modelling options. Our approach is similar to those used by past \lya\ BAO analyses (e.g. \cite{duMasdesBourboux:2020}), but note that those analyses were not blinded. Our focus here is on testing modelling variations that could not be tested with the synthetic data, either because those effects were not modelled in the mocks, or because their modelling was too simplistic. These tests allow us to understand how sensitive we are to potential imperfections in the model.

We perform robustness tests on \phis, the anisotropic parameter of the smooth component, because this parameter captures the new information we wish to measure. Furthermore, robustness tests on \alphap\ and \phip\ were done by \cite{duMasdesBourboux:2020} as part of their BAO analysis, using a different parametrization. In order to blind our \phis\ measurements, we added an unknown random number to \phis. This number was drawn from a uniform distribution centred on zero with width equal to three times the $68\%$ constraint on \phip\ obtained by \cite{duMasdesBourboux:2020}, which was the previous best measurement of AP at high redshift. We compare the blinded results obtained from our baseline analysis and alternative modelling options to test the robustness of our result to these choices.

\begin{figure}
\includegraphics[width=1.0\linewidth]{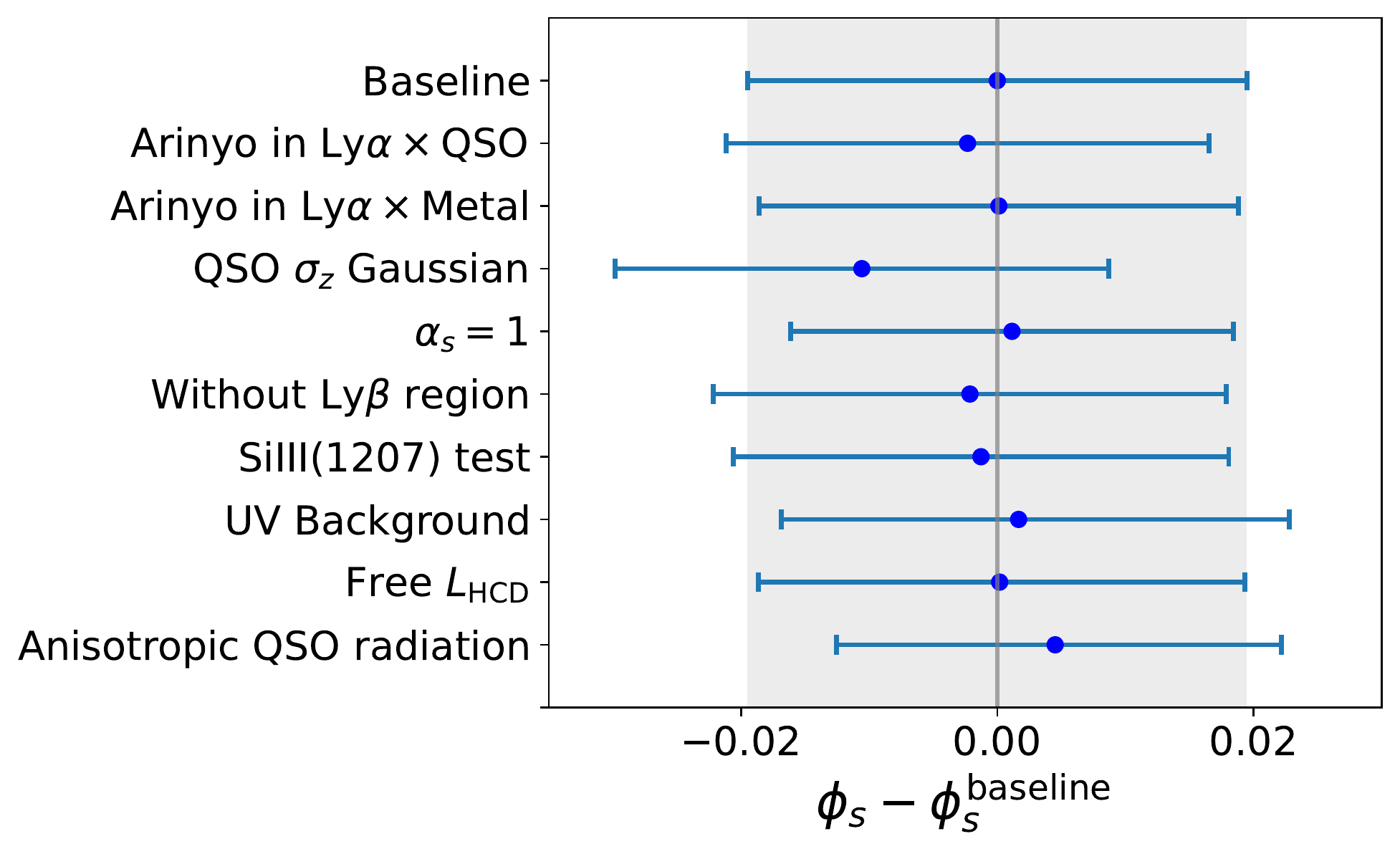}
\caption{The posterior value of \phis\ and the $68\%$ constraints under different modelling choices described in detail in the text, relative to the maximum posterior value of \phis\ obtained using the baseline model.}
\label{fig:bench}
\end{figure}

We present our results in \Cref{fig:bench}, where we show the $68\%$ constraints for each test, relative to the maximum posterior value of \phis\ obtained using the baseline model, $\phi_\mathrm{s}=1.037$. The first two tests expand the modelling of small scale non-linearities. For the \lya\ auto-correlation, we use a multiplicative correction proposed by \cite{Arinyo:2015} in order to model deviations from linear theory. This term was calibrated using hydrodynamic simulations by \cite{Arinyo:2015} and recently tested again by \cite{Givans:2022}. We refer to this correction as the Arinyo model. On the other hand, no such correction is included in the \lya-QSO cross-correlation, or in the \lya$\times$Metal correlations which model the metal contamination (see \cite{duMasdesBourboux:2020}). For the cross-correlation, this has been so far neglected because quasar redshift errors generally dominate on small scales \cite{Givans:2022}. Nevertheless, we add the Arinyo correction to the \lya-QSO cross-correlation and to \lya-metal correlations, and find that it does not have a significant impact on our result.

As noted above, the small scales of the \lya-QSO cross-correlation are dominated by quasar redshift errors. While quasar redshift errors were added to the synthetic data, they were drawn from a Gaussian distribution, whereas real quasar redshifts follow a more complex distribution with notably long tails \cite{Lyke:2020}. As in previous BAO analyses, our baseline model includes a Lorentzian damping term meant to model the redshift errors \cite{Percival:2009}, along with the finger-of-god (FOG) effect. In \cite{Cuceu:2022}, we found no significant difference between using this Lorentzian term or a Gaussian term when modelling QSO redshift errors and FOG in the mocks. However, the same test applied to eBOSS data results in $\sim0.5\sigma$ shift in the measured value of \phis. Note that the shift comes entirely from the cross-correlation, as the \lya\ auto-correlation is not affected. This indicates more attention is needed to the modelling of the small scales in the cross-correlation, and in particular future synthetic datasets should include realistic distributions of redshift errors. For the purposes of this work, we follow \cite{duMasdesBourboux:2020} and use the more realistic Lorentzian term in our baseline model.

As noted in the main text, our approach allows for two sets of $(\phi,\alpha)$ parameters, one for the BAO peak component, and one for the smooth (broadband) component. We extract cosmological information from \phis, \phip\ and \alphap, while \alphas\ is treated as a nuisance parameter, following \cite{Cuceu:2021}. In the baseline model we marginalize over \alphas\ due to its small correlation with \phis\ \cite{Cuceu:2021}. We could instead fix $\alpha_\mathrm{s}=1$, which would effectively set the broadband isotropic scale to that given by the Planck best fit cosmology. This does not produce a significant shift in the \phis\ measurement, but it does improve the precision by about $10\%$. However, the baseline approach is more robust, and therefore we marginalize over this parameter.

We use four correlation functions in our analysis. Combining the main auto-correlation, \lyalyalyalya, with the main cross-correlation, \lyalyaqso, is important for breaking parameter degeneracies, as noted by \cite{Cuceu:2021}. On the other hand, adding the two correlations using data from the \lyb\ region only results in a small improvement in precision \cite{Cuceu:2022}, while potentially adding new sources of contamination. We therefore test a measurement where we do not include these two correlations. The resulting \phis\ posterior is consistent with that obtained using all four correlations.

The result marked ``SiIII(1207) test" refers to whether we marginalize or fix the bias parameter corresponding to the SiIII(1207) line. This is relevant because this metal line produces a peak in the correlation function around a separation of $21$\hMpc, which is outside the range of scales we fit. In the baseline analysis we marginalize over this bias parameter, while in the test we instead fix its value to the best fit obtained by \cite{duMasdesBourboux:2020} where they fit scales down to $10$\hMpc. We find that this change does not have a significant impact on \phis.

In the baseline model we neglect the scale dependence of the \lya\ bias and RSD parameters, introduced by fluctuations in the ionizing UV background. This is because this effect has been tested in BAO analyses and not detected at a significant level \cite{Bautista:2017,deSainteAgathe:2019}. We follow previous works and include the model proposed by \cite{Gontcho:2014} for one of our tests. We find it has a negligible impact on the \phis\ posterior. Furthermore, we detect the effect at $<1.5\sigma$ significance, consistent with past results.

For the next test, we marginalize over the parameter $L_\mathrm{HCD}$, instead of fixing it to $10$\hMpc. This parameter is usually interpreted as the typical length scale of unmasked HCD systems, following the model introduced by \cite{deSainteAgathe:2019} based on the work of \cite{Rogers:2018}. We find that the \phis\ measurement is robust to this change.

Finally, we test a generalized version of the quasar ionizing flux model \cite{FontRibera:2013}. In the baseline analysis we use an isotropic model for this ionizing flux, following \cite{duMasdesBourboux:2020}. However, a more general version of this model that includes anisotropic and time-dependent emission has been tested in past BAO analyses \cite{duMasdesBourboux:2017}. We use the model introduced by \cite{duMasdesBourboux:2017} and sample two additional nuisance parameters representing the anisotropic and time-dependent emission, respectively. This results in a $\sim 10 \%$ tighter constraint on \phis\ that is consistent with the baseline measurement.

The tests performed here indicate that our \phis\ measurement is robust. The only noteworthy shift occurred when changing the model for quasar redshift errors. Furthermore, \cite{Youles:2022} recently found that large QSO redshift errors can also introduce spurious correlations in both the \lya\ auto and cross-correlation with quasars. Therefore, we conclude that this effect currently represents the largest source of systematic uncertainty for \lya\ full-shape analyses, and could become a significant problem for future datasets such as the Dark Energy Spectroscopic Instrument \cite{Abareshi:2022}.

\section{Fits of secondary correlations}

\begin{figure}
\includegraphics[width=1.0\linewidth]{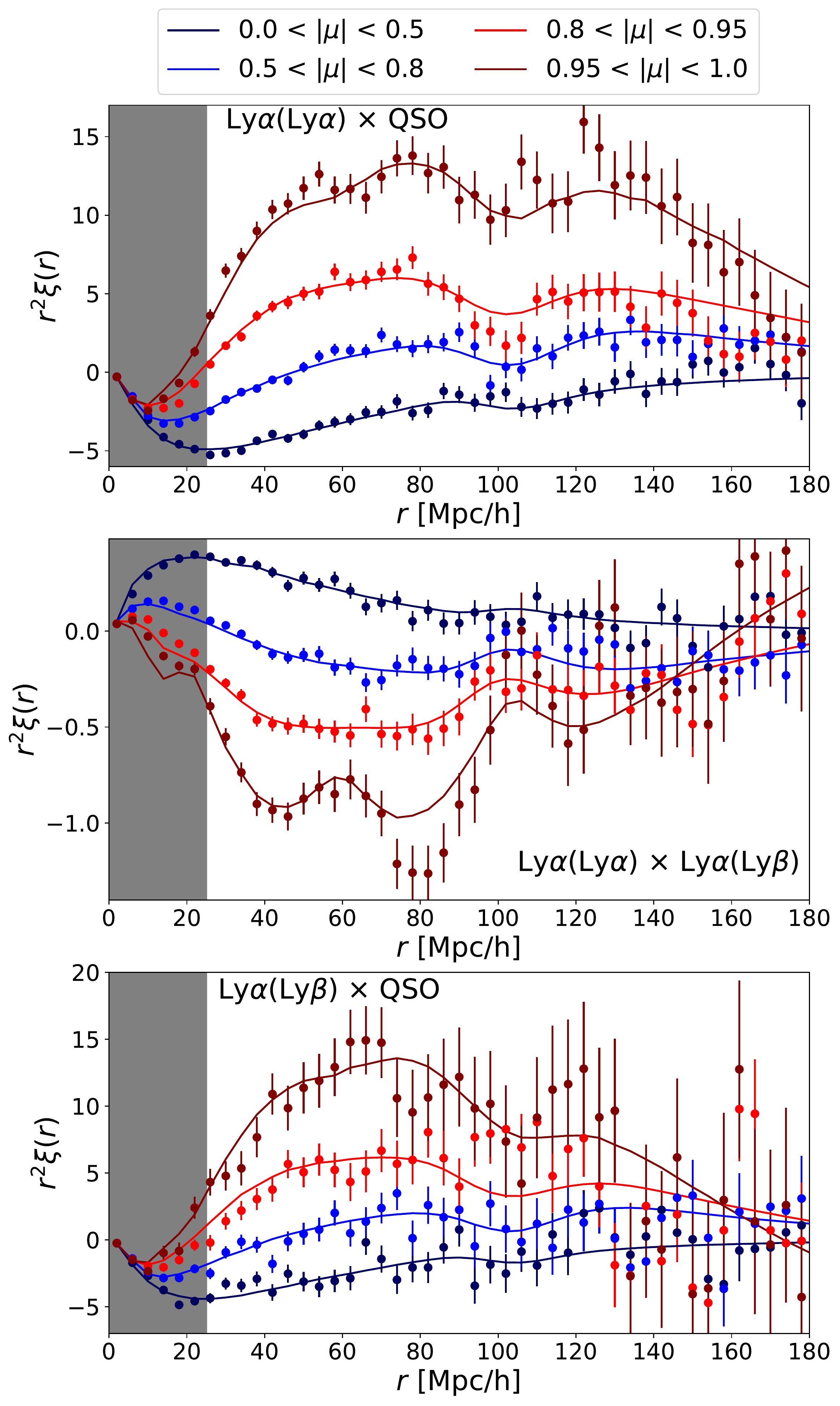}
\caption{The same as Figure 1 in the main text, but for the other 3 correlation functions. These include the cross-correlation between quasars and flux in the \lya\ region (top), the correlation of flux between the \lya\ and \lyb\ regions (middle), and the cross-correlation between quasars and flux in the \lyb\ region (bottom).}
\label{fig:other_corr}
\end{figure}

Our results are given by a joint fit of four correlation functions. The most important of these, the auto-correlation of flux in the \lya\ region, is shown in Figure 1 in the main text. In \Cref{fig:other_corr} here, we show the remaining 3 correlations in order of their contribution to the AP measurement from top to bottom \cite{Cuceu:2022}. The first is the cross-correlation between quasars and flux in the \lya\ region. The second is the correlation between \lya\ flux in the \lya\ and \lyb\ regions. The model for this is identical to the model for the main \lya\ auto-correlation \cite{duMasdesBourboux:2020,Cuceu:2022}. The final correlation is that between quasars and flux in the \lyb\ region. The model for this is identical to the model for the first \lya-QSO cross-correlation.

\section*{Data Availability}

The likelihood we used to perform cosmological inference is publicly available at \url{https://github.com/andreicuceu/cobaya_likelihoods}. This is made for the \texttt{Cobaya} package \cite{Torrado:2019,Torrado:2021}, but it can also be adapted to other cosmological inference frameworks. The \texttt{Vega} package for modelling and fitting \lya\ correlations is publicly available at \url{https://github.com/andreicuceu/vega}, and the eBOSS \lyaf\ correlation functions used in this work are publicly available at \url{https://svn.sdss.org/public/data/eboss/DR16cosmo/tags/v1_0_1/dataveccov/lya_forest/}. Other data supporting this research is available on request from the corresponding author.

\end{document}